\documentclass[12pt,a4paper]{article} 
\usepackage{jheppub} 
\usepackage{amsmath,amssymb}
\usepackage{hhline}
\usepackage{mathrsfs} 
\usepackage{longtable}
\usepackage{verbatim}
\usepackage{graphicx}
\usepackage{upgreek}
\DeclareMathAlphabet{\mathpzc}{OT1}{pzc}{m}{it}
\usepackage[small]{caption} 
\usepackage{color}
\usepackage{psfrag}
\usepackage{bm}

\newcommand{\al}{\left|}
\newcommand{\ar}{\right|}
\newcommand{\rr}{\right)}
\newcommand{\rl}{\left(}
\newcommand{\ccl}{\left[}
\newcommand{\ccr}{\right]}
\newcommand{\dd}{\partial}

\newcommand{\be}{\begin{equation}}
\newcommand{\ee}{\end{equation}}
\newcommand{\bea}{\begin{array}}
\newcommand{\eea}{\end{array}}
\newcommand{\ba}{\begin{eqnarray}}
\newcommand{\ea}{\end{eqnarray}}

\newcommand{\SU}[1]{\ensuremath{\mathrm{SU}(#1)}}

\newcommand{\ZZ}{\mathbb{Z}_2}
\newcommand{\Lag}{\mathscr{L}}
\newcommand{\mat}{\mathcal{M}}

\def\etmis{\not\!\!E_T}
\def\ptmis{\not\!\!P_T}

\DeclareMathOperator{\diag}{diag}

\numberwithin{equation}{section}
\numberwithin{table}{section}
\allowdisplaybreaks

\title{Discovering Minimal Universal Extra Dimensions (MUED) at the LHC}
\author[a,b]{Alexander Belyaev}
\author[a]{Matthew Brown}
\author[c]{Jes\'us M. Moreno}
\author[c]{Chlo\'e~Papineau}

\affiliation[a]{School of Physics and Astronomy, University of Southampton \\ Highfield, Southampton, SO17 1BJ, UK}

\affiliation[b]{Particle Physics Department, Rutherford Appleton Laboratory \\ Chilton, Didcot, Oxon, OX11 0QX, UK}

\affiliation[c]{Instituto de F\'{\i}sica Te\'orica, IFT-UAM/CSIC \\
Nicolas Cabrera 15, UAM, Cantoblanco, 28049 Madrid, Spain}

\emailAdd{a.belyaev@soton.ac.uk}
\emailAdd{m.s.brown@soton.ac.uk}
\emailAdd{jesus.moreno@csic.es}
\emailAdd{chloe.papineau@gmail.com}

\abstract{In this work we discuss our consistent implementation of the minimal model of Universal Extra Dimensions in CalcHEP.  We pay special  attention to the gauge invariance issues that  arise due to the incorporation of 5D quantum corrections. After validating the implementation we perform a complete study of the tri-lepton signature, including a realistic estimate of the backgrounds,  for the present LHC energy and luminosity. We also derive the expected LHC discovery reach for different luminosities, both at  $\sqrt{s}= 7$~TeV and 8~TeV.}

\arxivnumber{1212.4858}

\begin{document}

\begin{flushright}
\normalsize{IFT-UAM/CSIC-12-102}\\
\end{flushright}
\vskip 2cm
\maketitle

\section{Introduction}
\label{sec:intro}

Many beyond the Standard Model (SM) theories rely on the existence of extra spatial dimensions. For instance, the best candidate to unify  gravity and gauge interactions, string theory~\cite{Green:110204}, is only consistent in a ten dimensional space-time. On the other hand, the presence of extra dimensions could also have an impact on scales much lower than the Planck mass~\cite{ArkaniHamed:1998rs} (and this can be consistent with string theory~\cite{Antoniadis:1998ig}). In particular, extra dimensions provide a new perspective from which to address issues such as the nature of electroweak symmetry breaking,  the origin of the number of fermion generations~\cite{Dobrescu:2001ae} and their mass hierarchy~\cite{ArkaniHamed:1999dc}, and the supersymmetry breaking mechanism~\cite{ArkaniHamed:1999pv}. In any  case, the underlying physics will depend on several facts such as the size and the shape of the extra dimensions, the existence of space-time subspaces in which the different fields propagate, the presence of background fields, etc.  Let us concentrate on the size of these extra dimensions. Although present data constraint them, they could be as large as ${\cal O} (\text{TeV})^{-1}$. In this case they would be accessible at the Large Hadron Collider (LHC), which is already probing this scale. Furthermore, since this is also the typical mass of a weak interacting massive particle (WIMP), such models could be relevant for dark matter purposes.

The simplest of these models is the Universal Extra Dimension (UED) scenario, which was proposed a decade ago by Appelquist, Cheng and Dobrescu~\cite{Appelquist:2000nn}. In this model, the geometry consists of  the ordinary flat 4D space-time and an additional fifth dimension compactified on a circle of radius $R$. In its minimal version (MUED, defined in ref.~\cite{Cheng:2002ab}), the SM particles are promoted to 5D fields which propagate in the whole space-time. Notice that since gauge couplings are not dimensionless,  the model is non-renormalisable.
Therefore  it should be regarded as an effective theory valid up to some cut-off $\Lambda$, above which new degrees of freedom that complete the theory have to be incorporated. Since the extra dimension is compact, for small values of the radius R the theory may be reduced to an effective four-dimensional theory whose light spectrum contains the SM. The precise SM  spectrum is obtained by imposing a $Z_2$ symmetry that breaks the translational invariance along the fifth dimension and projects out half of the 4D fermions, allowing for chirality. This symmetry does not break Kaluza-Klein (KK) parity, a discrete symmetry that ensures the stability of the Lightest KK  Particle (LKP), the potential dark matter candidate of the model.

The spectrum is organised in towers of KK modes that get a common mass contribution proportional to their KK number. The degeneracy of these KK masses is lifted by the breaking of the electroweak symmetry and by the breaking of 5D Lorentz invariance caused by the compactification. It is possible in principle to introduce explicitly Lorentz-violating terms to the Lagrangian (such as terms localised at the boundaries of the compact dimension, or terms that are non-local in the compact coordinate). However, in MUED~\cite{Cheng:2002ab}, these terms are assumed to vanish at the theory's cuttoff scale. Even so, such terms will be induced again at lower scales by 5D quantum corrections.

In fact, a detailed evaluation of the 4D spectrum is crucial to explore the phenomenology of the model. From one side, it determines the sequence and the  kinematics of the  KK chain decays that will be produced at the LHC. On the other side, it  enters in the evaluation of  the LKP cosmological abundance~\cite{Servant:2002aq}, in which coannihilation processes are expected to play an important role.

It is then important to deal simultaneously with  the breaking of  both gauge and space-time symmetries and to understand their interplay with  quantum corrections. This question has been addressed in several works. In particular, analytic expressions for the mass spectrum without including radiative corrections were obtained in ref.~\cite{Muck:2001yv}. 
The effect of 5D quantum corrections was first evaluated in ref.~\cite{Cheng:2002iz}  in an approximation such that  the effects which involve both electro-weak symmetry breaking and radiative corrections were ignored. Since these terms are not gauge invariant, these approach is not fully satisfactory. A further step along this lines was made in ref.~\cite{Belanger:2010yx}, 
where the 5D corrections (both bulk and brane) were incorporated as a series of  wave function renormalisation factors  in the 4D expansion before considering the electroweak symmetry breaking. The consistency of this procedure has to be analysed since some of these terms cannot be derived from a  local 5D Lagrangian.

The aim of this work is to provide a consistent description of MUED, implement it in CalcHEP (a software package~\cite{Belyaev:2012qa} for calculating cross-sections and generating events) and
explore the signatures of this model to study its discovery potential at the LHC.

We have documented our implementation, collecting all the expressions 
(complete Lagrangian terms, masses and mixing values after EWSB, etc.) and provided detailed tables describing the notation used for all particles and parameters. We have also made the  model publicly available at the \emph{High Energy Model DataBase} (HEPMDB) in both Feynman-'t Hooft  and unitary gauges at \href{https://hepmdb.soton.ac.uk/}{https://hepmdb.soton.ac.uk/} under the name ``MUED-BBMP''
and have used the power of HEPMDB to perform scans over the MUED parameter space.
Using this setup we have studied the LHC potential to explore the MUED parameter space 
using a tri-lepton signature, which we found to be one of the most promising signatures sensitive to MUED.
There are several reasons for this. 
One is related to the compressed spectrum of MUED which leads to soft particles in the final state,
among which the leptons are the most convenient objects to be identified and dealt with.
Another reason for the advantage of the leptonic signatures, including the tri-lepton one, is 
the specific mass hierarchy of the KK particles in MUED which  provides
a large branching ratio of leptonic decays  from initial KK states; we 
demonstrate this below. Finally, the tri-lepton signature has quite a high signal-to-background ratio.
In addition, we have identified optimal kinematical cuts that further enhance 
the  statistical significance of the signal.
The results obtained are discussed in this article and compared with the previous studies~\cite{Bhattacharyya:2009br,Choudhury:2009kz,Datta:2011vg,Bhattacherjee:2010vm,Murayama:2011hj} devoted to the MUED collider phenomenology.
The analysis is done in the region compatible with present indirect constraints (electroweak precision measurements~\cite{Gogoladze:2006br},
 Higgs boson production and decay data~\cite{Nishiwaki:2011gk,Belanger:2012mc} and Dark Matter~\cite{Belanger:2010yx}).

The detailed content of the paper is the following: in section~\ref{sec:mued} we present our theoretical framework and provide the detailed MUED Lagrangian including the 5D quantum corrections. We also describe the spectrum for two benchmark points. In section~\ref{sec:imple} we  deal with the CalcHEP model implementation and its validation. The  phenomenology of the model is  presented  in section~\ref{sec:pheno}. We study in detail there the tri-lepton signature, paying special attention to  the background evaluation, and make a realistic estimate of the  LHC discovery reach. In section~\ref{sec:conclu} we summarise our results.  In appendix~\ref{app:u1u1} we give the explicit relations for masses and mixings in the neutral bosonic sector,  while in appendix~\ref{app:fourgluon} we describe the implementation of four gluon vertices in CalcHEP.



\section{Theoretical framework}
\label{sec:mued}

In this section, we start by reviewing the most important features of the MUED model \cite{Appelquist:2000nn,Cheng:2002ab}. We then explain our formalism in detail and show that it reconciles gauge invariance, electroweak symmetry breaking and 5D loop corrections.


\subsection{The MUED model at tree level}
\label{subsec:muedtree}

In the MUED Model, all SM particles are promoted to 5D fields which propagate in the bulk of an $S^1/\ZZ$ orbifold\footnote{We denote the extra dimension by $y$ or $x^5$ and the 4D coordinates by $x^{\mu}$. The metric is $\diag (+, -, -, -, -)$. Under a $\ZZ$-transformation,  $y \rightarrow -y$.} with length $\pi R$. Because of the periodicity of $S^1$, the fields can be expanded in Fourier modes $\exp \left(\frac{i n y}{R}\right)$ with $n$ an integer. However, half of these states are projected out under the action of the orbifold. This procedure allows for fermion chirality. At the end of the day, the $y$-independent $n=0$ modes reproduce the 4D SM particle content while in turn each SM particle is associated to a tower of Kaluza-Klein (KK) partners.


\subsubsection{Decomposition of the fields}

Under a $\ZZ$-transformation, a scalar field $\phi$ can be even or odd, i.e $\phi \rl -y \rr = \pm \phi \rl y \rr$, because its kinetic term contains two derivatives. Therefore its expansion in KK modes is either 
\ba
&&\phi \rl x , y \rr \ = \ \frac{1}{\sqrt{\pi R}} \ccl \phi^{(0)} (x)  + \sqrt{2} \, \sum_{n \geq 1} \phi^{(n)} (x) \cos \frac{n y}{R}  \ccr \quad \text{if} \ \phi \  \text{is even} , \nonumber \\
\text{or} \quad  &&\phi \rl x , y \rr \ = \ \sqrt{\frac{2}{\pi R}} \, \sum_{n \geq 1} \phi^{(n)} (x) \sin \frac{n y}{R} \quad \quad \quad \quad \quad \quad \quad \quad \text{if} \ \phi \  \text{is odd} . \label{phikk}
\ea
We see that an odd scalar does not have a zero mode and hence is not suitable to represent the Higgs field. Furthermore, the zero mode $\phi^{(0)}$ of an even scalar field comes accompanied by the infinite tower of $\phi^{(n)}$ states. We note that the factor $1/\sqrt{\pi R}$ accounts for the dimensionality of the field in 5D and ensures that the KK fields are canonically normalised in 4D.

The decomposition of a fermion can be deduced from its kinetic term $\bar{\psi} \gamma^M \dd_M \psi$, which contains $\bar{\psi} \gamma^5 \dd_5 \psi$. This term is invariant under $\ZZ$ if $\psi (-y) = \pm \gamma^5 \psi (y)$. If the sign is $+$ (resp. $-$), then the zero mode of the fermion is right-handed (resp. left-handed).
\ba
&&\psi_R  =  \frac{1}{\sqrt{\pi R}} \ccl \psi_R^{(0)} (x)  + \sqrt{2} \, \sum_{n \geq 1} \rl P_R \psi_R^{(n)} (x) \cos \frac{n y}{R}  + P_L \psi_R^{(n)} (x) \sin \frac{n y}{R}  \rr \ccr \  \label{psiRkk} \\
&&\psi_L   =  \frac{1}{\sqrt{\pi R}} \ccl \psi_L^{(0)} (x)  + \sqrt{2} \, \sum_{n \geq 1} \rl P_L \psi_L^{(n)} (x) \cos \frac{n y}{R}  + P_R \psi_L^{(n)} (x) \sin \frac{n y}{R}  \rr \ccr \  \label{psiLkk}
\ea
where the subscript $L, R$ refers to the zero mode chirality, and $P_{L, R} = \frac{1\pm \gamma^5}{2}$. Note that all the $\psi_{L, R}^{(n)}$ are Dirac fermions while the zero mode is chiral. Thus, there are four chiralities per KK level : $\psi_{L_L}^{(n)}$, $\psi_{L_R}^{(n)}$, $\psi_{R_L}^{(n)}$ and $\psi_{R_R}^{(n)}$. Only when necessary will we specify these chiralities. Otherwise, it should be understood that $\psi_{L, \, R}^{(n)}$ with $n \geq 1$ stands for a Dirac fermion.

Finally, a gauge boson $A_M$, $M=(\mu, 5)$, couples to a fermion via the term $\bar{\psi} \gamma^M D_M \psi \supset \bar{\psi} \gamma^M A_M \psi$. Consequently, $A_{\mu}$ is even while $A_5$ is odd
\ba
&&A_{\mu} \rl x , y \rr =  \frac{1}{\sqrt{\pi R}} \ccl A_{\mu}^{(0)} (x)  + \sqrt{2} \, \sum_{n \geq 1} A_{\mu}^{(n)} \cos \frac{n y}{R}  \ccr \ , \nonumber \\
&&A_5 \rl x , y \rr  =  \sqrt{\frac{2}{\pi R}} \, \sum_{n \geq 1} A_5^{(n)} (x) \sin \frac{n y}{R} \ . \label{Akk}
\ea


\subsubsection{The MUED Lagrangian}

It is simply obtained by embedding the SM Lagrangian into the 5D geometry:
\ba
\Lag_5 = && - \frac{1}{4} G^a_{MN}G^{a \, MN} - \frac{1}{4} B_{MN}B^{MN} - \frac{1}{4} W^i_{MN}W^{i \, MN} + \rl D_M H \rr^{\dagger} \rl D^M H \rr \nonumber \\
&&+ i \, \overline Q \Gamma^M D_M Q + i \, \overline L \Gamma^M D_M L + i \, \overline u \Gamma^M D_M u + i \, \overline d \Gamma^M D_M d + i \, \overline e \Gamma^M D_M e \nonumber \\
&&  + \mu^2 H^{\dagger} H - \lambda_H \rl H^{\dagger} H \rr^2 - \rl y_u \overline Q u \widetilde H + y_d \overline Q d H + y_e \overline L e H + \text{h.c.} \rr \ , \label{Ltree}
\ea
where $G$, $W$ and $B$ are the gauge bosons of respectively $\SU3$, $\SU2$ and ${\rm U}(1)$; $H$ is the Higgs field; $Q$ and $L$ are the quark and lepton doublets; and $u$, $d$ and $e$ are the up-type quarks, down-type quarks and electron-like singlets. All fields should have an understood $(x^{\mu},y)$ dependence and there is an implicit sum over fermion generations. The Lagrangian is defined at the cutoff scale, where Lorentz-violating terms are assumed to vanish.

The Dirac matrices are $\Gamma^M = \rl {\gamma^{\mu}, i \gamma^5} \rr$ and the covariant derivative is
\be
D_M \ = \ \dd_M - i \, Y g_1^{(5)} B_M - i \, g_2^{(5)} \frac{\sigma^i}{2} W^i_M - i \, g_3^{(5)} t^a G^a_M \ , \nonumber
\ee
with $Y$ the hypercharge, $\sigma^i$ the Pauli matrices, $t^a$ the Gell-Mann matrices and where the 5D gauge couplings are related to the 4D ones by $g^{(5)}_i = g_i  \sqrt{\pi R}$. Finally, the factors $y_u$, $y_d$ and $y_e$ are the Yukawa matrices and $\widetilde H = i \sigma^2 H^*$.

In principle, since the compactification explicitly breaks Lorentz invariance, one could introduce Lorentz-violating, yet gauge-invariant, terms into the above Lagrangian. This could be in the form of terms localised on the boundaries of the compact extra dimension (brane terms) or terms involving Wilson loops that are non-local in the $y$ coordinate (bulk terms). In the minimal model of UED \cite{Cheng:2002ab} these brane and bulk terms are assumed to be zero at the cutoff scale.

It is relatively straightforward to expand the various fields into KK modes and integrate the Lagrangian \eqref{Ltree} over the compact extra dimension in order to study the effective 4D theory: $\Lag_4 = \int_0^{\pi R} \Lag_5 (y) dy$. Due to the kinetic terms along the fifth dimension, the KK particles receive a geometrical contribution to their mass
\ba
m^2_{n} = m^2_{0} + \frac{n^2}{R^2} \label{mKK}
\ea
where $m_0$ stands for the mass of the corresponding zero mode, that is, the SM mass. This expression is modified when quantum corrections are included.


\subsection{MUED at one loop: mass corrections and gauge invariance}
\label{subsec:setup}

According to \eqref{mKK}, the lightest KK modes are the $n=1$ photon and gluon, which have the same mass. Other degenerate states are the two KK Dirac fermions associated with each quark or lepton. In fact, when the compactification scale $R^{-1}$ is much larger than the Higgs vev $v$, the entire spectrum at a given KK level is approximately degenerate. Under these circumstances, one expects radiative corrections to play an important role when studying the phenomenology of models with a small compact extra dimension\footnote{The importance of quantum corrections in this context was first pointed out by Hosotani  \cite{Hosotani:1983xw}.}.

Quantum corrections to the KK masses arise from the breaking of the 5D Lorentz invariance. They have been studied \cite{Cheng:2002iz,Georgi:2000ks,vonGersdorff:2002as,Puchwein:2003jq} in detail in the  $v \ll R^{-1}$ limit. There are two types of quantum corrections to the KK masses: the so-called bulk and brane contributions. The first ones are generic in a compactified theory since 5D Lorentz invariance is broken at long distances by the compactification. These  corrections are intrinsically non-local (in 5D), modify only the gauge boson masses and are independent of the KK level.

The second ones \cite{Georgi:2000ks} have their origin in the breaking of the translational symmetry at the orbifold fixed points. Brane  corrections are divergent in the cutoff scale $\Lambda$ and require brane-localised counter-terms to renormalise the action. In the effective 4D theory, these corrections are proportional to the usual Kaluza-Klein mass $\frac{n}{R}$ and depend logarithmically on $\Lambda$.

Summing both bulk and brane contributions, one has
\ba
&&\delta m^2_{\text{vector} \, (n)} \ = \  a_v \, \frac{n^2}{R^2} \ + \ b_v  \ , \nonumber  \\
&&\delta m_{\text{fermion} \, (n)} \ = \  a_f \, \frac{n}{R}  \ , \nonumber \\
&&\delta m^2_{\text{scalar} \, (n)} \ = \  a_s \, \frac{n^2}{R^2} \ , \label{deltam}
\ea
for $n \geq 1$, where the factors $a_v, \, a_f, \, a_s, \, b_v$ are dictated by the quantum numbers of the particle (see \cite{Cheng:2002iz} for the complete set of corrections). The first terms represent the brane corrections, while the constants $b_v$ stand for the bulk corrections. For example, for the quarks doublets, one has
\be
a_f \ = \ \frac{1}{16 \pi^2} \, \rl 3 g_3^2 + \frac{27}{16} g_2^2 + \frac{1}{16} g_1^2 \rr \, \ln  \frac{\Lambda^2}{\mu^2} \nonumber
\ee
where the $g_i$ are the gauge couplings for the different SM gauge groups, $\Lambda$ is the UV cutoff, and $\mu$ is the running scale. Comparatively, the bulk corrections for the ${\rm U}(1)$ gauge boson $B$ are
\ba
b_v \ = \ - \frac{39}{2} \, \frac{\zeta \rl 3 \rr \, g_1^2}{16 \pi^4} \, \frac{1}{R^2} \ . \nonumber \\ \nonumber
\ea

As mentioned in the Introduction, adding these corrected mass terms to the Lagrangian \eqref{Ltree} explicitly breaks gauge invariance in 5D. In the remainder of this section, we model the brane corrections with wave-function normalisation factors as described in \cite{Belanger:2010yx}, and proceed with the compactification. The strategy is to focus on the quadratic part of the Lagrangian in order to identify the mass eigenstates, the Goldstone bosons and Fadeev-Popov ghosts. Note that the quadratic part of the 5D Lagrangian is diagonal in the Kaluza-Klein basis, see e.g \cite{Muck:2001yv}. We comment on the bulk corrections at the end of this section.


\subsubsection{The case of a scalar field}

Before turning to the full MUED Lagrangian at one loop, let us consider a scalar field $\Phi$ in 5D. The key point of our setup is to recognise that the brane corrections are proportional to the KK mass $\frac{n^2}{R^2}$ and hence can be incorporated into the kinetic term along the fifth dimension. We introduce normalisation functions by splitting the 5D part from the 4D part in the Lagrangian
\be
\Lag_5 \ = \ \frac{1}{2} \, \al \dd_{\mu} \Phi \ar^2 - \frac{1}{2} \, Z_{\Phi} \, \al \dd_5 \Phi \ar^2  + m_0^2 \al \Phi \ar^2 \ . \label{L5phi}
\ee
Following section~\ref{subsec:muedtree}, we decompose the field in KK modes and integrate over the fifth dimension, $\Lag_4 = \int_0^{\pi R} \Lag_5 \rl y \rr dy$. In 4D, the spectrum consists in a tower of KK scalar fields $\phi^{(n)}$ with masses $m_n^2 = m_0^2 + Z_{\Phi} \frac{n^2}{R^2}$. It is straightforward to identify $Z_{\Phi} = 1 + a_s$ so that the total mass matches the tree-level mass \eqref{mKK} plus the brane correction \eqref{deltam}.

Let us comment on this procedure. First of all, the Lagrangian \eqref{L5phi} is not Lorentz covariant. However, since the orbifold breaks translational invariance in 5D, we emphasise that this Lagrangian is phenomenologically relevant as far as brane corrections go\footnote{Alternatively, this is equivalent to rescaling $R \rightarrow R/\sqrt{Z_{\Phi}}$.}. Also, in principle, loop diagrams of the ``brane" type yield KK-number violating mass mixings since $p_5$ is not conserved in these processes (see \cite{Cheng:2002iz}, eq. (34)). This means that in the 4D effective theory, the KK states are no longer the mass eigenstates. One should then diagonalise the infinite mass matrix. However, the off-diagonal KK-mixing terms in this matrix are small compared to the diagonal mass terms and so, to a very good approximation, we can disregard them and retain only KK-number preserving mass corrections. In 5D, this approximation manifests itself through the fact that our $Z_{\Phi}$ function is independent on the extra dimension, and hence KK number is conserved. In other words, instead of considering brane-localised counter-terms, we have effectively spread them throughout the compact dimension.

\subsubsection{The MUED Lagrangian at one loop}

The same idea of splitting the fifth direction from the Minkowski 4D can be applied to the whole Lagrangian \eqref{Ltree}. Whenever there is a contraction of Lorentz indices, that is for all the kinetic terms, we introduce a $Z$ function.

This procedure preserves gauge invariance. For instance, consider a field strength $F^{a \, MN}F^a_{MN}$. It is gauge invariant in 5D by construction. Due to its antisymmetry, it reduces to $F^a_{\mu \nu}F^{a\, \mu \nu} + 2 F^a_{\mu 5}F^{a\, \mu 5}$. Each of these two terms is clearly gauge invariant on its own, therefore splitting them explicitly in the Lagrangian and adding a factor $Z$ to the second one does not affect gauge invariance in 5D. The same applies to the Higgs and the fermion kinetic terms.

We are now ready to write our 5D MUED Lagrangian and analyse it:
\ba
\Lag_5 = && - \frac{1}{4} G^a_{\mu \nu}G^{a \, \mu \nu} - \frac{1}{4} B_{\mu \nu}B^{\mu \nu} - \frac{1}{4} W^i_{\mu \nu}W^{i \, \mu \nu}  + \rl D_{\mu} H \rr^{\dagger} \rl D^{\mu} H \rr\nonumber \\
&& + \frac{Z_G}{2} G^a_{\mu 5}G^{a \, \mu}_{\ \ 5} + \frac{Z_B}{2} B_{\mu 5}B^{\mu}_{\ 5} + \frac{Z_W}{2} W^i_{\mu 5}W^{i \, \mu}_{\ \ 5}  - Z_H \rl D_5 H \rr^{\dagger} \rl D_5 H \rr \nonumber \\
&&+ i \, \overline Q \gamma^{\mu} D_{\mu} Q + i \, \overline L \gamma^{\mu} D_{\mu} L  + i \, \overline u \gamma^{\mu} D_{\mu} u   + i \, \overline d \gamma^{\mu} D_{\mu} d + i \, \overline e \gamma^{\mu} D_{\mu} e \nonumber \\
&& - Z_Q \, \overline Q \gamma^5 D_5 Q - Z_L \, \overline L \gamma^{5} D_{5} L  - Z_u \,  \overline u \gamma^{5} D_{5} u   - Z_d \, \overline d \gamma^{5} D_{5} d - Z_e \, \overline e \gamma^{5} D_5 e  \nonumber \\
&& + \mu^2 H^{\dagger} H - \lambda_H \rl H^{\dagger} H \rr^2 - \rl y_u \overline Q u \widetilde H + y_d \overline Q d H + y_e \overline L e H + \text{h.c.} \rr  \ . \label{L1L}
\ea
In Ref.~\cite{Cheng:2002iz}, expressions are given for brane mass corrections (ignoring electroweak symmetry breaking) in their equation~(44). The $Z$ functions in the above Lagrangian can be found by simply matching
\[
m_{\text{boson}(n)}^2 = \left(\frac{n}{R}\right)^2 + \delta m_{\text{boson}(n)}^2 = Z \frac{n^2}{R^2}
\]
for bosons and
\[
m_{\text{fermion}(n)} = \frac{n}{R} + \delta m_{\text{fermion}(n)} = Z\frac{n}{R}
\]
for fermions (neglecting bulk corrections), which results in the following:
\begin{equation}
    \begin{split}
    Z_{Q} &= 1 + \left(3\frac{g_3^2}{16\pi^2}+ \frac{27}{16}\frac{g_2^2}{16\pi^2} + \frac{1}{16}\frac{g_1^2}{16\pi^2}\right) \ln\frac{\Lambda^2}{\mu^2} \\
    Z_{u} &= 1 + \left(3\frac{g_3^2}{16\pi^2}+ \frac{g_1^2}{16\pi^2}\right)  \ln\frac{\Lambda^2}{\mu^2} \\
    Z_{d} &= 1 + \left(3\frac{g_3^2}{16\pi^2}  + \frac{1}{4}\frac{g_1^2}{16\pi^2}\right)  \ln\frac{\Lambda^2}{\mu^2},\\
    Z_{L} &= 1 + \left(\frac{27}{16}\frac{g_2^2}{16\pi^2} + \frac{9}{16}\frac{g_1^2}{16\pi^2}\right)  \ln\frac{\Lambda^2}{\mu^2},\\
    Z_{e} &= 1 + \frac{9}{4}\frac{g_1^2}{16\pi^2}\ln\frac{\Lambda^2}{\mu^2},\\
    Z_{B} &= 1 + \left(-\frac{1}{6}\right)\frac{g_1^2}{16\pi^2}\ln\frac{\Lambda^2}{\mu^2},\\
    Z_{W} &= 1 + \frac{15}{2}\frac{g_2^2}{16\pi^2}\ln\frac{\Lambda^2}{\mu^2},\\
    Z_{G} &= 1 + \frac{23}{2}\frac{g_3^2}{16\pi^2}\ln\frac{\Lambda^2}{\mu^2},\\
    Z_{H} &= 1 + \left(\frac{3}{2}g_2^2   + \frac{3}{4} g_1^2 - \frac{m_H^2}{v^2} \right)\frac{1}{16\pi^2}\ln\frac{\Lambda^2}{\mu^2}, 
    \end{split}
\end{equation}
where $\Lambda$ is the 5D momentum cutoff (typically taken to be around $20R^{-1}$) and $\mu$ is the renormalisation scale.

We will focus on the quadratic part of the Lagrangian in order to identify the mass eigenstates of the 4D theory. Also note that the quadratic part of the Lagrangian is diagonal in the KK indices. Therefore, the strategy is the same as for MUED at tree level, see \cite{Muck:2001yv}: we KK expand the fields and perform the compactification. We then diagonalise the quadratic part of the Lagrangian in 4D for a given level $n$. After identifying the Goldstone bosons, we perform the gauge fixing and incorporate the Fadeev-Popov Lagrangian.


\paragraph{The bosonic sector\\}

For the reader interested in the details of the computation, an example of diagonalisation of the Lagrangian \eqref{L1L} is presented in appendix \ref{app:u1u1} where we consider the subgroup U(1)$\times$U(1) spontaneously broken by the Higgs. As shown therein, the identification of the Goldstone bosons is tedious, though not conceptually difficult, and the gauge fixing proceeds as usual. We list here the key results of the diagonalisation procedure.

Once the Higgs field acquires a vev, the bosons $B$ and $W^3$ mix. However, the KK states $B^{(n)}$ and $W^{3 \, (n)}$ mix with an angle that depends on $n$, $Z_B$ and $Z_W$. Consequently, the $n \geq 1$ mass eigenstates cannot be identified with a KK photon or a KK $Z$-boson. We call the mass eigenstates $P_{\mu}^{(n)}$ and $V_{\mu}^{(n)}$ and their masses are
\ba
m^2_{\left\{P, V\right\}(n)} &=&  \frac{1}{2} \left[ \frac{v^2}{4} \rl g_1^2 + g_2^2 \rr + \frac{n^2}{R^2}  \rl Z_B + Z_W \rr \right.  \nonumber \\ &\quad& \quad \quad \left. \mp \sqrt{ \rl \frac{v^2}{4} \rl g_2^2 - g_1^2 \rr + \frac{n^2}{R^2}  \rl Z_W - Z_B \rr \rr^2 + \frac{1}{4} g_1^2 g_2^2 v^4} \, \right]  . \label{mPVn}
\ea
Likewise, the fifth components $B_5^{(n)}$, $W^{3 \, (n)}_{5}$ and the Higgs pseudo-scalar $\chi^{3 \, (n)}$ also mix. Two of the mass eigenstates are the Goldstone bosons associated with $P_{\mu}^{(n)}$ and $V_{\mu}^{(n)}$. The third scalar mass eigenstate is a Higgs neutral field $a_0^{(n)}$ which mass is
\be 
 m^2_{a_0(n)} = Z_H \ccl \frac{n^2}{R^2} + \frac{v^2}{4} \rl \frac{g_1^2}{Z_B} + \frac{g_2^2}{Z_W} \rr \, \ccr \label{ma0n}  .
\ee

On the other hand, the gauge bosons $W^1_M$ and $W^2_M$ give rise to a tower of KK states $W_{\mu}^{\pm \, (n)}$ since they are orthogonal to the breaking of $\SU2$. Their mass is simply 
\be
m^2_{W(n)} \ = \ m_W^2 \ + \ Z_W \frac{n^2}{R^2} \ . \label{mWn}
\ee
The associated Goldstone boson $G_W^{\pm \, (n)}$ is a combination of their fifth component $W^{\pm \, (n)}_5$ with the Higgs pseudo-scalars $\chi^{1 \, (n)}$ and $\chi^{2 \, (n)}$. The other linear combination is a physical charged Higgs scalar $a_{\pm}^{(n)}$ whose mass is
\be
m^2_{a(n)} \ = \ \frac{Z_H}{Z_W} \ m^2_{W(n)} \ . \label{mapmn}
\ee
The only remaining degree of freedom of the Higgs, which we call $H^{(n)}$, acquires a mass
\be
m^2_{H(n)} \ = \ m_H^2 \ + \ Z_H \, \frac{n^2}{R^2} \ . \label{mHn}
\ee
Its zero mode corresponds to the SM Higgs boson.

Finally, the gluon has a KK tower $G_{\mu}^{a \, (n)}$ with mass
\be
m^2_{G(n)} \ = \ Z_G \, \frac{n^2}{R^2} \ , \label{mGn}
\ee
and its associated Goldstone bosons are the fifth components $G_5^{a \, (n)}$.

Therefore, at the n$^{\rm th}$ KK level, there are five gauge bosons: $G_{\mu}^{(n)}$, $P_{\mu}^{(n)}$, $V_{\mu}^{(n)}$, $W_{\mu}^{\pm \, (n)}$; and four scalars: the Higgs $H^{(n)}$ and the three scalars $a_0^{(n)}$ and $a_{\pm}^{(n)}$. We point out that among the scalars, only the Higgs field possesses a zero mode.

Once we have identified the Goldstone bosons of the theory, the gauge-fixing Lagrangian is determined very easily, see for example the end of appendix \ref{app:u1u1}. The next step is to include the ghosts fields. As an illustrative example, let us focus on the $\SU3$ sector. The 4D gauge-fixing Lagrangian
\be
\Lag_{GF}^{(n)} =  - \frac{1}{2 \xi} \, \rl \dd^{\mu} G_{\mu}^{a \, (n)} - \xi Z_G \dd_5 G_5^{a \, (n)} \rr^2 \nonumber
\ee
can be promoted to a 5D Lagrangian
\be
\Lag_{GF} = - \frac{1}{2 \xi} \, \rl \dd^{\mu} G_{\mu}^{a} - \xi Z_G \dd_5 G_5^{a} \rr^2 = - \frac{1}{2 \xi} F^a_G F^a_G  \ ,\label{LGFGn}
\ee
where we have introduced the gauge-fixing functions $F^a_G$. Because this Lagrangian is quadratic, it cannot give rise to any KK-number mixing. The next step is to introduce the 5D ghosts fields $c^a$ and ${\overline c}^a$, which are $\ZZ$-even. The Fadeev-Popov Lagrangian is
\be
\Lag_{FP} \ = \ - {\overline c}^a \frac{\delta F^a_G}{\delta \alpha^b} c^b \nonumber
\ee
where $\alpha^a$ is the parameter of the gauge transformation. We find
\be
\Lag_{FP} =  {\overline c}^a \ccl - \dd^{\mu} \dd_{\mu} + \xi Z_G \dd_5^2 \ccr c^a - g_3^{(5)} f^{abc} \dd^{\mu}  {\overline c}^a G_{\mu}^c c^b + g_3^{(5)} \xi Z_G f^{abc} \dd_5  {\overline c}^a G_{5}^c c^b \ . \nonumber
\ee
For the $\SU2 \times$U(1) sector of the theory, particular care has to be taken since the symmetry is spontaneously broken. We emphasise however that the Fadeev-Popov procedure in this case does not bring any additional complication compared to the SM.


\paragraph{The fermionic sector\\}

We now proceed to the diagonalisation of the fermionic sector of the theory. The left and right fermions do not receive the same corrections since they are respectively doublets and singlets of $\SU2$.
Recall that each fermion of the SM has two associated infinite towers of KK Dirac fermions $\psi_R^{(n)}$ and $\psi_L^{(n)}$. After electroweak symmetry breaking, they mix with the following matrix
\ba
&&\begin{pmatrix}  -Z_R \frac{n}{R} & \quad m_{\psi} \cr \cr  m_{\psi} & \quad Z_L \frac{n}{R}  \end{pmatrix}  \label{massmixU}
\ea
where $m_{\psi}$ is the zero-mode mass and we labelled the left and right radiative corrections as $Z_{L, \, R}$. At the n$^{\text{th}}$ KK level, the mixing is
\be
\begin{pmatrix} \psi_R^{(n)} \cr \cr \psi_L^{(n)} \end{pmatrix}  \ = \ \begin{pmatrix} - \gamma^5 \cos \alpha_{\psi(n)} & \sin \alpha_{\psi(n)} \cr \cr \gamma^5 \sin \alpha_{\psi(n)} & \cos \alpha_{\psi(n)} \end{pmatrix} \begin{pmatrix}  \psi_1^{(n)} \cr \cr \psi_2^{(n)} \end{pmatrix}  \ , \label{mixU}
\ee
with 
\be
\tan \rl 2 \alpha_{\psi(n)} \rr \ = \ \frac{2 m_\psi }{  \rl Z_L + Z_R \rr n/R } \quad ,   \quad \text{for} \  n \geq 1 \ , \label{alphaU}
\ee
and where $\psi_i^{(n)}$ are the mass eigenstates after EWSB, with masses
\be
m_{\left\{1, \, 2 \right\} (n)} \ = \ \frac{1}{2}  \ccl \, \frac{Z_L + Z_R}{\cos \rl 2 \alpha_{\psi(n)} \rr} \mp \rl Z_L - Z_R \rr \, \ccr \, \frac{n}{R} \ . \label{m12ferm}
\ee
The same applies to all fermions except for the neutrinos.

Let us briefly comment on the corrections to the top quark mass. Because the Yukawa coupling is large, the top receives an additional correction through the Higgs, see \cite{Cheng:2002iz}. These corrections can be included as extra contributions to the wave-function normalisation. However, because of gauge invariance, the left-handed bottom quark also receives a Higgs-induced correction. The corresponding terms in the Lagrangian are
\be
 - Z_{TL} \, \overline T \gamma^5 D_5 T  - Z_{tR} \, \overline t \gamma^5 D_5 t  - Z_{d} \, \overline b \gamma^5 D_5 b \ , \nonumber
\ee
where $Z_d$ is the common correction for right-handed quarks, and $Z_{TL}$ and $Z_{tR}$ contain the new contributions. Consequently, the bottom masses depend on $Z_{TL}$ instead of the naive expectation ($Z_Q$).

In summary, at the n$^{\rm th}$ KK level, there are two Dirac fermions labelled by an index ($1$ or $2$). Their masses are given by \eqref{m12ferm} and depend on the left and right corrections.


\subsubsection{Comments on the bulk corrections}

As mentioned at the beginning of this section, the bulk mass corrections arise when one of the internal lines of the 5D loop diagram winds around the extra dimension. They affect only the gauge boson masses and are independent of the KK number, that is, they shift the entire tower of KK states except for the zero mode. Interestingly, the bulk corrections computed in \cite{vonGersdorff:2002as, Cheng:2002iz} seem to be independent of the regularisation scheme \cite{Gies:2003ic, Alvarez:2006sf, Bauman:2007rt, Bauman:2008rr}. For instance, in latticised (or deconstructed) versions of 5D theories \cite{Hill:2000mu, Falkowski:2003iy, Kunszt:2004ps}, which are often considered UV complete, the same results were found.

Because they do not depend on the KK number, the bulk corrections cannot be incorporated within a wave-normalisation function in the manner of the previous paragraphs. As it turns out\footnote{We warmly thank A. Pomarol for very valuable discussion on this matter.}, they can be modelled by non-local operators (Wilson lines) in 5D. The reason is that since one internal line winds around the fifth dimension, such loops cannot be shrunk to a point and hence they are non-local from the 5D point of view (but local in the effective 4D theory). Even though we believe that this issue deserves further attention, it goes beyond the scope of this work to consider non-local operators.

Quantitatively, the bulk corrections are negligible. The results derived in \cite{Cheng:2002iz} are
\ba
&&\delta m^2_{B \, (n)} \ = \  - \frac{39}{2} \, \frac{\zeta \rl 3 \rr \, g_1^2}{16 \pi^4} \, \frac{1}{R^2}  \nonumber \\
&&\delta m^2_{W \, (n)} \ = \  - \frac{5}{2} \, \frac{\zeta \rl 3 \rr \, g_2^2}{16 \pi^4} \, \frac{1}{R^2} \quad \quad \quad \quad \quad \text{for} \ n \neq 0 \ . \nonumber \\
&&\delta m^2_{G \, (n)} \ = \  - \frac{3}{2} \, \frac{\zeta \rl 3 \rr \, g_3^2}{16 \pi^4} \, \frac{1}{R^2}  \nonumber 
\ea
Numerically, for all $n \geq 1$ and for phenomenologically acceptable values of $R$ ($\sim 1$ TeV), the bulk corrections are very small compared to the brane ones with the only exception of the $B^{(1)}$ boson. However, because $g_1$ is so small, it turns out that both brane and bulk corrections are in any case negligible compared to the tree-level mass $1/R^2$ for $B^{(1)}$. Therefore, in our implementation and phenomenological studies, we have disregarded the bulk corrections.

%


\subsection{Detailed analysis of the spectrum}
\label{subsec:spectrum}

\begin{figure}[h!t!]
    {
    \begin{center}
    \includegraphics[scale=0.285]{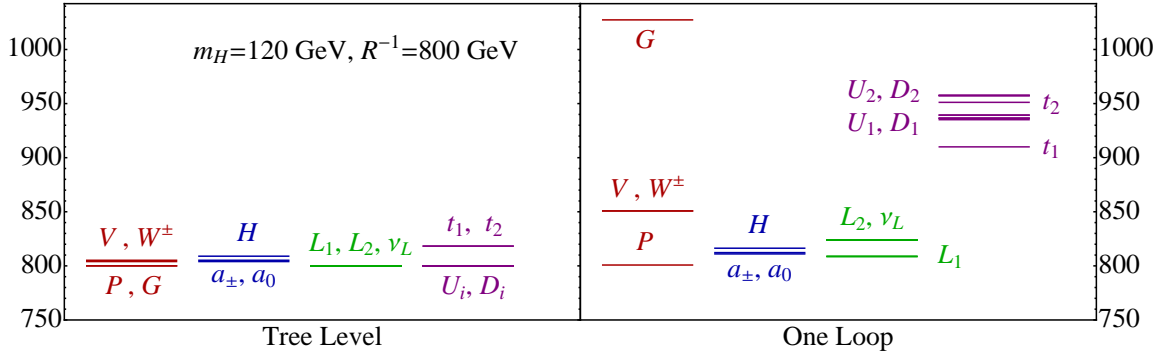}
     \end{center}
    }
    \vspace{-0.5cm}
  \caption{The first KK level of the MUED spectrum for $R^{-1} = 800$ GeV, $m_H = 120$ GeV, $\Lambda R= 20$ and  $\mu R= 1$, at tree level (left) and one loop (right).}
  \label{fig:spectrum800}
\end{figure}

\begin{figure}[h!t!]
    {
    \begin{center}
    \includegraphics[scale=0.285]{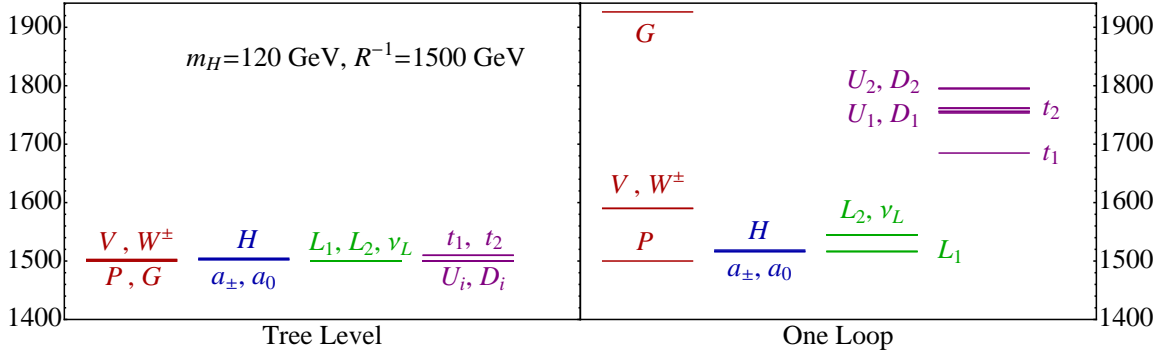}
     \end{center}
    }
    \vspace{-0.5cm}
  \caption{The first KK level of the MUED spectrum for $R^{-1} = 1500$ GeV, $m_H = 120$ GeV, $\Lambda R= 20$ and  $\mu R= 1$, at tree level (left) and one loop (right).}
  \label{fig:spectrum1500}
\end{figure}

Armed with the masses computed in section \ref{subsec:setup}, we are ready to plot the spectrum and analyse its features. For each 5D field, the $Z$ function is matched with the results obtained in \cite{Cheng:2002iz}. In order to efficiently compare our results with the existing literature, we fix the cutoff $\Lambda = 20 R^{-1}$ and the running scale $\mu = R^{-1}$. Furthermore, we choose two benchmark values for the compactification scale $R^{-1} = 800$ GeV and $R^{-1} = 1500$ GeV. As we will explain in section \ref{sec:pheno}, these points are close to the edge of the allowed region. The spectrum of the first KK level is shown in figures~\ref{fig:spectrum800} and \ref{fig:spectrum1500} for a Higgs mass $m_H = 120$ GeV and these two values of $R^{-1}$.

First of all, the degeneracy of the tree-level spectrum is significantly lifted once we include corrections. As expected, the lightest KK particle (LKP) is the gauge boson $P^{(1)}$ which is at $99\%$ composed of the $B^{(1)}$ boson because the electroweak scale is much smaller than the radius. It is stable and hence is a good dark matter candidate. The next-to-lightest KK particle (NLKP) for the parameters of figure~\ref{fig:spectrum800} is a charged KK lepton, while as seen in figure~\ref{fig:spectrum1500}, the KK lepton and the scalar $a_0$ are almost degenerate. Also note that the $V^{(1)}$ boson consists essentially of the $W^{3 \, (1)}$ boson and hence the three $\SU2$ bosons are still degenerate at one loop.

The heaviest KK state is the gluon because $g_3$ is large and the brane corrections are positive. This is also the reason why the quark masses are lifted by a greater amount than the leptons. The LHC being a hadron collider, as KK quarks and gluons are produced we expect them to decay into the several lighter KK species plus SM particles. Notice however that the corrections in the top quark masses by the Higgs are negative and proportional to the top Yukawa coupling thus pushing these masses down. Remarkably, for $R^{-1} = 1500$ GeV there is just enough gap for the second top $t_2^{(1)}$ to decay into a SM top plus $P^{(1)}$ while at $R^{-1} = 800$ GeV this decay is forbidden.
\begin{figure}[h!t!]
    {
    \begin{center}
     \includegraphics[width=0.8\textwidth]{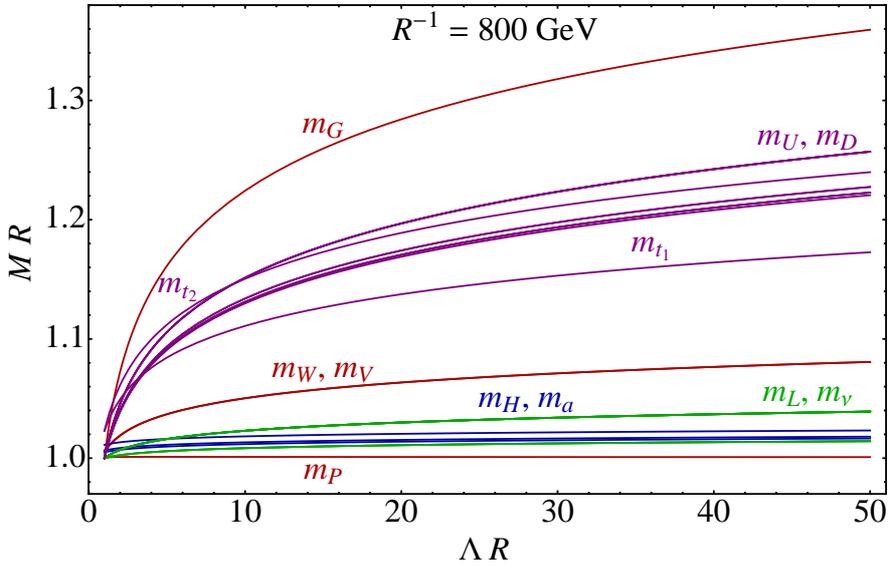}
     \end{center}
    }
    \vspace{-0.4cm}
      \caption{Evolution of the $n=1$ masses with the cutoff $\Lambda R$ at fixed $R^{-1} = 800$ GeV and $\mu R = 1$.}
\label{fig:lambda-dependence}
\end{figure}

It is interesting to study the dependence of the $n=1$ spectrum on the external parameters of the model, namely the cutoff $\Lambda$ and the radius $R$. In figure~\ref{fig:lambda-dependence}, we show the logarithmic evolution of the rescaled masses $M (\Lambda) \times R$ with the cutoff $\Lambda R$ for a fixed $R^{-1}=800$ GeV and $\mu R = 1$. Because the corrections scale with the gauge couplings the gluon mass varies over $20\%$ from $\Lambda R = 5$ to $40$ while the $P$ mass is almost independent on $\Lambda$. Notice that there are some crossings in the spectrum. This means that depending on which value we choose for the cutoff, some decay channels may appear that do not exist otherwise. Let us emphasise again that the cutoff dependence cannot be subtracted from the theory since the MUED model is non-renormalisable. Therefore, there is no theoretical motivation for choosing one value of $\Lambda$ over another.

\begin{figure}[h!t!]
    {
    \begin{center}
     \includegraphics[width=0.8\textwidth]{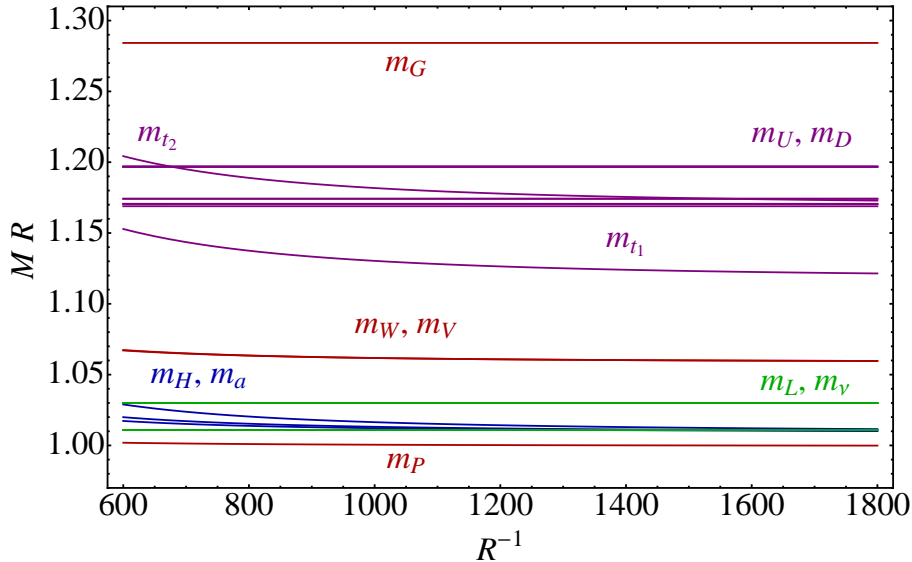}
     \end{center}
    }
    \vspace{-0.4cm}
\caption{Evolution of the $n=1$ masses with the compactification scale $R^{-1}$ at fixed $\Lambda / \mu = 20$.}
\label{fig:R-dependence}
\end{figure}
In figure~\ref{fig:R-dependence}, we plot the rescaled masses $M (R) \times R$ as functions of $R^{-1}$ for a fixed ratio $\Lambda / \mu = 20$.  Because we incorporated the quantum corrections in a gauge invariant way, we see that the rescaled masses exhibit some dependence on $R$, whereas there wouldn't be any had we added the corrections $\delta m^2$ by hand. Most notably it is the electroweak sector of the theory, the Higgs, the $W$ bosons, the tops and the scalars $a_0$ and $a_{\pm}$, that acquires the strongest dependence. When $R^{-1}$ is close to the electroweak scale, the interplay between loop corrections and symmetry breaking is considerable and leads to some crossing in the spectrum. For much larger values of $R^{-1}$, both scales decouple, the spectrum flattens and we recover the behaviour found in \cite{Cheng:2002iz}.

\section{Model implementation\label{sec:imple}}


\subsection{Implementation details and notation\label{subsec:files}}


We have implemented the MUED model presented in section \ref{subsec:setup} into the LanHEP package for automatic evaluation of  Feynman rules \cite{Semenov:2008jy}. Its output files are compatible with the CalcHEP/CompHEP matrix element generator \cite{Belyaev:2012qa}. To date, there exist two public implementations of the MUED model corrected at one loop but none of them contains a KK Higgs sector. The first one is realised as a FeynRules model file \cite{Christensen:2009jx}, while the other consists in a CalcHEP input file created manually in the unitary gauge \cite{Datta:2010us}. The LanHEP package was updated in \cite{Semenov:2010qt} in order to treat extra-dimensional models. Consequently, we were able to implement the model directly in 5D and have LanHEP integrate out the extra dimension automatically. Our model file contains three main parts: the declaration of parameters, variables and fields, the reconstruction of the 5D fields in terms of the mass eigenstates and finally the 5D Lagrangian.

The first step is to declare the external variables of our model, which we list in table \ref{tab:param}. Except for the compactification scale $R^{-1}$, the cutoff $\Lambda$ and the running scale $\mu$, the rest of the parameters are those of the SM. These parameters can be modified in LanHEP, or directly within CalcHEP once the file has been compiled. The values given for the parameters in table \ref{tab:param} are the ones which were used to generate the cross-sections of table \ref{tab:proc} as we will explain later on.

\begin{table}[h!t!]
\begin{center}
\begin{tabular}{|l|l|l|} \hline
Parameter & Value & Comment \\ \hline
$e$   &0.31343             &Electromagnetic coupling constant ($\alpha (M_Z) \equiv$1/128) \\  
$g_s$    &1.21978             &Strong coupling constant ($Z$ point)  (PDG-2010) \\
$\sin \theta_W$   &0.48094             &sine of the Weinberg angle (PDG-2010) \\
$R^{-1}$  &800                 &Inverse radius of the fifth dimension \\
$\Lambda R$    & 20                  &UV cutoff  multiplied by R\\
$\mu R$ &1                    &running scale multiplied by R \\
$m_Z$    &91.1876             &mass of the $Z$-boson \\
$m_e$    &0.000511            &mass of the electron \\
$m_{\mu}$   &0.10566             &mass of the muon \\
$m_{\tau}$  &1.77682             &mass of the tau-lepton \\
$m_u$    &0.0025              &mass of the $u$-quark \\
$m_d$    &0.00495             &mass of the $d$-quark \\
$m_c$    &1.27                &mass of the $c$-quark \\
$m_s$    &0.101               &mass of the $s$-quark \\
$m_{\rm top}$  &172                 &mass of the $t$-quark \\
$m_b$    &4.67                &mass of the $b$-quark \\
$m_H$    &120                 &mass of the Higgs \\
 \hline
\end{tabular}
\caption{Parameters of our MUED implementation. Mass scales are in GeV.}
\label{tab:param}
\end{center}
\end{table}
We next define the variables that are functions of the fundamental parameters of the model. For example, the vev of the Higgs is defined as a function of the $Z$-boson mass, the Weinberg angle and the electric charge. The first main set of 5D variables are the wave-function normalisation factors $Z_{i}$ introduced in section \ref{subsec:setup} and which are chosen to match the results of \cite{Cheng:2002iz}. They depend on the gauge couplings $g_i$, $R^{-1}$, $\Lambda$ and $\mu$. From these functions, we can declare the different mixing angles found in section \ref{subsec:setup} and appendix \ref{app:u1u1}. For the $B^{(n)}$-$W^{3 \, (n)}$ as well as the $B_5^{(n)}$-$W^{3\, (n)}_5$-$\chi^{(n)}$ mixings, we also declare the mixing matrix entries of eq.\eqref{M2V} and \eqref{M2S}. These sets of variables are crucial to reconstruct the 5D fields in terms of the mass eigenstates.

Once all the parameters are defined, we can declare the particles of the model (after compactification of the extra dimension) by their type (vector, spinor or scalar), name, mass and colour charge. Our file contains up to the second KK level. The reason is that, even though only the first KK level is expected to be within LHC reach, $n=2$ KK particles can appear as virtual particles in the processes. Particles with $n>2$ cannot appear in tree-level diagrams involving $n=0,1$ particles in the external state, so long as there are no more than five $n=1$ external particles, but they would need to be considered if one wished to go to higher loop order or to consider higher KK modes in the external states. In general, if the model included KK particles up to KK number $n_{\text{max}}$, one could safely consider tree-level processes with any number of external particles, so long as the simple sum of the external particles' KK numbers did not exceed $2n_\text{max} + 1$.\footnote{A version of our model including up to the fourth KK level is available upon request, but the publicly available $n_\text{max}=2$ model should suffice for all conceivable LHC processes.}

In tables \ref{tab:particles0}, \ref{tab:particles1} and \ref{tab:particles2}, we list the names given to the mass eigenstates in the file.
\begin{table}[h!t!]
\begin{center}
\begin{tabular}{|l|l|l|l|} \hline
Name & Common name & Particle & Mass \\ \hline
\texttt{G}   & $G$  & Gluon & - \\
\texttt{A}   & $\gamma$, $A$  & Photon & - \\
\texttt{Z}   & $Z$  & $Z$ boson & \texttt{MZ} \\
\texttt{W+}/\texttt{W-}   & $W^+$/$W^-$  & $W^{\pm}$ bosons & \texttt{MW} \\
\texttt{e1}, \texttt{e2}, \texttt{e3}   & $e$, $\mu$, $\tau$  &Electron, Muon, Tau & \texttt{Me}, \texttt{Mmu}, \texttt{Mtau} \\
\texttt{n1}, \texttt{n2}, \texttt{n3}   & $\nu_e$, $\nu_{\mu}$, $\nu_{\tau}$  & Neutrinos & - \\
\texttt{u}, \texttt{d}, \texttt{c}, \texttt{s}, \texttt{t}, \texttt{b}   & $u$, $d$, $c$, $s$, $t$, $b$  & Quarks & \texttt{Mu}, \texttt{Md}, \texttt{Mc}, \texttt{Ms}, \texttt{Mb}, \texttt{Mtop}  \\
\texttt{H}   & $H$  & Higgs boson & \texttt{MH}\\
\hline
\end{tabular}
\caption{The zero-mode (SM) particle names of our MUED implementation.}
\label{tab:particles0}
\end{center}

\end{table}
\begin{table}[h!t!]
\begin{center}
\begin{tabular}{|l|l|l|l|} \hline
Name & Common name & Particle & Mass \\ \hline
\texttt{$\sim$G\_1}   & $G^{(1)}$  & Gluon & \texttt{MG1} \\
\texttt{$\sim$P\_1}   & $P^{(1)}$  & $P$ boson & \texttt{MP1} \\
\texttt{$\sim$V\_1}   & $V^{(1)}$  & $V$ boson & \texttt{MV1} \\
\texttt{$\sim$W+\_1}/\texttt{$\sim$W-\_1}   & $W^{+(1)}$/$W^{-(1)}$  & $W^{\pm}$ bosons & \texttt{MW1} \\
\texttt{$\sim$e1\_1}, \texttt{$\sim$e2\_1}   & $e_1^{(1)}$, $e_2^{(1)}$  &  Electron 1 \& 2 & \texttt{Me11},  \texttt{Me21} \\
\texttt{$\sim$mu1\_1}, \texttt{$\sim$mu2\_1}   & $\mu_1^{(1)}$, $\mu_2^{(1)}$  &  Muon 1 \& 2 & \texttt{Mmu11}, \texttt{Mmu21} \\
\texttt{$\sim$tau1\_1}, \texttt{$\sim$tau2\_1}   & $\tau_1^{(1)}$, $\tau_2^{(1)}$  & Tau 1 \& 2 & \texttt{Mtau11}, \texttt{Mtau21} \\
\texttt{$\sim$n1\_1}, \texttt{$\sim$n2\_1}, \texttt{$\sim$n3\_1}   & $\nu_e^{(1)}$, $\nu_{\mu}^{(1)}$, $\nu_{\tau}^{(1)}$  & Neutrinos & \texttt{MneL1}, \texttt{MnmL1}, \texttt{MntL1} \\
\texttt{$\sim$u1\_1}, \texttt{$\sim$u2\_1}   & $u_1^{(1)}$, $u_2^{(1)}$  & Up 1 \& 2 & \texttt{Mu11}, \texttt{Mu21} \\
\texttt{$\sim$d1\_1}, \texttt{$\sim$d2\_1}   & $d_1^{(1)}$, $d_2^{(1)}$  & Down 1 \& 2 & \texttt{Md11}, \texttt{Md21} \\
\texttt{$\sim$H\_1}   & $H^{(1)}$  & Higgs boson & \texttt{MH1} \\
\texttt{$\sim$a0\_1}   & $a^{0(1)}$  & $a^0$ scalar & \texttt{Ma1} \\
\texttt{$\sim$a+\_1}/\texttt{$\sim$a-\_1}   & $a^{+(1)}$/$a^{-(1)}$  & $a^{\pm}$ scalars & \texttt{Mac1} \\
\hline\end{tabular}
\caption{First KK level particle names of our MUED implementation. The second and third generation of quarks are declared likewise  by replacing \texttt{u} $\rightarrow$ \texttt{c}, \texttt{t} and \texttt{d} $\rightarrow$ \texttt{s}, \texttt{b}. The notation with ``\texttt{$\sim$}" for the first KK level is standard for dark matter studies.}
\label{tab:particles1}
\end{center}
\end{table}

\begin{table}[h!t!]
\begin{center}
\begin{tabular}{|l|l|l|l|} \hline
Name & Common name & Particle & Mass \\ \hline
\texttt{G\_2}   & $G^{(2)}$  & Gluon & \texttt{MG2} \\
\texttt{P\_2}   & $P^{(2)}$  & $P$ boson & \texttt{MP2} \\
\texttt{V\_2}   & $V^{(2)}$  & $V$ boson & \texttt{MV2} \\
\texttt{W+\_2}/\texttt{W-\_2}   & $W^{+(2)}$/$W^{-(2)}$  & $W^{\pm}$ bosons & \texttt{MW2} \\
\texttt{e1\_2}, \texttt{e2\_2}   & $e_1^{(2)}$, $e_2^{(2)}$  & Electron 1 \& 2 & \texttt{Me12},  \texttt{Me22} \\
\texttt{mu1\_2}, \texttt{mu2\_2}   & $\mu_1^{(2)}$, $\mu_2^{(2)}$  & Muon 1 \& 2 & \texttt{Mmu12},  \texttt{Mmu22} \\
\texttt{tau1\_2}, \texttt{tau2\_2}   & $\tau_1^{(2)}$, $\tau_2^{(2)}$  & Tau 1 \& 2 & \texttt{Mtau12},  \texttt{Mtau22} \\
\texttt{n1\_2}, \texttt{n2\_2}, \texttt{n3\_2}   & $\nu_e^{(2)}$, $\nu_{\mu}^{(2)}$, $\nu_{\tau}^{(2)}$  & Neutrinos & \texttt{MneL2}, \texttt{MnmL2}, \texttt{MntL2} \\
\texttt{u1\_2}, \texttt{u2\_2}   & $u_1^{(2)}$, $u_2^{(2)}$  &Up 1 \& 2 & \texttt{Mu12},  \texttt{Mu22} \\
\texttt{d1\_2}, \texttt{d2\_2}   & $d_1^{(2)}$, $d_2^{(2)}$  & Down 1 \& 2 & \texttt{Md12},  \texttt{Md22} \\
\texttt{H\_2}   & $H^{(2)}$  & Higgs boson & \texttt{MH2} \\
\texttt{a0\_2}   & $a^{0(2)}$  &  $a^0$ scalar & \texttt{Ma2} \\
\texttt{a+\_2}/\texttt{a-\_2}   & $a^{+(2)}$/$a^{-(2)}$  &  $a^{\pm}$ scalars & \texttt{Mac2} \\
\hline
\end{tabular}
\caption{Second KK level particle names of our MUED implementation.}
\label{tab:particles2}
\end{center}
\end{table}
Having defined the 4D particles of the theory, we reconstruct the 5D interacting fields simply by inverting the several mixing matrices presented in section \ref{subsec:setup} and appendix \ref{app:u1u1}. We can then implement the Lagrangian exactly as it is written in \eqref{L1L}. Upon compilation, the program integrates over the extra dimension and calculates the masses and vertices of the theory for the particles that were defined. Needless to say, that the Lagrangian is implemented in a 5D way ensures that gauge invariance is preserved through compactification.

Let us comment on the gauge-fixing and Fadeev-Popov ghost terms in the Lagrangian. LanHEP allows us to write a model in both unitary and Feynman-'t Hooft gauges by adding an option ``gauge" to the declaration of the gauge bosons. The Goldstone boson and ghost of a given field \texttt{A} are implicitly defined in the program as \texttt{A.f} and \texttt{A.c} respectively. In order to implement the gauge fixing and ghost Lagrangian terms, the simplest way is to reconstruct the 5D Goldstone bosons and ghosts of the fundamental $G$, $B$ and $W^i$ fields out of these ``\texttt{.f}" and ``\texttt{.c}". For instance, one may use \texttt{$\sim$P\_1.f}, \texttt{$\sim$V\_1.f} and \texttt{$\sim$a0\_1} in order to reconstruct $B_5^{(1)}$ (see appendix \ref{app:u1u1}). The full 5D gauge-fixed Lagrangian with Fadeev-Popov ghosts is then easily written in the fundamental basis.

Finally, let us mention that we have taken particular care of the important four-point vertices involving KK gluons. Indeed, neither LanHEP nor CalcHEP recognise the colour structure of these interactions. The common trick is to split the four-point vertices and introduce non-physical auxiliary fields  \cite{Belyaev:2005ew, Datta:2010us}. The splitting implemented in the DKM model only works for the SM gluons: it is incorrect for the KK gluons. We have derived a correct form of the splitting for all four-point gluon vertices (see appendix \ref{app:fourgluon} for details) and incorporated it into our model.


\subsection{Consistency checks\label{subsec:valid}}


At the LanHEP level, a first important validation is that of the masses: an error is generated if the masses declared within the model (when particles are declared) differ from those calculated after integration. This step ensures that the whole diagonalisation procedure is correct. Furthermore, within CalcHEP, we have checked that our model is gauge invariant by calculating several dozen $2 \rightarrow 2$ and $2 \rightarrow 3$ processes in both unitary and Feynman-'t Hooft gauges.

Let us now compare some cross-sections produced within the model described in \cite{Datta:2010us} which we denote by DKM, and our implementation (BBMP). 
In table \ref{tab:proc}, we list processes involving KK gauge bosons. 
The parameters used are given in table \ref{tab:param} and the centre-of-mass energy is $\sqrt{s}=2$ TeV.
We also require the transverse momenta of the final state particles to each be greater than 100~GeV in order to remove possible infrared divergences.
The DKM model by default includes bulk corrections to the KK masses, which we neglect in our BBMP model.
Therefore, for the purposes of comparison in table~\ref{tab:proc}, we set the variable \texttt{zeta3} equal to zero in the DKM model file in order to remove the bulk corrections (which are all proportional to $\zeta(3)$).

Although the processes shown in the table are unrealistic for phenomenology, they are useful to understand the differences between the two models. 
For instance, since the Higgs sector is modified, so are the gauge boson eigenstates. 
Instead of the KK-photon $\gamma^{(n)}$ and the KK $Z^{(n)}$-boson, our model contains the gauge bosons $P^{(n)}$ and $V^{(n)}$ introduced in section \ref{subsec:setup}. 
Whenever there is an ambiguity concerning the initial states, we replace $\gamma^{(1)} \rightarrow P^{(1)}$ and $Z^{(1)} \rightarrow V^{(1)}$. 
If the ambiguity is over the final states, we simply sum over all possible combinations.

\begin{table}[h!]
\begin{center}
\begin{tabular}{|r@{}c|l|r@{}l|r@{}l|}
  \hline
  &   & Process & \multicolumn{2}{|r|}{DKM $\sigma$ [pb]} &  \multicolumn{2}{|l|}{BBMP $\sigma$ [pb]} \\
\hline
  &  1 & $G^{(1)}\, G^{(1)} \rightarrow G \, G$                                & $3.952$ & $\times 10^{ 1}$ & $3.952$ & $\times 10^{ 1}$ \\
  &  2 & $G^{(1)}\, G \rightarrow G^{(1)} \, G$                                & $7.600$ & $\times 10^{ 3}$ & $7.600$ & $\times 10^{ 3}$ \\
* &  3 & $G^{(1)}\, G^{(1)} \rightarrow G^{(1)} \, G^{(1)}$                    & $8.619$ & $\times 10^{ 3}$ & $8.600$ & $\times 10^{ 3}$ \\
* &  4 & $G^{(1)}\, Z^{(1)} \rightarrow c \, \bar{c}$                          & $2.132$ & $\times 10^{-1}$ & $2.037$ & $\times 10^{-1}$ \\
* &  5 & $G^{(1)}\, \gamma^{(1)} \rightarrow b \, \bar{b}$                     & $3.651$ & $\times 10^{-2}$ & $3.249$ & $\times 10^{-2}$ \\
* &  6 & $\gamma^{(1)} \, \gamma^{(1)} \rightarrow t \, \bar{t}$               & $2.641$ & $\times 10^{-2}$ & $2.758$ & $\times 10^{-2}$ \\
* &  7 & $Z^{(1)} \, Z^{(1)} \rightarrow d \, \bar{d}$                         & $9.098$ & $\times 10^{-2}$ & $9.165$ & $\times 10^{-2}$ \\
* &  8 & $Z^{(1)}\, Z^{(1)} \rightarrow W^+ \, W^-$                            & $9.293$ & $\times 10^{ 0}$ & $9.288$ & $\times 10^{ 0}$ \\
* &  9 & $W^{+ \, (1)}\, W^{- \, (1)} \rightarrow Z \, Z$                      & $2.744$ & $\times 10^{ 0}$ & $2.761$ & $\times 10^{ 0}$ \\
  & 10 & $W^{+ \, (1)}\, W^{- \, (1)} \rightarrow Z \, \gamma$                 & $1.653$ & $\times 10^{ 0}$ & $1.653$ & $\times 10^{ 0}$ \\
* & 11 & $W^{+ \, (1)}\, W^{- \, (1)} \rightarrow W^+ \, W^-$                  & $3.152$ & $\times 10^{ 0}$ & $3.081$ & $\times 10^{ 0}$ \\
  & 12 & $W^{+ \, (1)}\, W^{- \, (1)} \rightarrow \gamma \, \gamma$            & $2.489$ & $\times 10^{-1}$ & $2.489$ & $\times 10^{-1}$ \\
  & 13 & $Z\, \gamma \rightarrow W^{+ \, (1)}\, W^{- \, (1)}$                  & $1.028$ & $\times 10^{ 0}$ & $1.028$ & $\times 10^{ 0}$ \\
* & 14 & $Z^{(1)}\, Z^{(1)} \rightarrow  W^{+\,(1)}\, W^{- \, (1)}$            & $7.240$ & $\times 10^{ 2}$ & $7.210$ & $\times 10^{ 2}$ \\
* & 15 & $Z \, Z^{(1)} \rightarrow W^{+}\, W^{- \, (1)}$                       & $2.045$ & $\times 10^{ 2}$ & $2.029$ & $\times 10^{ 2}$ \\
* & 16 & $W^{+ \, (1)}\, W^{- \, (1)} \rightarrow W^{+ \, (1)}\, W^{- \, (1)}$ & $3.663$ & $\times 10^{ 2}$ & $3.661$ & $\times 10^{ 2}$ \\
\hline     
  & 17 & $W^{+}\, W^{- \, (1)} \rightarrow Z^{(1)}\, \gamma$                   & $5.290$ & $\times 10^{ 1}$ & \multicolumn{2}{|l|}{\begin{tabular}{lr@{}l} $P^{(1)}$ & $1.016$ & $\times 10^{-1}$ \\
                                                                                                                                                           $V^{(1)}$ & $5.280$ & $\times 10^{ 1}$ \\
                                                                                                                                                           total     & $5.290$ & $\times 10^{ 1}$  \end{tabular}} \\
\hline
* & 18 & $W^{+}\, W^{- \, (1)} \rightarrow Z^{(1)}\, Z$                        & $2.041$ & $\times 10^{ 2}$ & \multicolumn{2}{|l|}{\begin{tabular}{lr@{}l} $P^{(1)}$ & $3.940$ & $\times 10^{-1}$ \\ 
                                                                                                                                                           $V^{(1)}$ & $2.026$ & $\times 10^{ 2}$ \\ 
                                                                                                                                                           total     & $2.029$ & $\times 10^{ 2}$  \end{tabular}} \\
\hline
\end{tabular} 
\caption{Sample of processes with two gauge bosons for cross-section comparison (in pb) between previous implementation (\cite{Datta:2010us}, DKM) and our implementation (BBMP) for partonic $\sqrt{s} = 2~\text{TeV}$. 
All final state particles are required to have transverse momenta greater than 100~GeV. 
Cross-sections are accurate to the number of significant figures shown.
Asterisks highlight cross-sections that differ between models; 
differences are explained in the text.}
\label{tab:proc}
\end{center}
\end{table}

As we can see, the two implementations are in good agreement for all processes considered. There are however some slight differences, marked with asterisks. First, processes 1 and 2 that involve both the SM and the first KK gluons give the same cross-sections, while process 3, with four $G^{(1)}$ bosons, disagrees. This is due to the way the four-gluon vertices are split, as explained at the end of the last subsection and in detail in appendix \ref{app:fourgluon}: when a second KK level is present in the theory, the splitting cannot be the same as in the DKM implementation. Processes 4--7 involve KK gauge bosons into SM quark pairs. The differences are due to electroweak symmetry breaking not being implemented for KK particles in the DKM model. Processes 4 and 5 proceed only by exchange of KK quarks. In the DKM model, these quarks are purely left- and right-handed but in our BBMP model the chiralities mix after electroweak symmetry breaking. For processes 6 and 7 there is an additional difference: in our model a (SM) Higgs can be exchanged as well.

The remaining processes 8--18 considered in table \ref{tab:proc} involve only $\SU2 \times$U(1) gauge bosons. The differences observed are again easily explained by the exchange of Higgs bosons $H$, $H^{(1)}$ or $a^{0(1)}$. Conversely, the processes which agree perfectly are those containing a SM photon simply because no Higgses can be exchanged at tree level.

Another consequence of the missing KK Higgs sector in the DKM implementation is that scattering of gauge bosons will not respect unitary if KK modes are involved. With our implementation of the KK Higgs sector, unitarity is preserved.
\begin{figure}[htb]
    \begin{center}
        \includegraphics[width=0.49\textwidth]{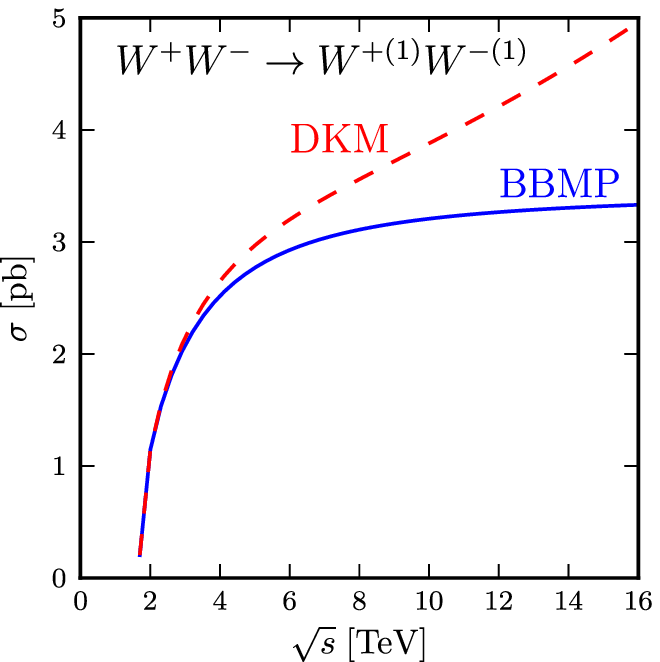}
        \includegraphics[width=0.49\textwidth]{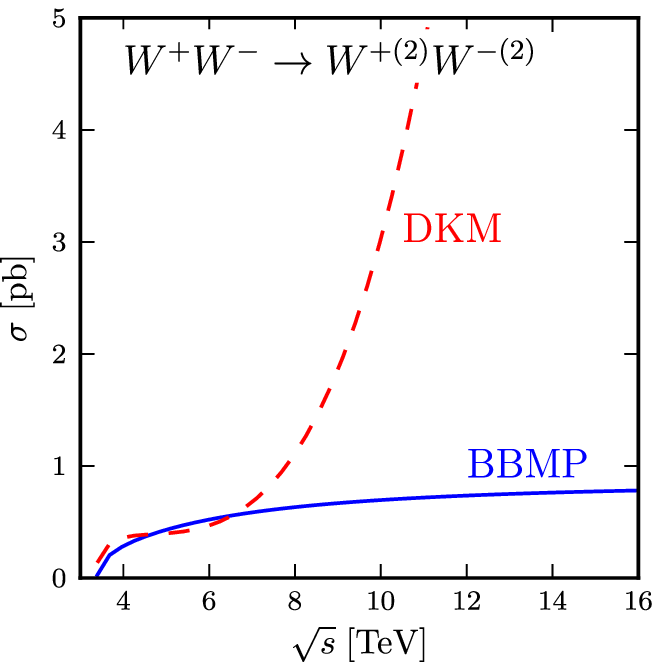}
        \caption{Cross-section for $W^{+}W^{-} \to W^{+(1)}W^{-(1)}$ (left) and $W^{+}W^{-} \to W^{+(2)}W^{-(2)}$ (right) scattering as a function of centre of mass energy $\sqrt{s}$ for our model implementation and the DKM~\cite{Datta:2010us} implementation, demonstrating improved unitary behaviour. All parameters are as in table~\ref{tab:param}.}
        \label{fig:WWscattering}
    \end{center}
\end{figure}
In figure~\ref{fig:WWscattering}, the scattering cross-section for $W^{+}W^{-} \to W^{+(1)}W^{-(1)}$ and $W^{+}W^{-} \to W^{+(2)}W^{-(2)}$ is plotted as a function of centre of mass energy $\sqrt{s}$, demonstrating the improved unitary behaviour of our model.

\section{Phenomenology\label{sec:pheno}}

\subsection{MUED parameter space,  signal rates and signature under study}

Neglecting the weak dependence on the cutoff $\Lambda$ of the theory, the parameter space of MUED is unambiguously defined by the inverse compactification radius $R^{-1}$ and the Higgs boson mass $m_H$. Moreover, this parameter space is constrained by experiment. Cosmological measurements of the density of dark matter limit $R^{-1}$ (related to the mass of the lightest Kaluza Klein particle) to be below about 1.6 TeV~\cite{Belanger:2010yx}. In addition, tests of the Higgs sector  provide a very powerful constraint on MUED. In cosmology, for example, the Higgs boson mass is bound from {\it above} by the requirement that the dark matter candidate should be neutral~\cite{Belanger:2010yx}. This limits the Higgs mass to be below 230 GeV.

Constraints on the MUED parameter space also come from collider experiments. As we know, the discovery of a Higgs-like 
particle was claimed on 4th of July 2012, based on data released
by the CMS~\cite{:2012gu} and ATLAS~\cite{:2012gk} collaborations.
With detailed information on individual Higgs boson production and decay
processes provided by the CMS and ATLAS experiments, one can understand much better the nature of the Higgs boson, interpret it within MUED
and set limits on the model.
In~\cite{Nishiwaki:2011gk,Belanger:2012mc}, which established such limits, 
it was stressed that the signals from the Higgs boson from MUED are {\it always} enhanced as compared to those of the SM. Therefore, the LHC sensitivity to the SM Higgs boson would lead to even better sensitivity to  the Higgs boson within the MUED scenario.

One should also mention that universal extra dimensions can be limited via
precision electroweak constraints. 
\begin{figure}[htb]
\includegraphics[width=0.52\textwidth]{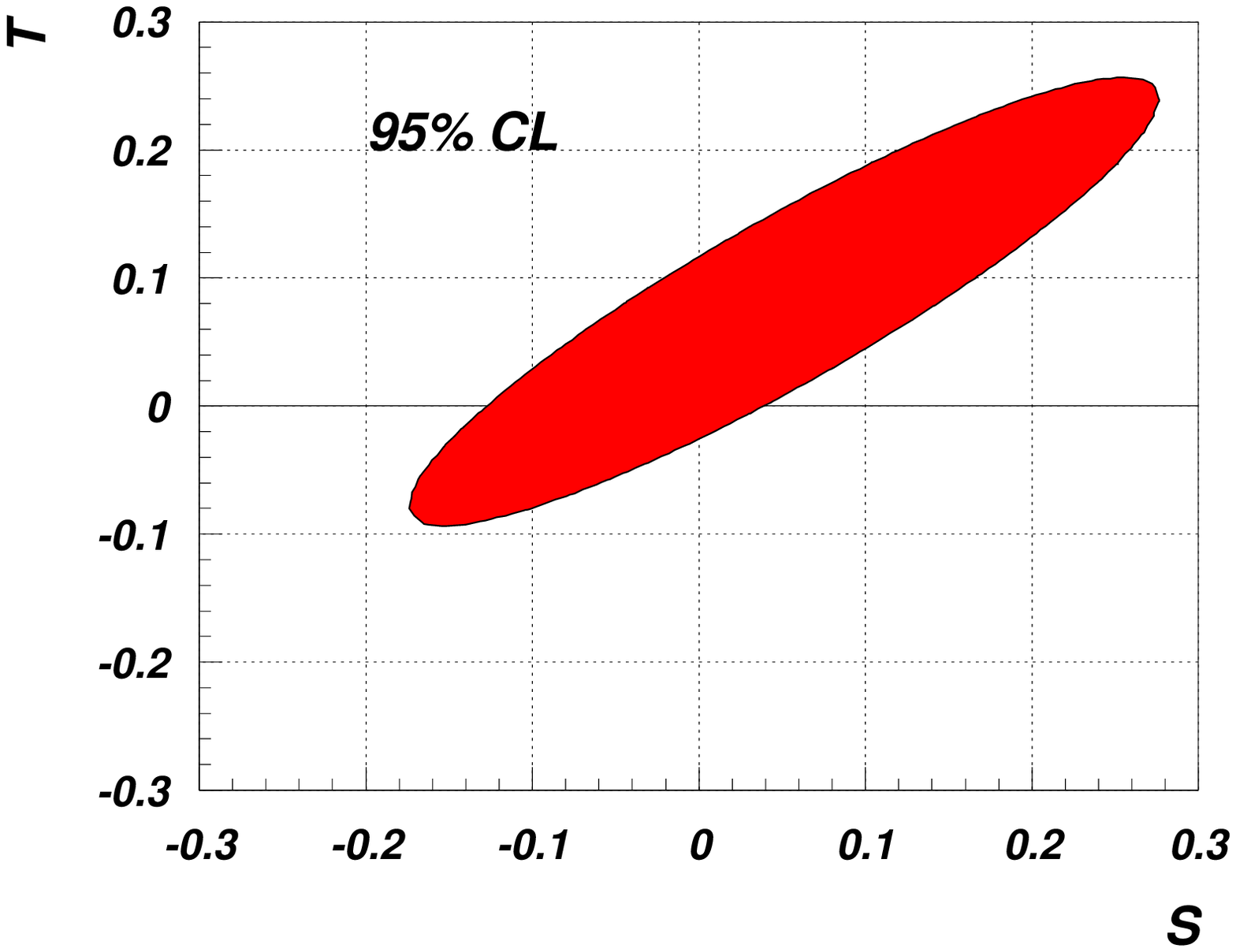}%
\includegraphics[width=0.52\textwidth]{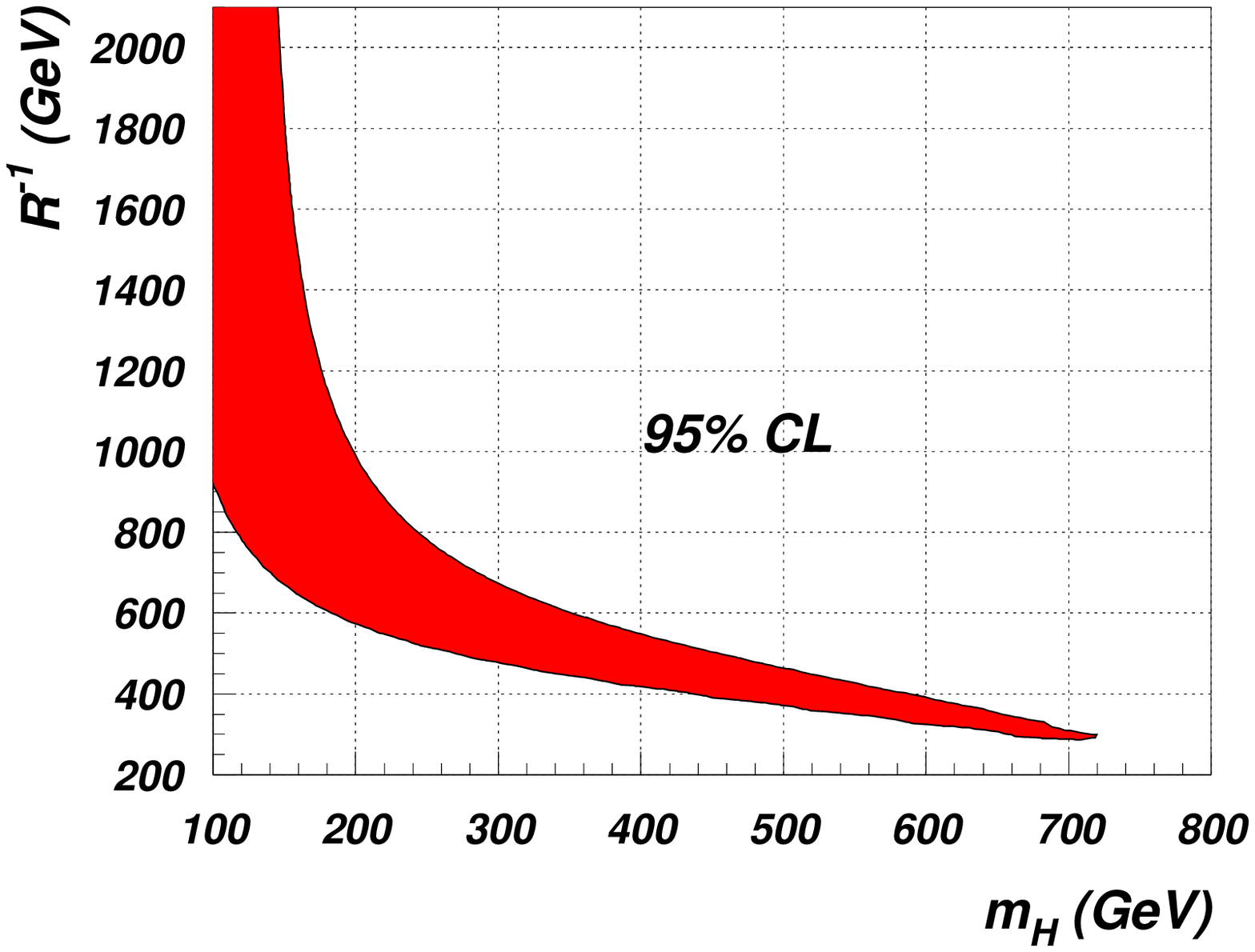}
\caption{\label{fig:ew-fits} Left:  $S$-$T$ parameter space allowed by experimental constraints at $95\%$ CL
(using $S_{|U=0}=0.05\pm0.09$, $T_{|U=0}=0.8\pm0.07$ and $S$-$T$ correlation $+0.91$). Right:
$95\%$ CL for  $m_H-R^{-1}$ parameter space.}
\end{figure}
Using results for the MUED contribution 
to the oblique $S$, $T$ and $U$ parameters from~\cite{Gogoladze:2006br}
and recent the electroweak fit of the Standard Model after the
discovery of the Higgs-like boson at the LHC~\cite{Baak:2012kk}
($S_{|U=0}=0.05 \pm 0.09$ and 
 $T_{|U=0}=0.08  \pm 0.07$ with a correlation
coefficient of $+0.91$) in figure~\ref{fig:ew-fits}
we present the $S$-$T$ parameter space allowed by experimental constraints at 95\% confidence level (CL)
(left) as well as 95\% CL for the $m_H$--$R^{-1}$ parameter space (right).

\begin{figure}[htb]
\includegraphics[width=0.52\textwidth]{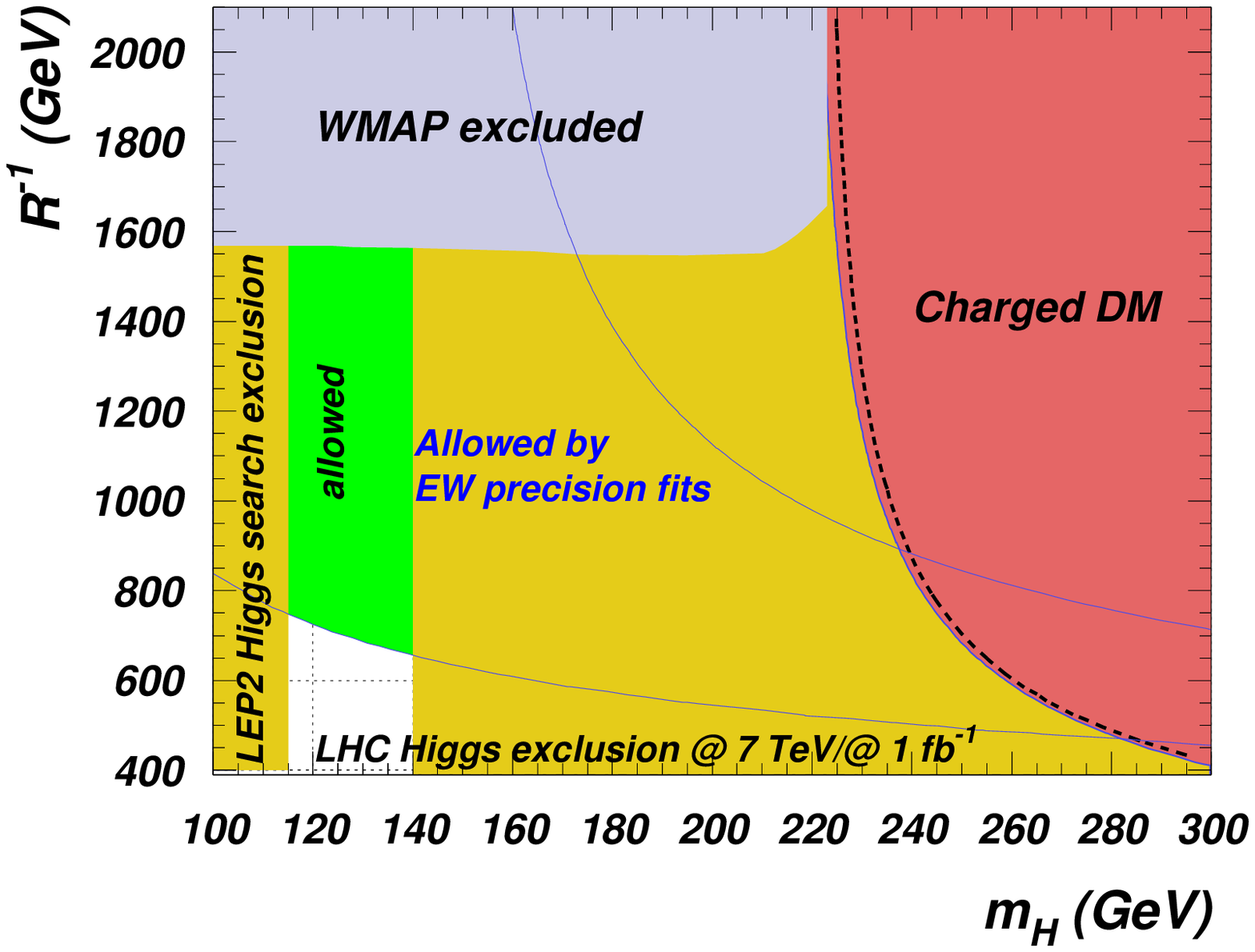}%
\includegraphics[width=0.52\textwidth]{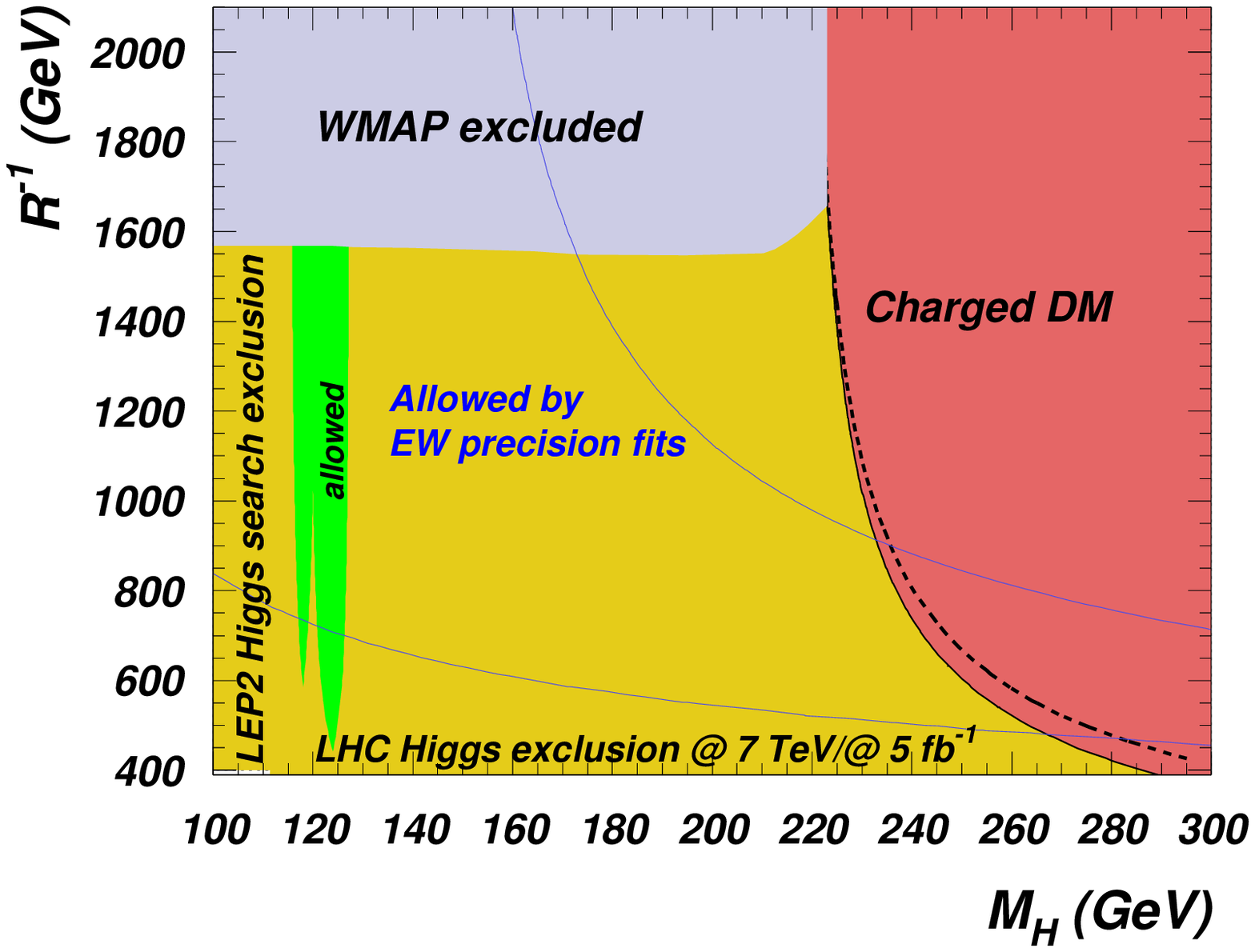}%
\caption{\label{fig:mued-para-space}Combination of Higgs constraints from 1 fb$^{-1}$ (left) and 5 fb$^{-1}$ (right)
of LHC Higgs data~\cite{Belanger:2012mc}, electroweak precision (from figure~\ref{fig:ew-fits}) and dark matter relic density~\cite{Belanger:2010yx} limits on the MUED parameter space.}
\end{figure}

The combined effect of all the constraints mentioned above is shown  in figure~\ref{fig:mued-para-space} including constraints from measurements of dark matter relic density (taken directly from ref.~\cite{Belanger:2010yx}),
electroweak precision tests (described in the previous paragraph) and limits from the Higgs  boson search (taken from ref.~\cite{Belanger:2012mc}).
This figure  also demonstrates how fast constraints are evolving from the LHC side
from the Higgs search: compare 1 fb$^{-1}$ (left) to 5 fb$^{-1}$ (right)
of data.   

\begin{figure}[htb]
\begin{center}
\includegraphics[width=0.85\textwidth]{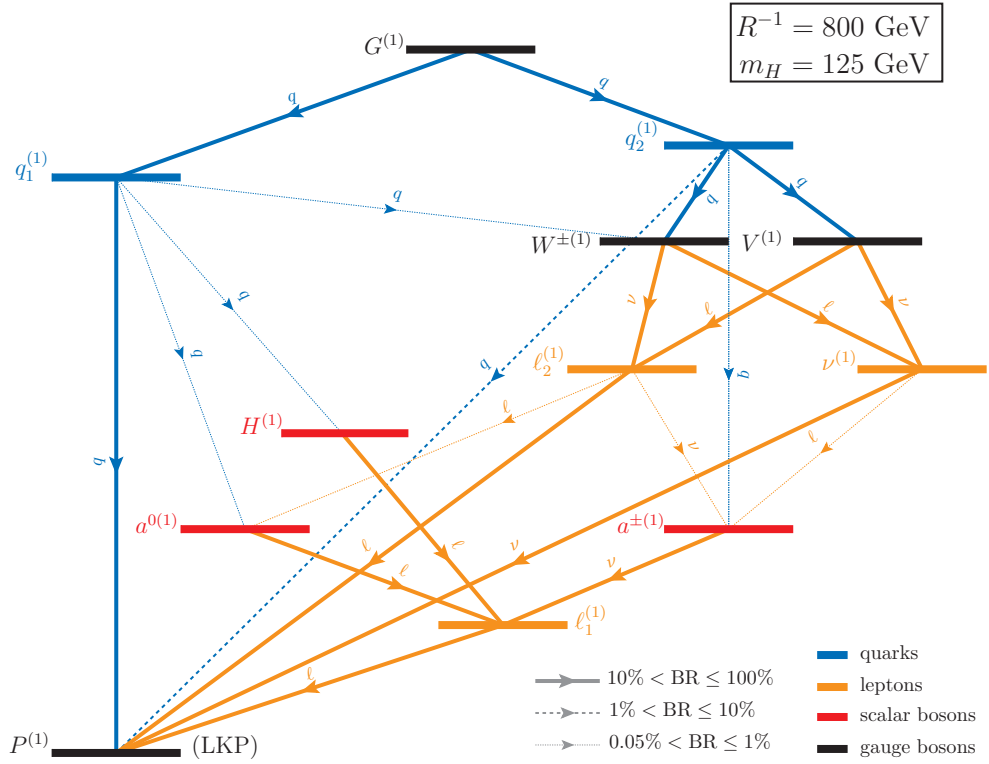}
\\ \vspace{5mm}
\includegraphics[width=0.85\textwidth]{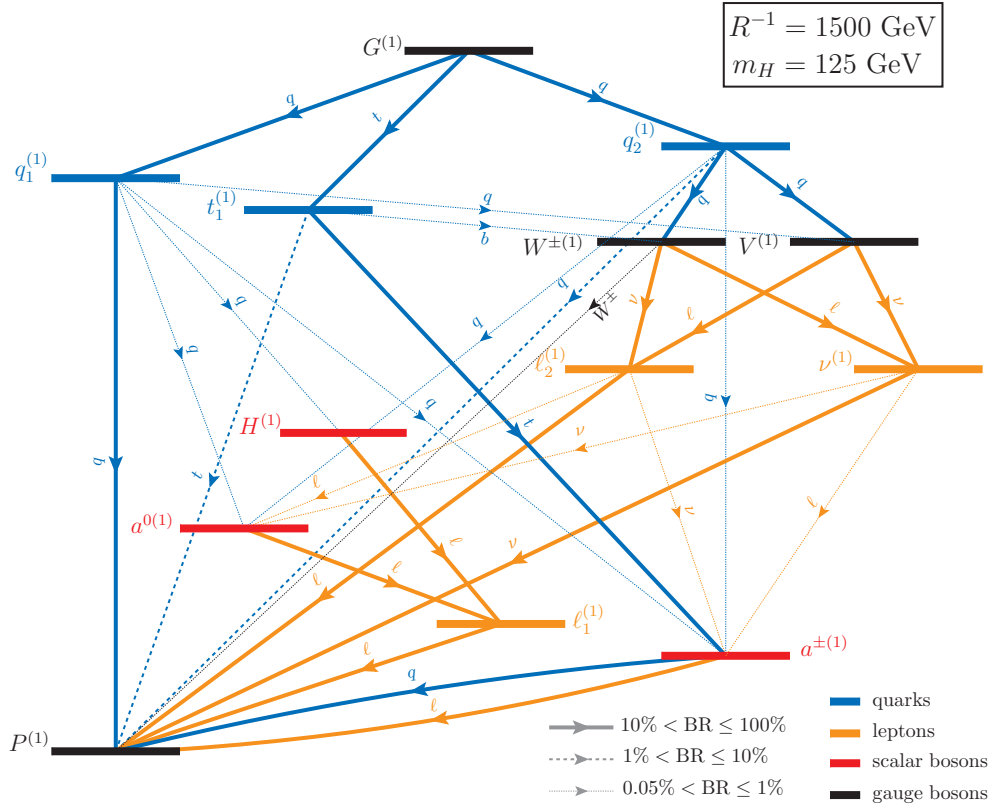}
\caption{\label{fig:decay}The principal decay channels
of the  $n=1$ KK gluon as well as decay channels of  its subsequent decay products
for $R^{-1}=800$~GeV (top) and $R^{-1}=1500$~GeV (bottom).}
\end{center}
\end{figure}

In this paper we explore the LHC potential of {\it directly} testing MUED
through production of KK-particles. Because of the small mass split of MUED KK-particles, which is below about 30\% of $R^{-1}$ (as illustrated in figures~\ref{fig:spectrum1500} and \ref{fig:lambda-dependence}), the main signal will come from strongly produced particles -- KK-gluons and KK-quarks. In figure~\ref{fig:decay} we present the principal decay chains of the  $n=1$ KK-gluon for $R^{-1}=800$~GeV (top) and $R^{-1}=1500$~GeV (bottom). 

One can see that for sufficiently large $R^{-1}$ (e.g. $R^{-1}=1500$~GeV as compared to $R^{-1}=800$ GeV) the  additional decay channel of $G^{(1)}\to t t^{(1)}_1$ opens up due to increased mass split between $G^{(1)}$ and $t^{(1)}_1$. This channel, followed by the  
 $t^{(1)}_1 \to t a^{(1)}_{\pm} \to t \ell \nu$ decay chain, would lead to a quite striking signature
 of four top quarks, two leptons and missing transverse momentum (remembering that $G^{(1)}$ is pair produced).

A generic feature of MUED is the high lepton multiplicity in KK quark and gluon decays,
which can be even higher than the quark multiplicity. This can be seen by looking at the decay products of KK quarks and gluons in figure~\ref{fig:decay}
which are dominated by orange colour, indicating the lepton decay channels.
The reason for this pattern is in the specific mass hierarchy:
\begin{equation}
\label{eq:massh}
m_{q^{(1)}_{1,2}} > m_{W^{(1)}}, m_{V^{(1)}} > m_{\ell_{1,2}^{(1)}}, m_{\nu^{(1)}}
\end{equation}
\begin{equation}
\label{eq:massh2}
m_{q^{(1)}_{1,2}} > m_{W^{(1)}, V^{(1)}} > m_{\nu^{(1)},\ell_{1,2}^{(1)}}
\end{equation}
which is defined by the MUED radiative corrections.

The mass hierarchy noted in  Eq.~(\ref{eq:massh})
tells us that the $W_\pm^{(1)}$ and $V^{(1)}$ bosons will {\it always} decay
to leptons (SM or KK)  and {\it never} to quarks.
This happens because decays of $W^{\pm(1)}$ and $V^{(1)}$ to real $W^\pm$ and $Z$ bosons
are either forbidden, due to the insufficient mass split between $W^{\pm(1)},V^{(1)}$
and $P^{(1)}$, or---when $W^{\pm(1)} \to W^\pm P^{(1)}$ or $V^{(1)} \to Z P^{(1)}$ are 
kinematically allowed (for large values $R^{-1} \gtrsim 1.4$~TeV)---they are suppressed by the phase space and the small $P^{(1)}$--$V^{(1)}$ mixing which governs this decay.

This is a very specific and unique pattern for MUED contrary to, for example,
SUSY GUT theories. In those theories, sleptons are typically heavier than
gauginos and therefore gauginos primarily decay to quarks
because of the high SM $W^\pm$ and $Z$ boson branching ratios to quarks.

The lepton multiplicity in MUED can be easily as large as four (from the pair production of KK gluons or quarks). In principle, it can be as large as eight, but the probability of this is very 
low since such a process would involve rare decays of KK leptons into  KK Higgs bosons.
Moreover, leptons produced in this way will be quite soft  because of the very small mass difference between
$\ell_2^{(1)}$ and $a^{0(1)}$ and so they are unlikely to pass the experimental lepton selection
cuts which we discuss below.

It should also be stressed that the small mass split (by up to about 40\%
relative to the LKP mass)
of the MUED mass spectrum, and its specific pattern, unambiguously defines
the dominant production channel of MUED particles at the LHC and therefore
dictates the search strategy: looking for evidence of MUED in the form of strongly produced KK quarks and gluons.
The production rate of KK-gluons and quarks is determined by their masses which are functions of 
the  $R^{-1}$ and $\Lambda$ parameters.

\begin{figure}[htb]
\includegraphics[width=0.52\textwidth]{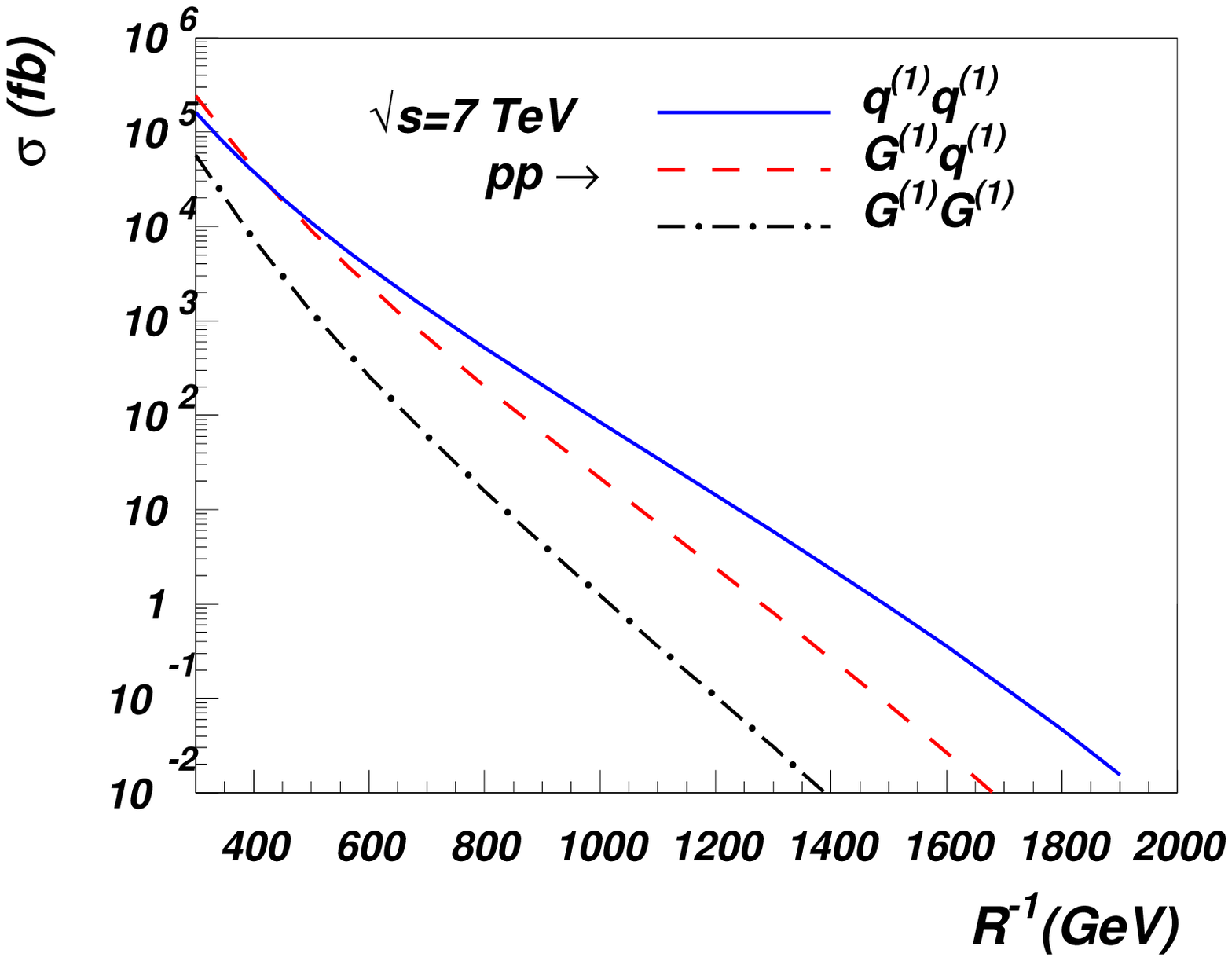}%
\includegraphics[width=0.52\textwidth]{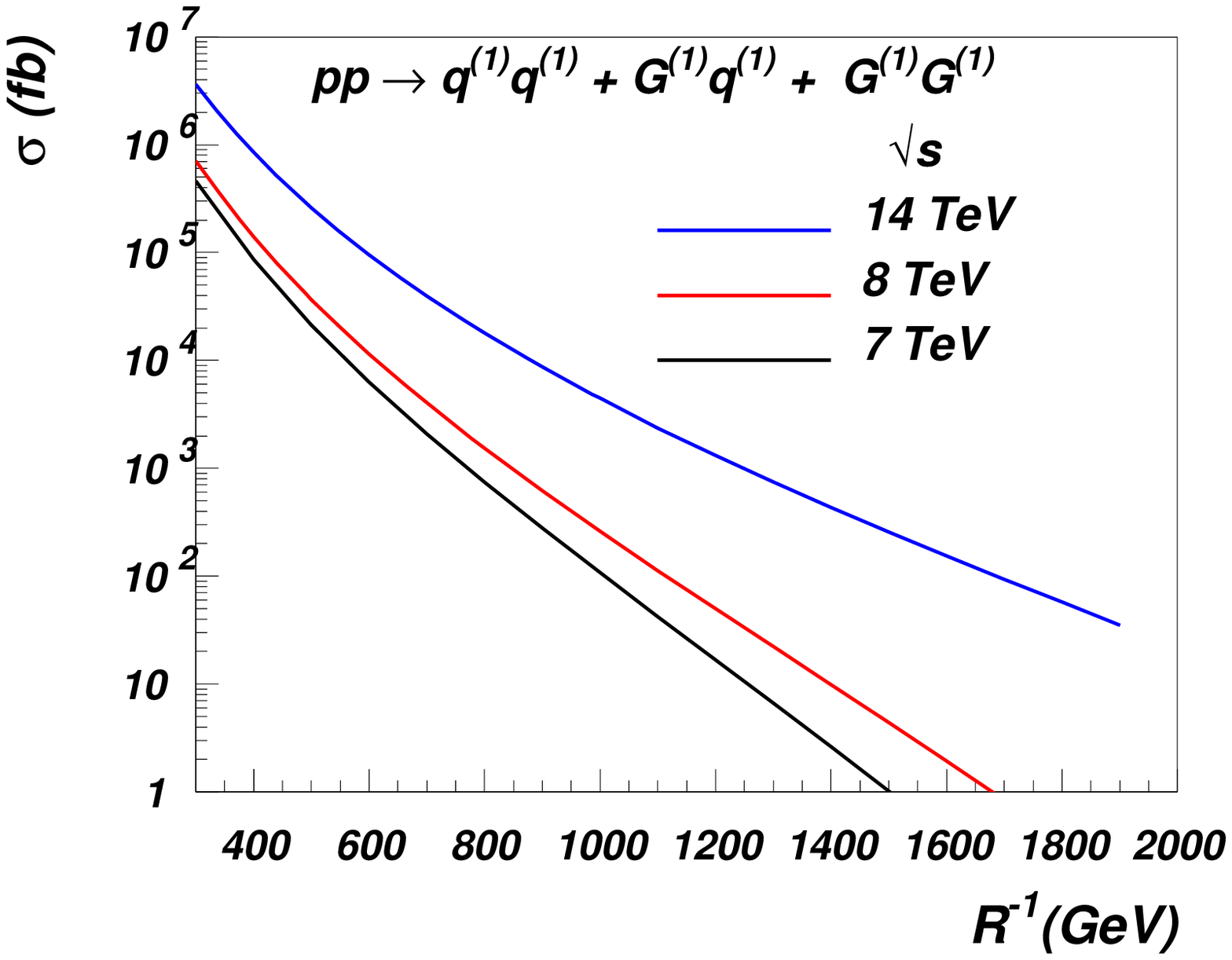}
\caption{Cross section of strongly produced KK quarks and gluons:
left -- for $\sqrt{s}=7$~TeV with the decomposed contribution from 
$G^{(1)}G^{(1)}$, $G^{(1)} q^{(1)}$ and  $q^{(1)} q^{(1)}$ production;
right -- the total $G^{(1)}G^{(1)}+G^{(1)} q^{(1)}+q^{(1)} q^{(1)}$ production cross section for
$\sqrt{s}=7$, 8 and 14~TeV
collider energies.
The CTE6L PDF  was used, with QCD scale $Q=R^{-1}$. $\Lambda=20R^{-1}$.}
\label{fig:cs}
\end{figure}

Figure~\ref{fig:cs} presents 
the cross sections for the $G^{(1)}G^{(1)}$, $G^{(1)} q^{(1)}$ and  $q^{(1)}q^{(1)}$
processes. (Here, $q^{(1)}$ represents all $n=1$ KK quarks, remembering that there are two KK quarks for each SM quark at each KK level.) One can see that  $q^{(1)} q^{(1)}$ plays a dominant role because it receives contributions from the multiple KK-quark flavours. The $G^{(1)} q^{(1)}$
production cross section  is about factor 2--10 below the  $q^{(1)} q^{(1)}$ one,
while the  $G^{(1)} G^{(1)}$
is below the $G^{(1)} q^{(1)}$ by another factor 5--20.
In our analysis we sum over all three subprocesses.
The calculation  of the total cross section has been done at tree-level,
so our conclusions on the LHC potential to explore MUED are on the conservative side since one would expect 
the NLO and NLL K-factors to be of the order of 2 and above,
similar to the  SUSY case studied in~\cite{Beenakker:1996ch,Kulesza:2009kq,Beenakker:2009ha,Beenakker:2011fu}.
In our estimates we are using CTE6L parton density functions (PDF) evaluated at the QCD scale equal to $R^{-1}$
while  $\Lambda$ was chosen to be $20R^{-1}$.

We will show below that the LHC is sensitive to the process $pp\to G^{(1)}G^{(1)}+G^{(1)} q^{(1)}+q^{(1)} q^{(1)}$ in the tri-lepton channel if it has a cross-section as large as around 100~fb. It can be seen from figure~\ref{fig:cs} that this sensitivity is reached for $R^{-1}$ below about 1~TeV, 1.1~TeV and 1.8~TeV
for the LHC with $\sqrt{s}=7,8$ and 14~TeV respectively. In this paper we have focused on $\sqrt{s}=7,8$~TeV studies relevant to present LHC
energies; however, we would like to stress that the 14 TeV LHC will significantly extend the limit on $R^{-1}$, thus completely covering the MUED parameter space~\cite{14TeV-Mued} in combination with the Dark Matter relic density constraints. 

\begin{figure}[htb]
\includegraphics[width=0.52\textwidth]{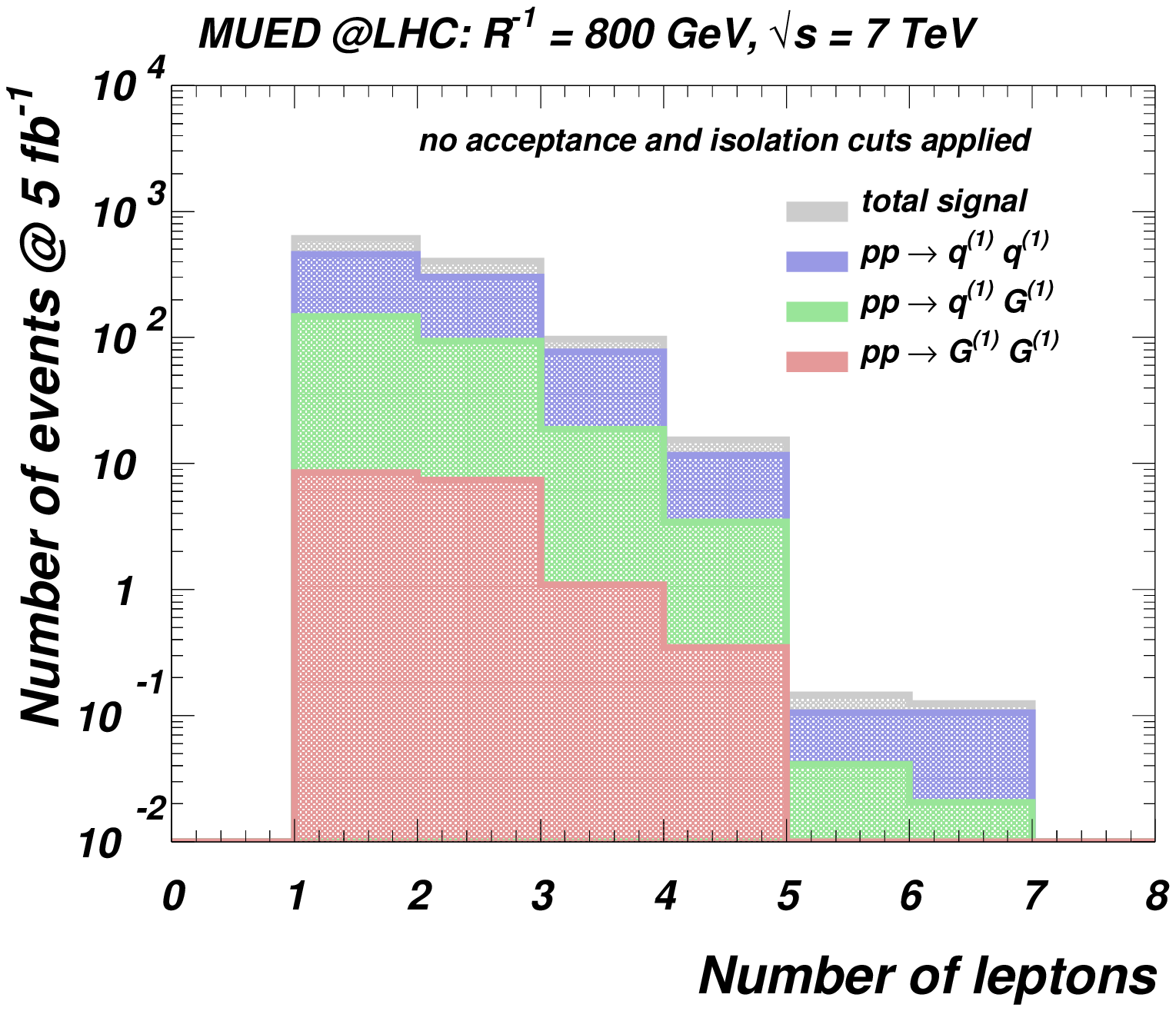}%
\includegraphics[width=0.52\textwidth]{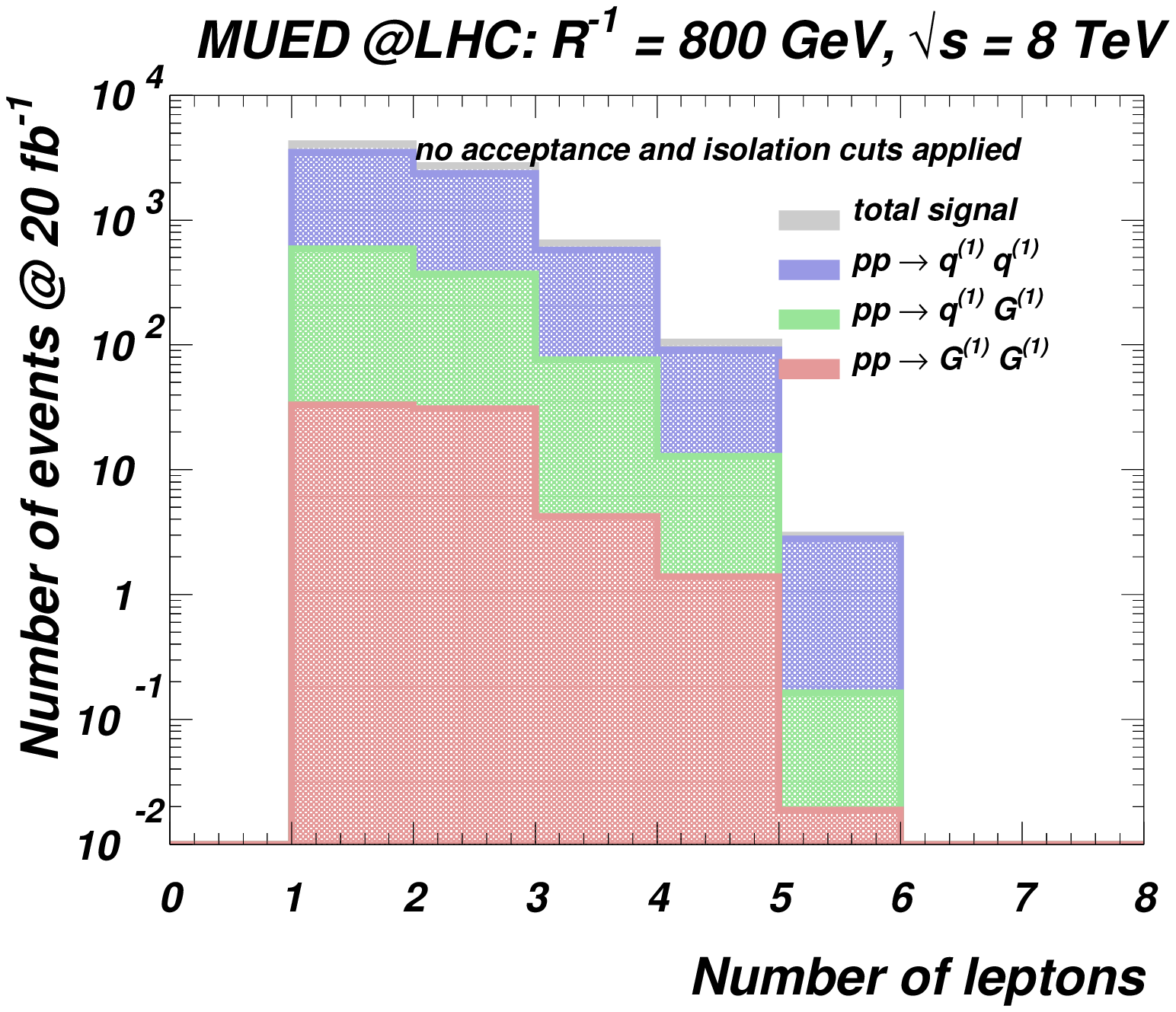}\\
\includegraphics[width=0.52\textwidth]{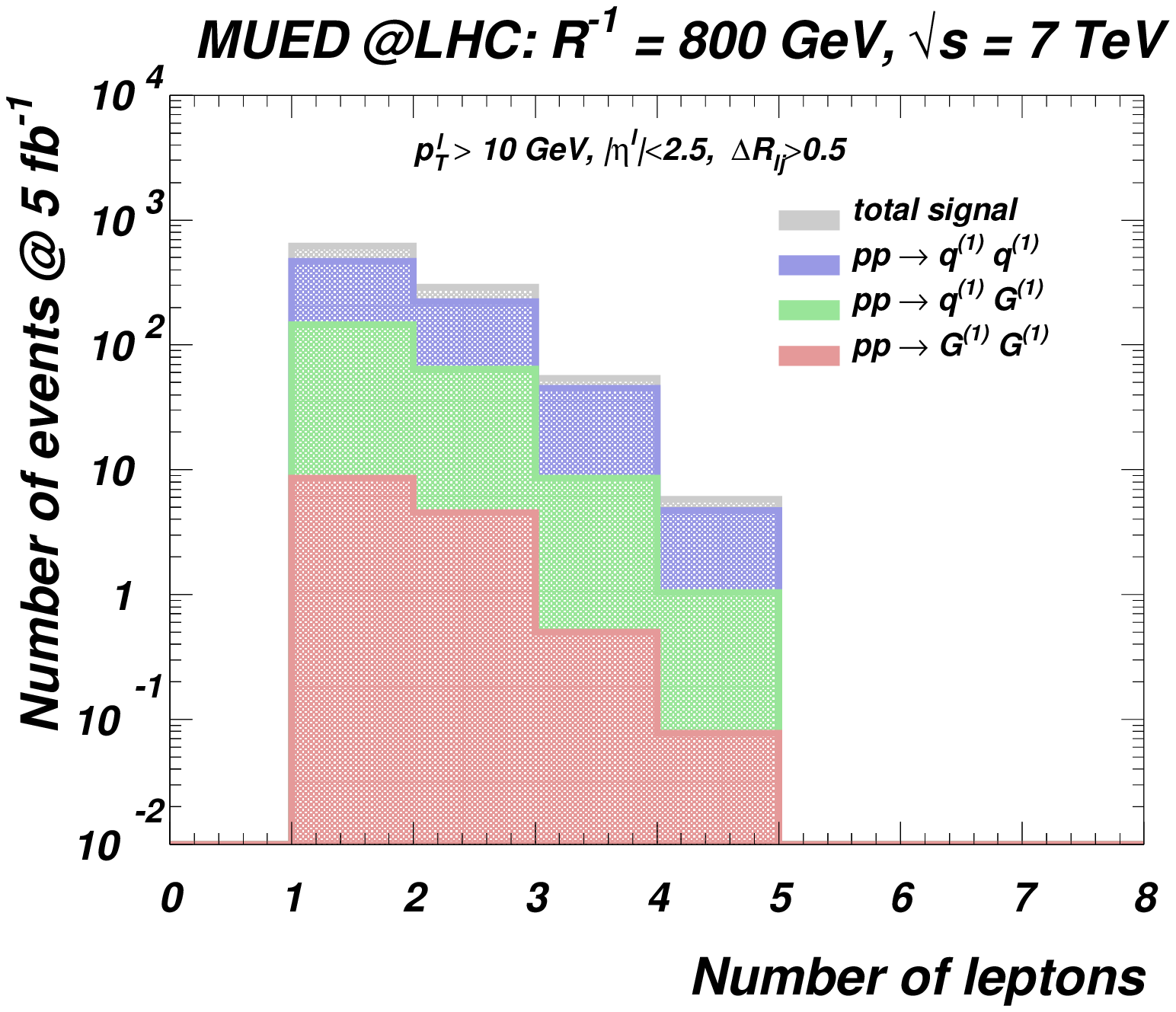}%
\includegraphics[width=0.52\textwidth]{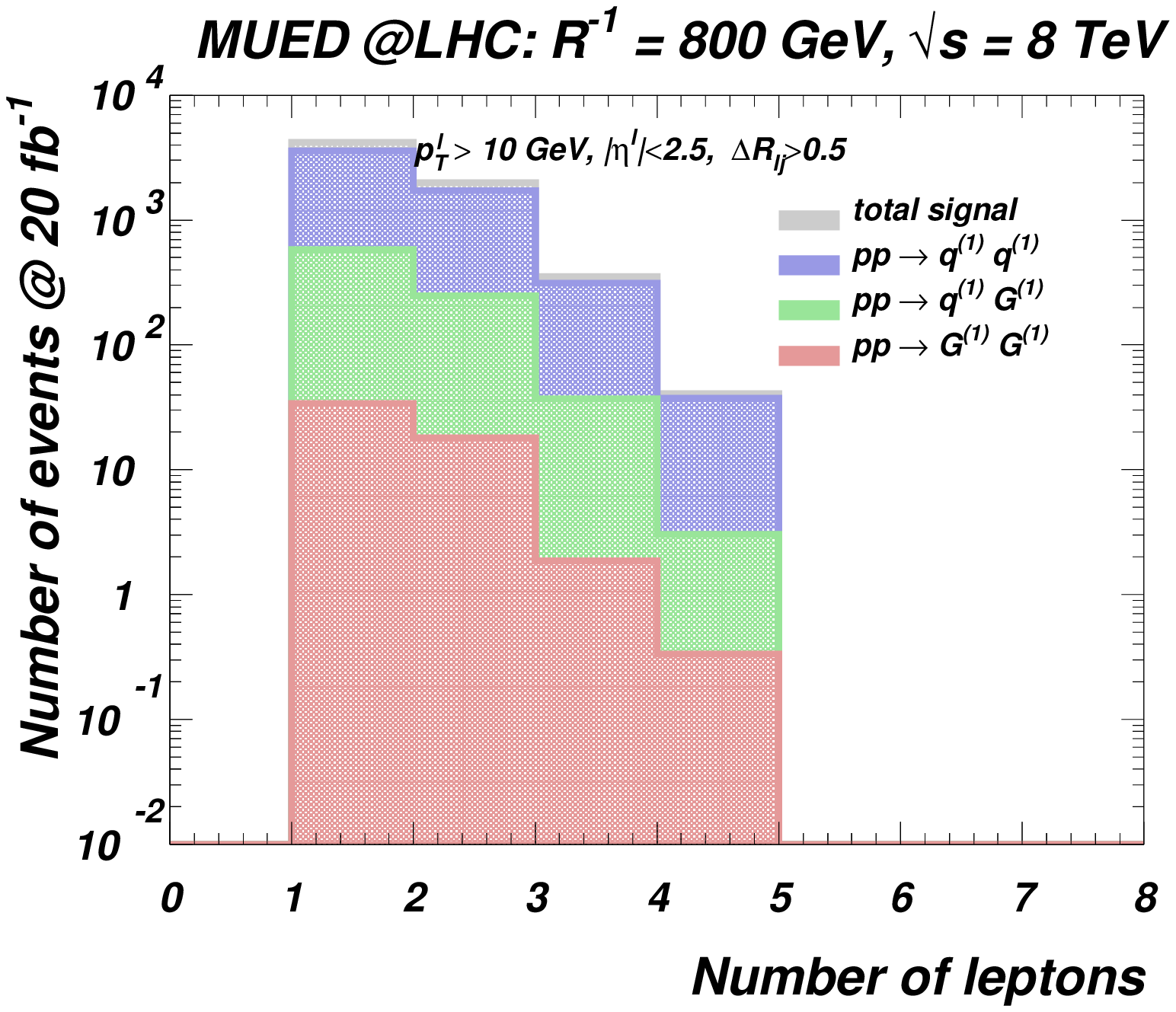}
\caption{Lepton multiplicity distribution in MUED for production and subsequent decay of KK quarks and gluons for $\sqrt{s} = 7$ (left), 8 (right)~TeV, before (top) and after (bottom) acceptance and isolation cuts. $R^{-1}$=800 GeV and $\Lambda = 20R^{-1}$.
The cut of $P_T^{\ell_1}>20$~GeV on the most energetic lepton
was required for all distributions.}
\label{fig:nl-s}
\end{figure}

In figure~\ref{fig:nl-s}
we present the charged lepton multiplicity distributions (calculated in CalcHEP using our MUED model implementation)
for signal sub-processes $pp\to G^{(1)}G^{(1)}/G^{(1)} q^{(1)}/q^{(1)} q^{(1)}$
and their sum for $\sqrt{s}=7$ (left frames) and 
$\sqrt{s}=8$ (right frames) TeV
after the following $P_T^{\ell_1}$ cut for the most energetic lepton:
\begin{equation}
P_T^{\ell_1}> 20\text{ GeV}.
\label{eq:cut-ptlmax}
\end{equation}
The top and bottom frames of figure~\ref{fig:nl-s} present results respectively
before and after the following acceptance and selection cuts:
\begin{equation}
P_T^{\ell_1}> 20\text{ GeV}, \ \ \ P_T^\ell\text{(all)}>10\mbox{\ GeV}, \ \ \ |\eta_\ell|<2.5, \ \ \ 
\Delta R_{\ell j}= \sqrt{\Delta \phi_{\ell j}^2+\Delta \eta_{\ell j}^2}>0.5,
\label{eq:cut-acc}
\end{equation}
where $\ell=e^\pm, \mu^\pm$, which we have summed together.

For the chosen benchmark $R^{-1}=800\,\text{GeV}, \Lambda=20 R^{-1}$,
the mass of the KK gluon is 978~GeV while mass of the different KK quarks is
in the range 904--925 GeV, depending on quark flavour.
The  tree-level cross sections are given in Table~\ref{tab:cs}.
\begin{table}[htb]
\centering
\begin{tabular}{|l|l|l|}
\hline
& \multicolumn{2}{c|}{Cross section (fb)}           \\
\hline
Process  &    $\sqrt{s}=7$ TeV & $\sqrt{s}=8$ TeV	\\
\hline
$pp\to q^{(1)} q^{(1)}$	&   523	    & 1000		 	\\
$pp\to G^{(1)} q^{(1)}$	&   205	    & 493		 	\\
$pp\to G^{(1)} G^{(1)}$	&   15.8    & 45.2		 	\\
Total                   &   744     & 1538          \\
\hline
\end{tabular}
\caption{\label{tab:cs}
Tree-level cross sections for 
 $pp\to G^{(1)}G^{(1)}/G^{(1)} q^{(1)}/q^{(1)} q^{(1)}$ production
 for $\sqrt{s}=7$ and 8 TeV.}
\end{table}
One can see that the total cross section
doubles with the collider energy increase from 7 to 8 TeV.

It is worth stressing that jets and leptons are quite soft because of the
small mass split of the KK spectrum,
so one should try to keep the $P_T$ selection cuts as moderate as possible.
The lepton cuts can certainly be kept much softer than those for jets --
this is related to both identification and trigger cuts.
Even after quite moderate acceptance/selection cuts, as one can see from 
figure~\ref{fig:nl-s}, the lepton multiplicity does not go beyond 4.

In this paper we restrict ourselves to only analysis of the tri-lepton signature
for several reasons:
1) it is a clean signature with a comparatively low background level as we demonstrate below;
2) the signal rate is quite high for this signature
in the range of $R^{-1}$ that we have focused on;
3) in this paper we restrict ourselves to parton-level analysis
which is actually is quite reliable for this signature --
this is especially important for the background estimation.

Although we restrict ourself to parton-level analysis, we do take into account realistic electromagnetic energy resolution, using a value of
$0.15/\sqrt{E(\text{GeV})}$, which is typical for the ATLAS and CMS detectors, as well as their typical hadronic energy resolution of $0.5/\sqrt{E(\text{GeV})}$,
and perform the respective Gaussian smearing for lepton and  quarks.

\subsection{Signal-versus-background analysis}

We consider the following   backgrounds, 
ordered according to their relative contribution to the tri-lepton signature 
under study.
\begin{enumerate}
\item
Di-boson production ($VV$, $V=W^\pm,Z$)
where we took into account the off-shellness
of one of the bosons decaying into lepton pairs;
we have also included $VV$+jet production with $P_T^j>30$ GeV
to take into account the leading-log QCD corrections and 
correct lepton $P_T$ spectrum.
\item
$t\bar{t}$ production
when both top-quarks decay leptonically,
while the third jet could come from the semi-leptonic
$B$-meson decays. The probability of this is quite low (we found it to be of the order of $10^{-3}$) 
since
the lepton from $B$-meson decays tends not to be isolated from 
hadrons coming from the same decay because of the large momentum
of the b-quark
\item
$t\bar{t}\ell\bar{\ell}$ and $t\bar{t}\ell\bar{\nu} (t\bar{t}\bar{\ell}\nu)$
production
\item
triple gauge boson production processes ($VVV$)
\item
$t\bar{t}VV$ production
\item
background coming from 4-gauge boson production $VVVV$.
\end{enumerate}

\begin{figure}[htb]
\includegraphics[width=0.52\textwidth]{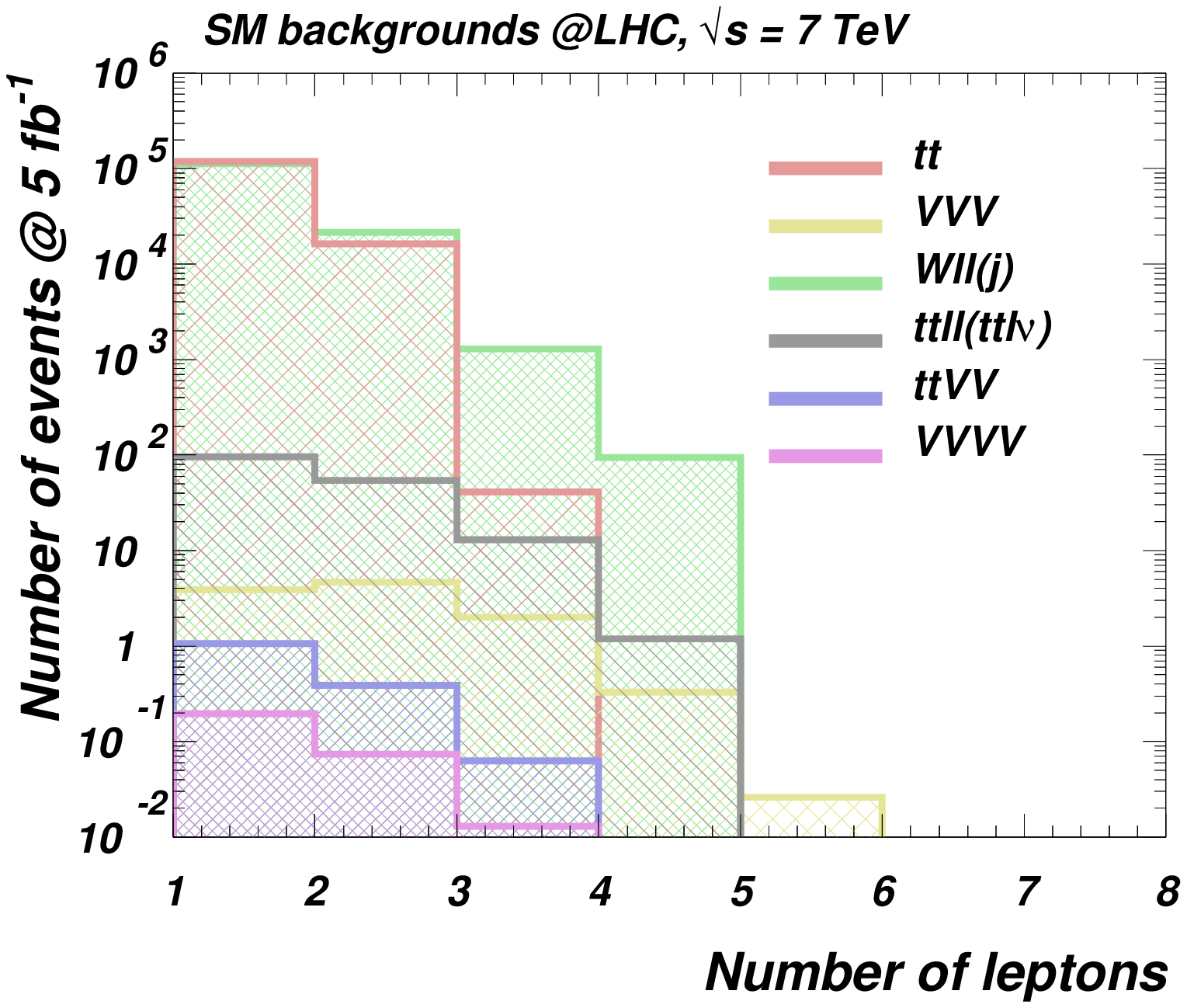}%
\includegraphics[width=0.52\textwidth]{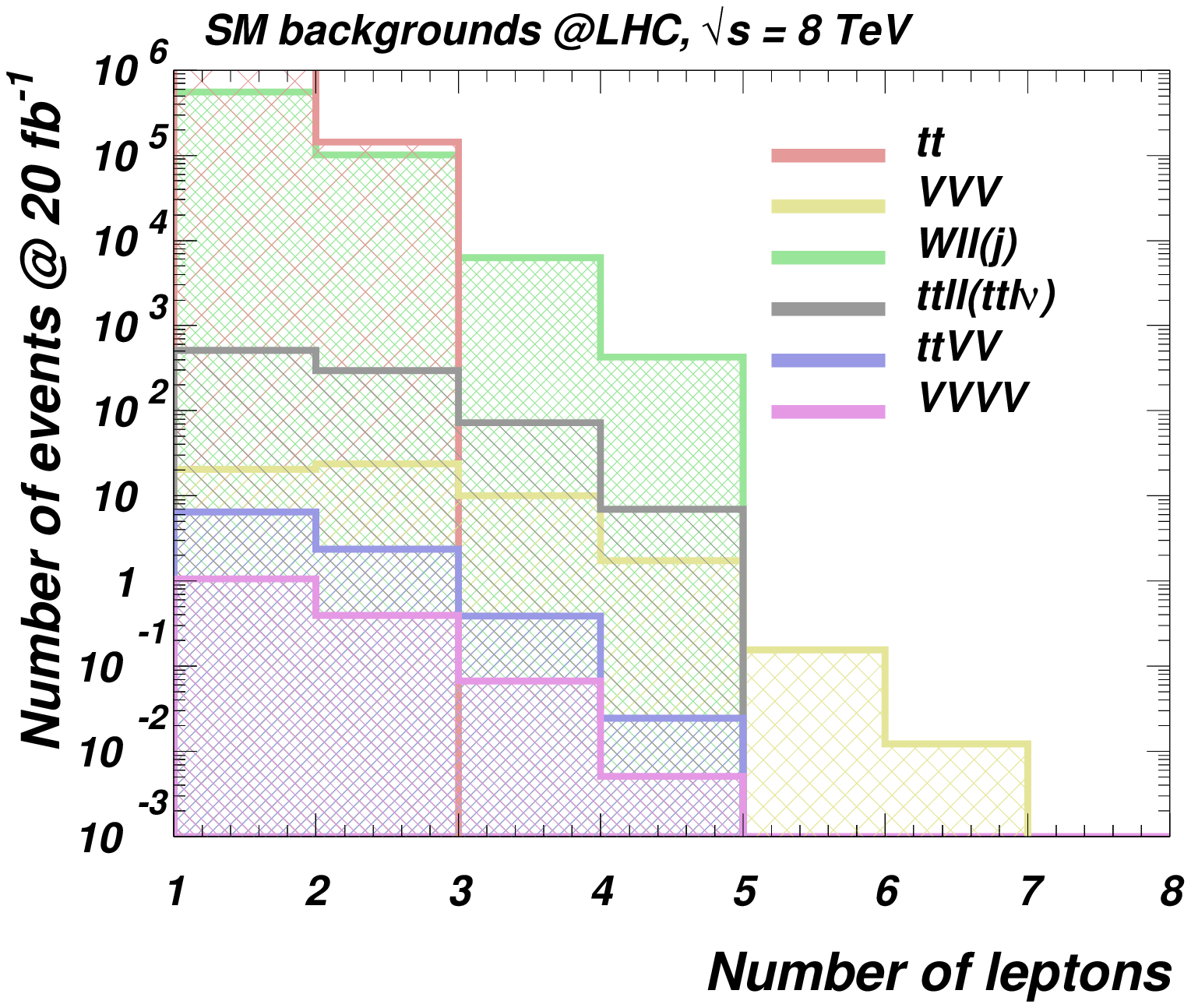}%
\caption{\label{fig:nl-b}Lepton multiplicity distribution for main backgrounds
contributing to the tri-lepton signature.}
\end{figure}

The lepton 
multiplicity of backgrounds contributing to the tri-lepton signature
is presented in  figure~\ref{fig:nl-b} after applying the cuts given in \eqref{eq:cut-ptlmax}--\eqref{eq:cut-acc} for $\sqrt{s}=7$ (left) and 8~TeV (right). As for the signals, we calculated these backgrounds in CalcHEP, using our MUED model implementation. 
One can see that indeed the $VV(j)$ background is the dominant one.
Its contribution to  the tri-lepton signature is about one order of magnitude above the
$t\bar{t}$, $t\bar{t}\ell\ell$ and $t\bar{t}\ell\nu$ which in turn are bigger
than the background from $VVV$ production. The
$t\bar{t}VV$ and  $VVVV$ backgrounds are virtually negligible.

\begin{figure}[htb]
\includegraphics[width=0.52\textwidth]{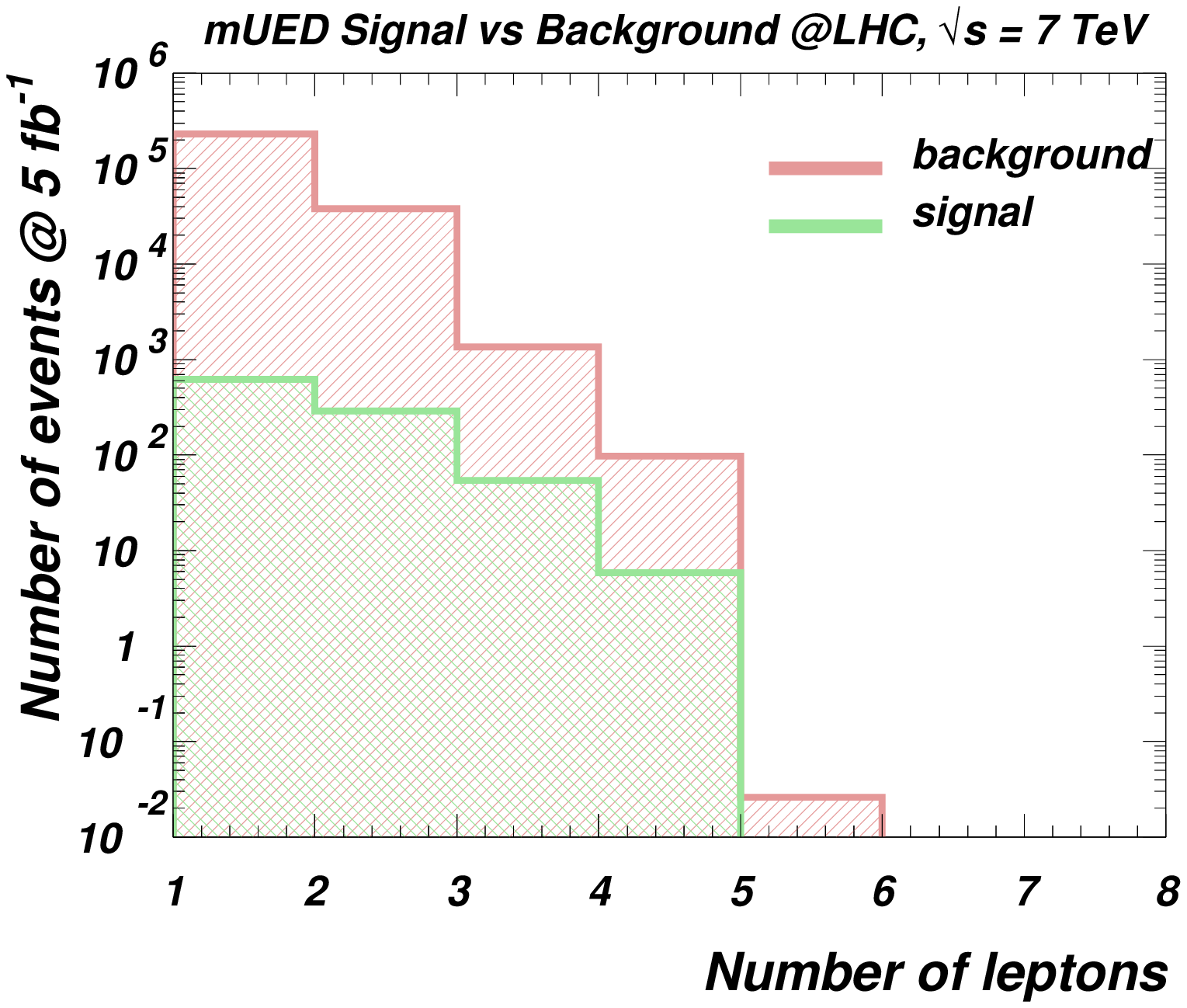}%
\includegraphics[width=0.52\textwidth]{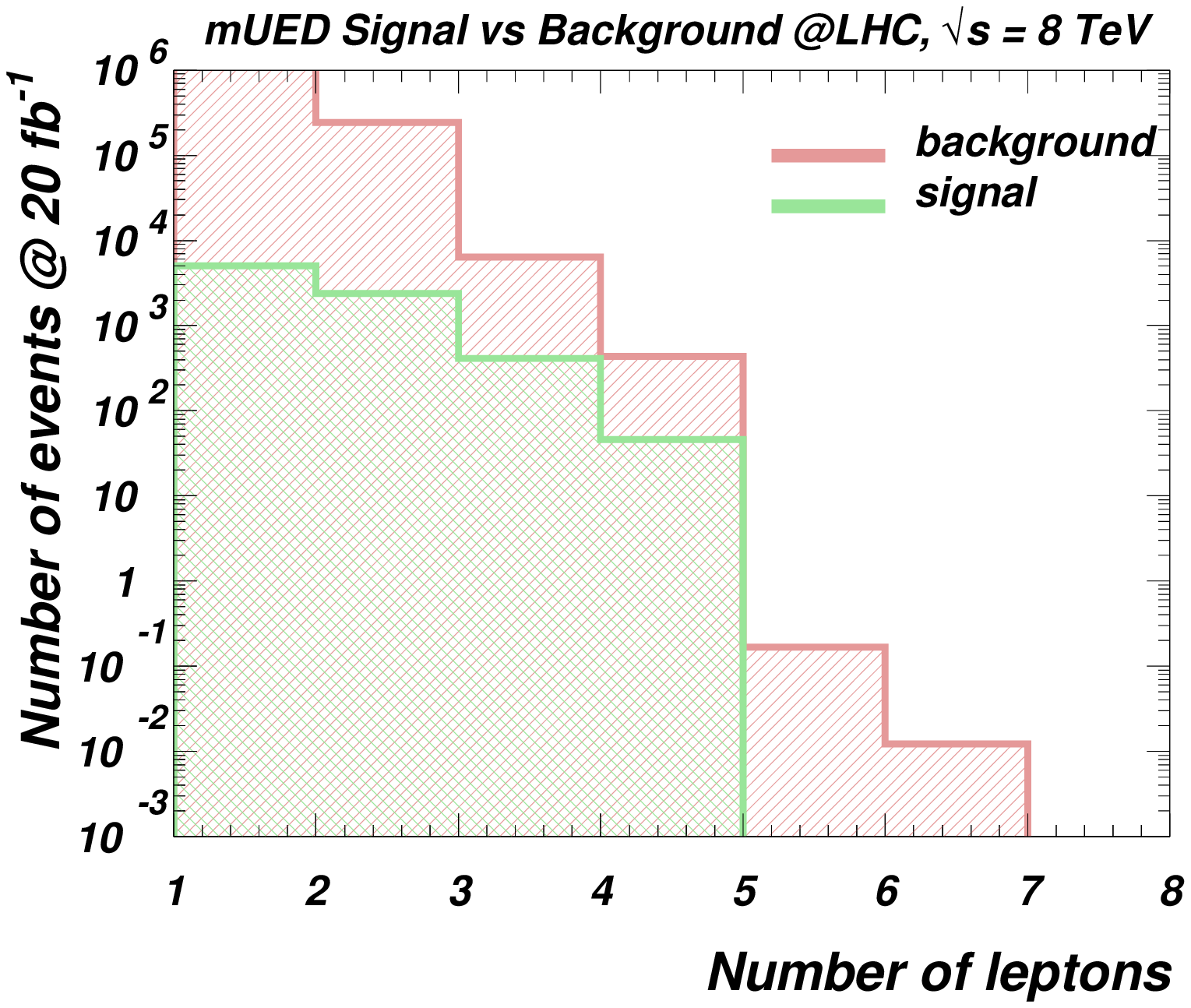}%
\caption{Lepton multiplicity distribution for  background versus signal after the initial cuts \eqref{eq:cut-ptlmax}--\eqref{eq:cut-acc}, using $R^{-1} = 800$~GeV and $\Lambda = 20 R^{-1}$.}
\label{fig:nl-sb}
\end{figure}

In figure~\ref{fig:nl-sb} we present the lepton multiplicity for the total background
and the signal after the cuts in \eqref{eq:cut-ptlmax}--\eqref{eq:cut-acc} for $\sqrt{s}=7$ (left) and 8~TeV (right).
One can see that after basic cuts the number of signal events is about
a factor of 20 below the background level
for the chosen benchmark of $R^{-1}=800$~GeV and $\Lambda = 20R^{-1}$:
for $\sqrt{s}=7$~TeV and 5 fb$^{-1}$ integrated luminosity
one expects about 60 signal versus 1200 background events in tri-lepton channel,
while 
for $\sqrt{s}=8$~TeV and 20 fb$^{-1}$ integrated luminosity
one expects about 400 signal versus 8000 background events.

\begin{figure}[htb]
\includegraphics[width=0.52\textwidth]{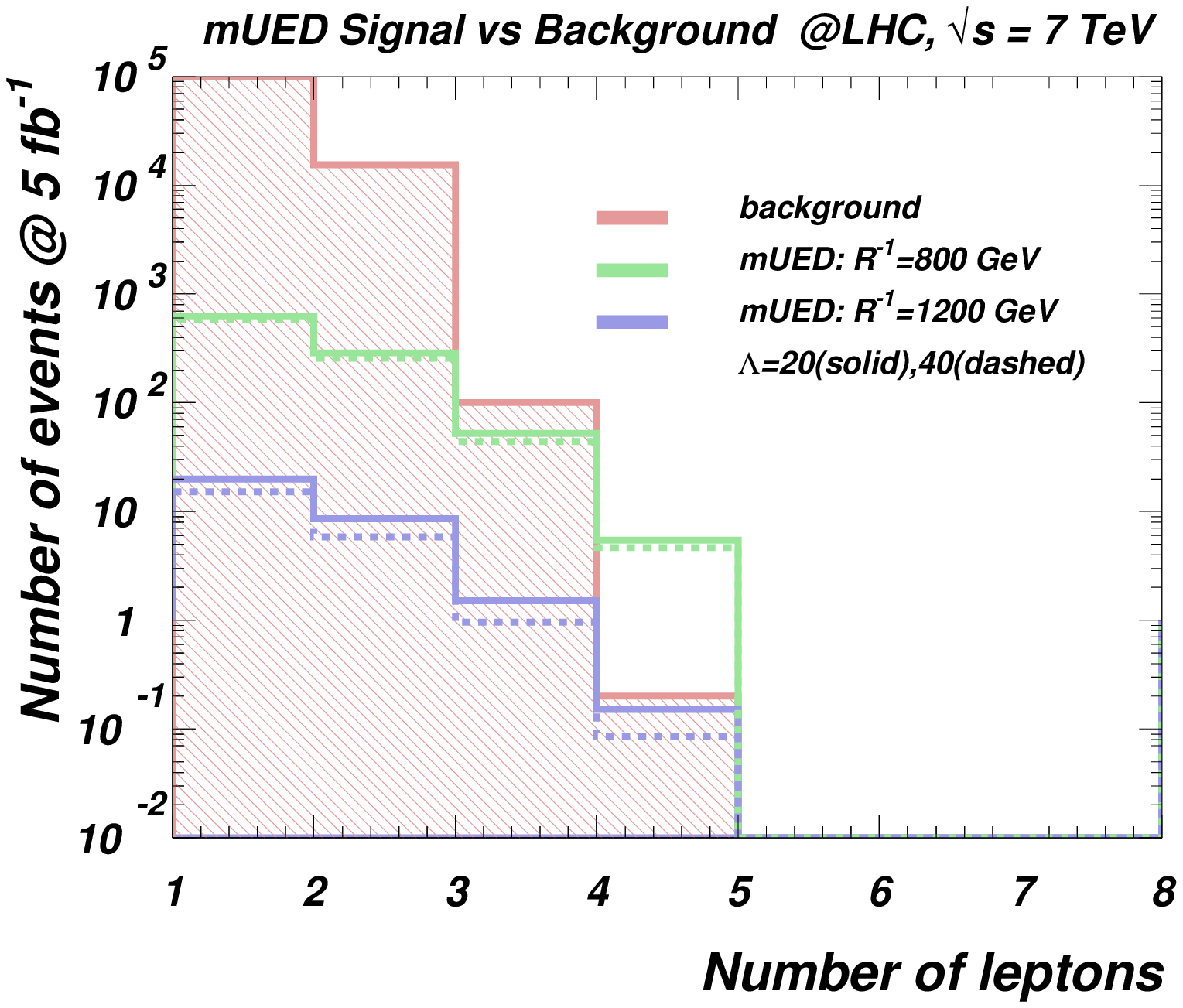}%
\includegraphics[width=0.52\textwidth]{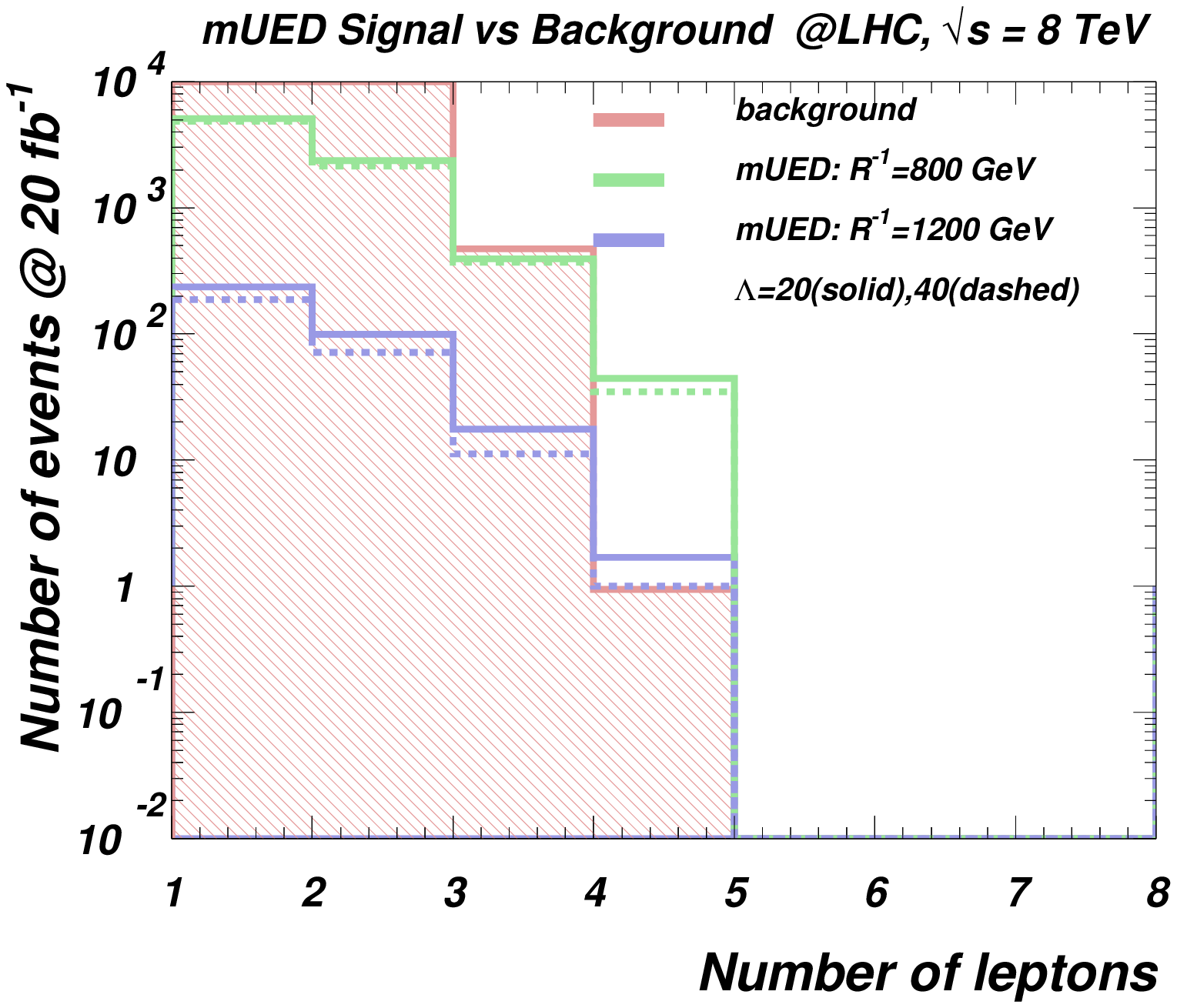}%
\caption{\label{fig:nl1-sb}Lepton multiplicity distribution for  background and signal after \eqref{eq:cut-ptlmax}--\eqref{eq:cut-wc} cuts.}
\end{figure}

Our task now is to explore kinematical variables in order to suppress background while leaving the signal intact.
One of them is the invariant mass $M_{\ell\bar\ell}$ of the most energetic opposite-sign leptons,
presented in figure~\ref{fig:mll-sb} for signal and background.
In this and following figure we present cases for
$R^{-1}=800$~GeV and $R^{-1}=1200$ GeV. For each case we present results for 
$\Lambda=20R^{-1}$ and $40R^{-1}$ to give the reader an idea of how distributions evolve with $\Lambda$. One can see that increase of $\Lambda$ slightly 
hardens the $M_{\ell\bar\ell}$ spectrum since the absolute values of  KK particle masses as well as the mass split among them increases as  $\Lambda$ increases; this was shown above in  figure~\ref{fig:lambda-dependence}.
\begin{figure}[htb]
\includegraphics[width=0.52\textwidth]{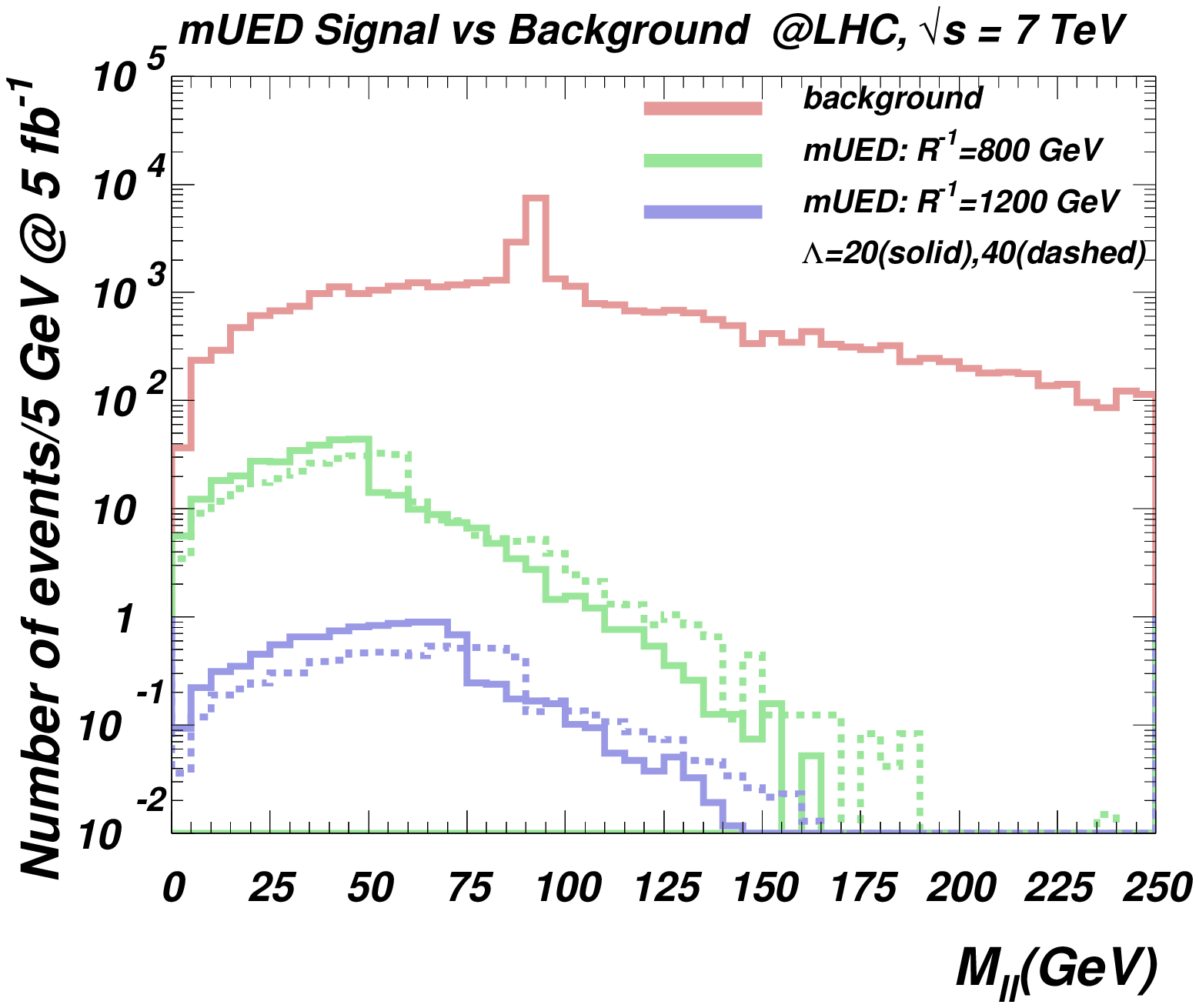}%
\includegraphics[width=0.52\textwidth]{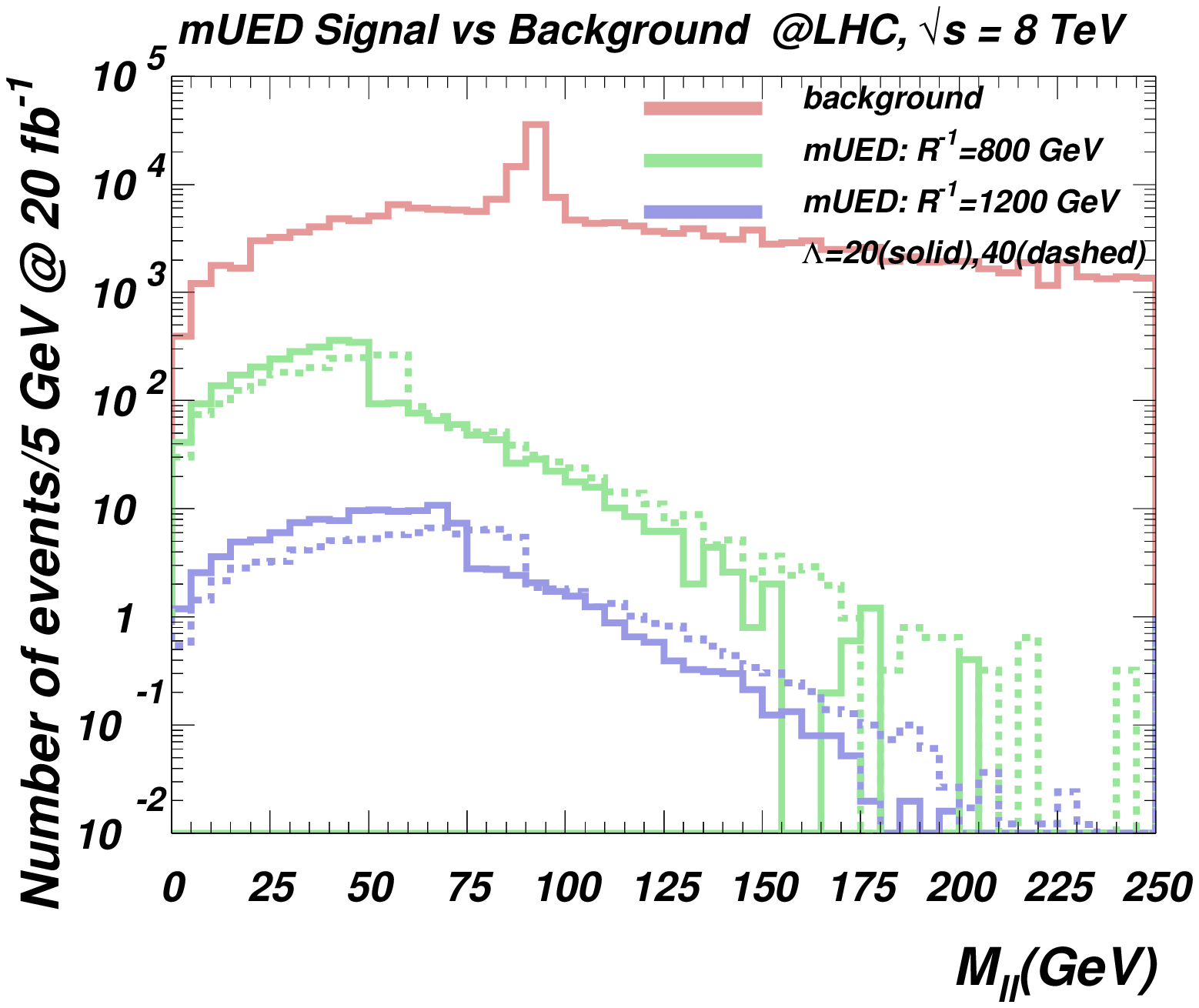}%
\caption{Invariant mass of the two most energetic (highest $P_T$)
opposite-sign leptons.}
\label{fig:mll-sb}
\end{figure}
One can clearly see that the background receives a substantial contribution from the real $Z$-boson while signal does not.
We thus veto $M_{\ell\bar\ell}$ around $m_Z$ with the cut
\begin{equation}
|m_Z-M_{\ell\bar\ell}|>10 \, \text{GeV}.
\label{eq:cut-wc}
\end{equation}
In figure~\ref{fig:nl1-sb} we present the lepton multiplicity distribution
for signal and background after cuts \eqref{eq:cut-ptlmax}--\eqref{eq:cut-wc}. 
The background for the tri-lepton signature
is reduced by a factor of about 20 down to around the level of the signal for
$R^{-1}=800$~GeV.
 
\begin{figure}[htb]
\includegraphics[width=0.52\textwidth]{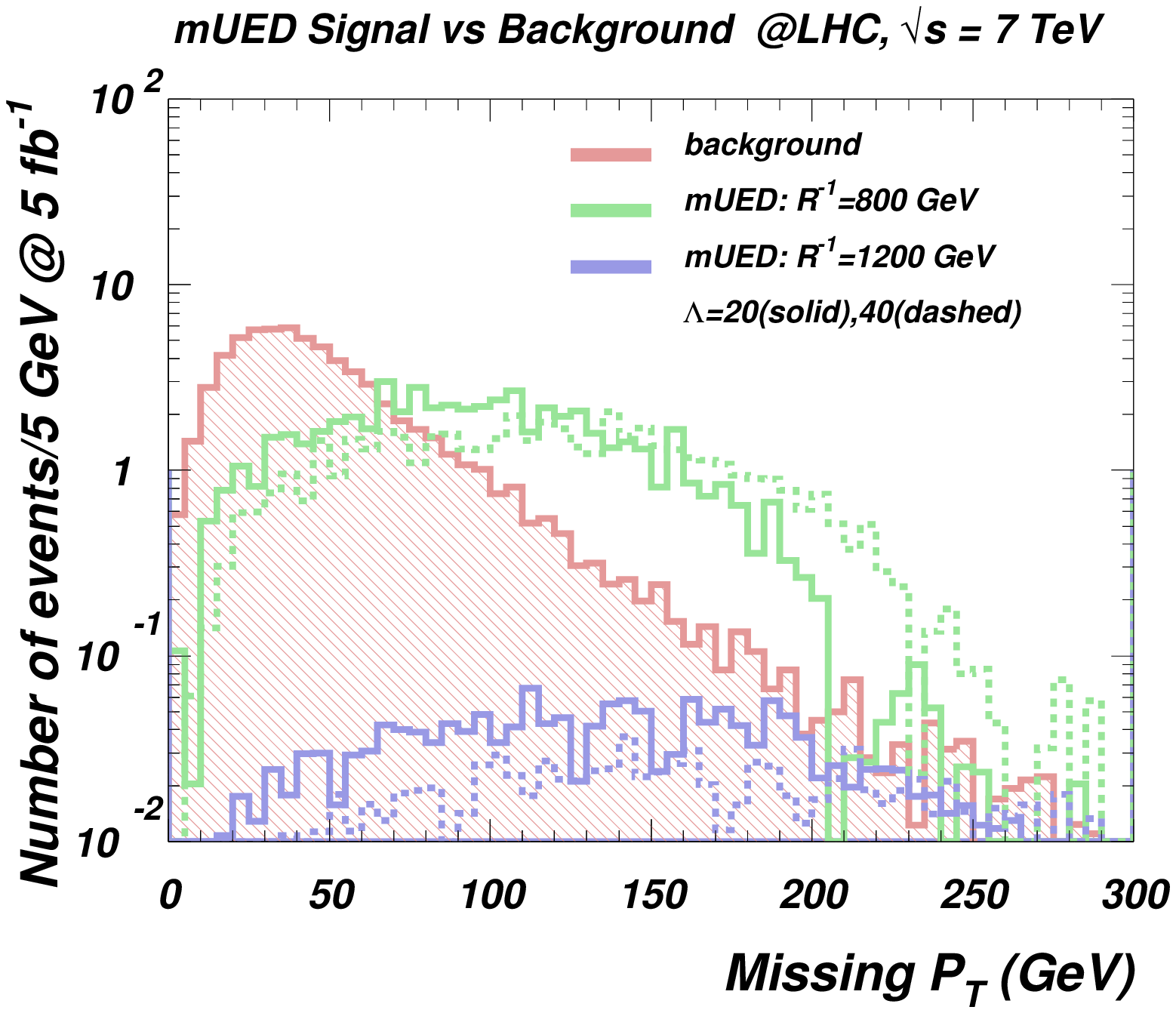}%
\includegraphics[width=0.52\textwidth]{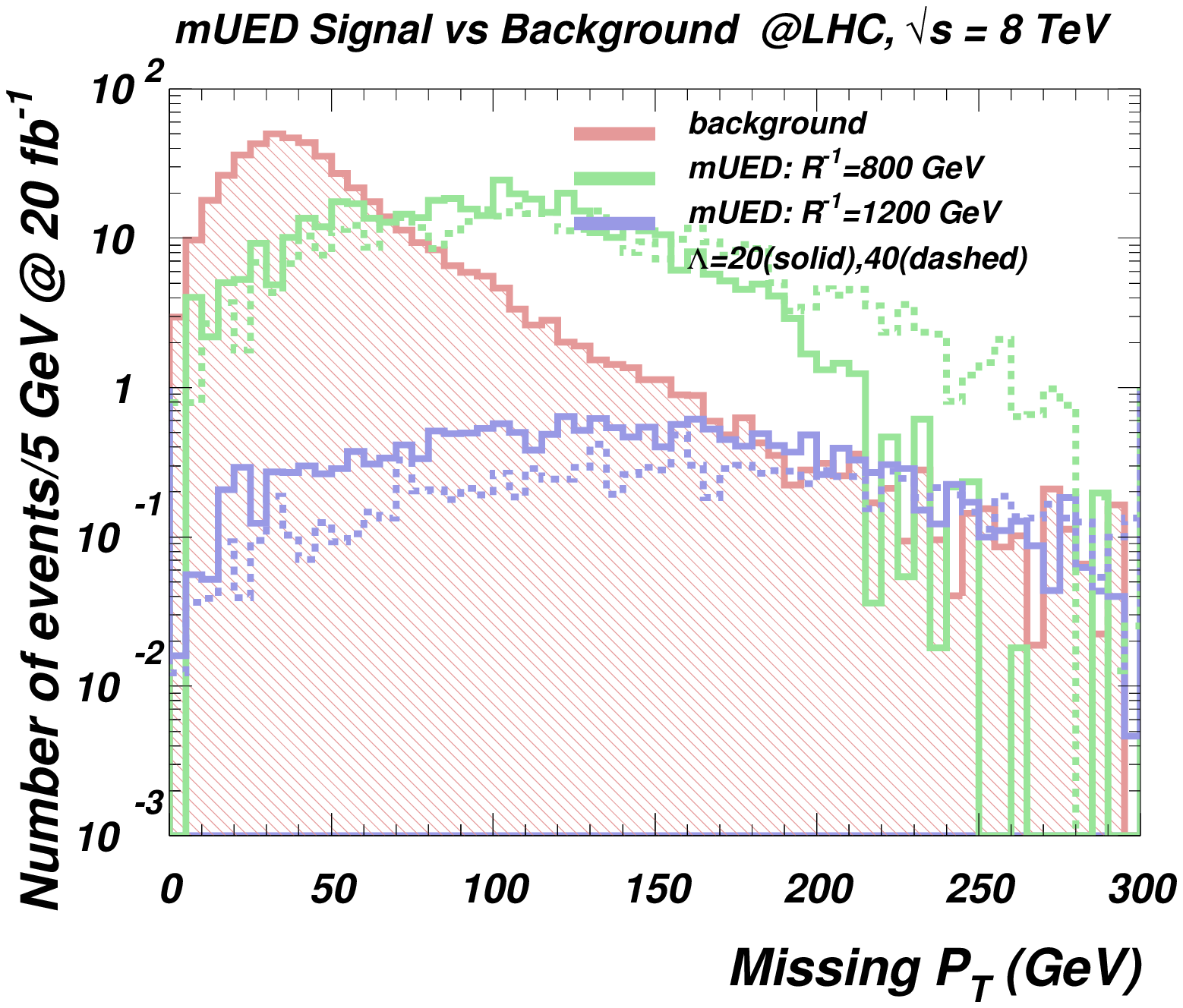}%
\caption{Missing transverse momentum $\ptmis$ distribution for   
background versus signal after  \eqref{eq:cut-ptlmax}--\eqref{eq:cut-wc} cuts.}
\label{fig:mpt-sb}
\end{figure}

For the next step we analyse the missing transverse momentum distributions 
$\ptmis$ for signal and background
which are presented in figure~\ref{fig:mpt-sb}.
One can see that the maximum  $\ptmis$ takes place at a smaller value of 
$\ptmis$ for the background than for signal. We find that the cut
\begin{equation}
\ptmis>50\,\text{GeV}
\label{eq:cut-ptm}
\end{equation}
is quite safe for the signal while helping to  further reduce the background.

\begin{figure}[htb]
\includegraphics[width=0.52\textwidth]{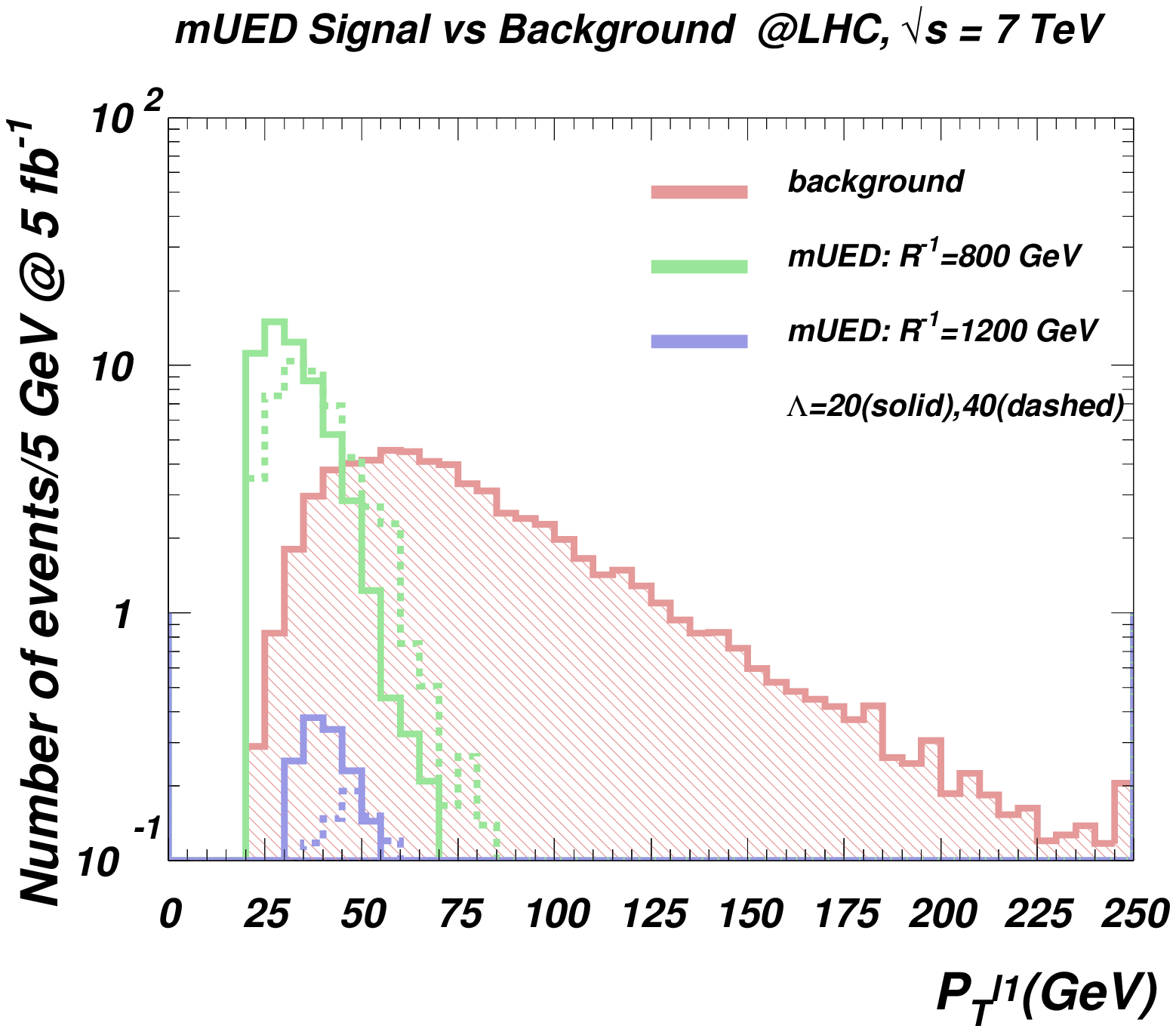}%
\includegraphics[width=0.52\textwidth]{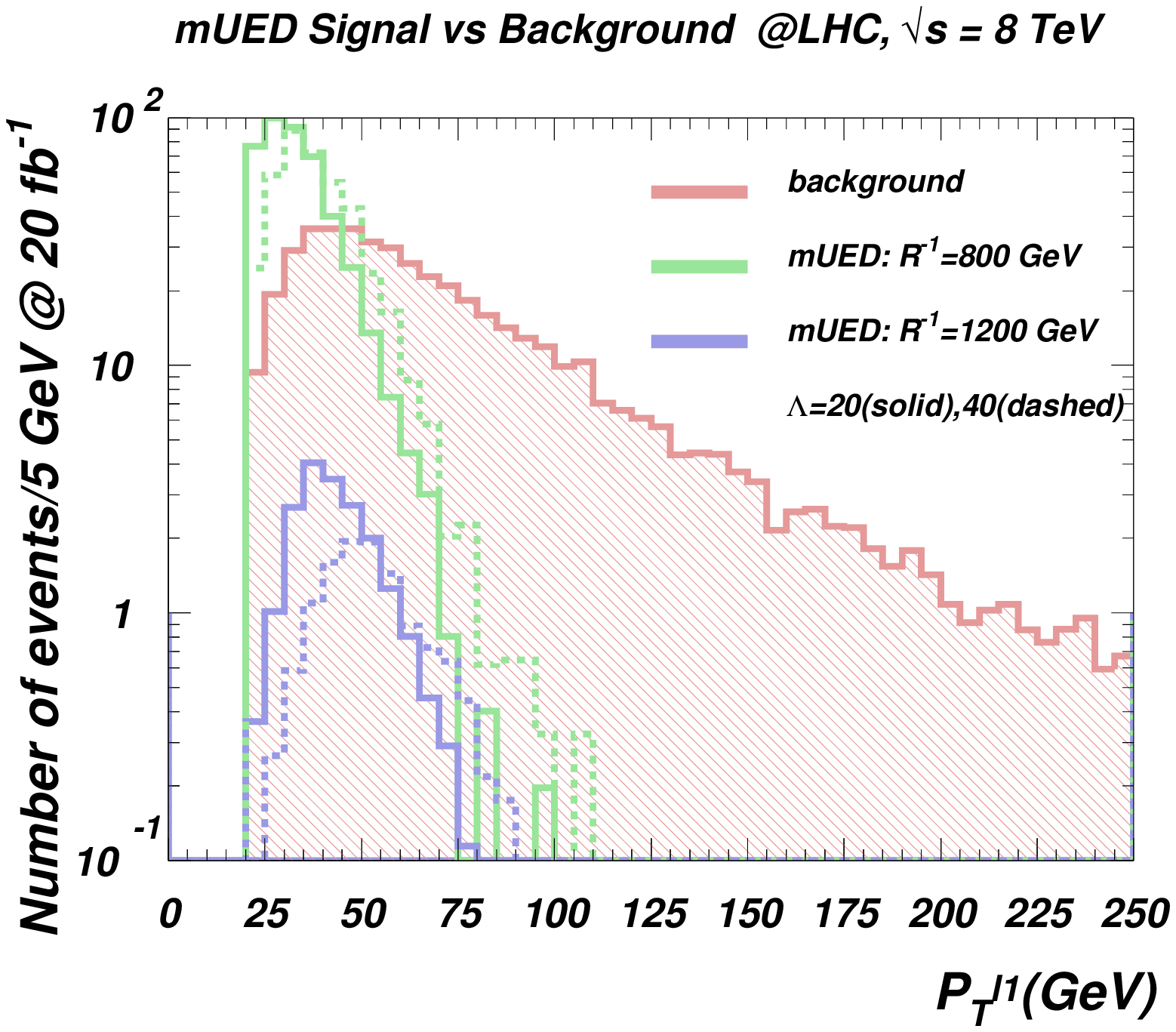}\\
\includegraphics[width=0.52\textwidth]{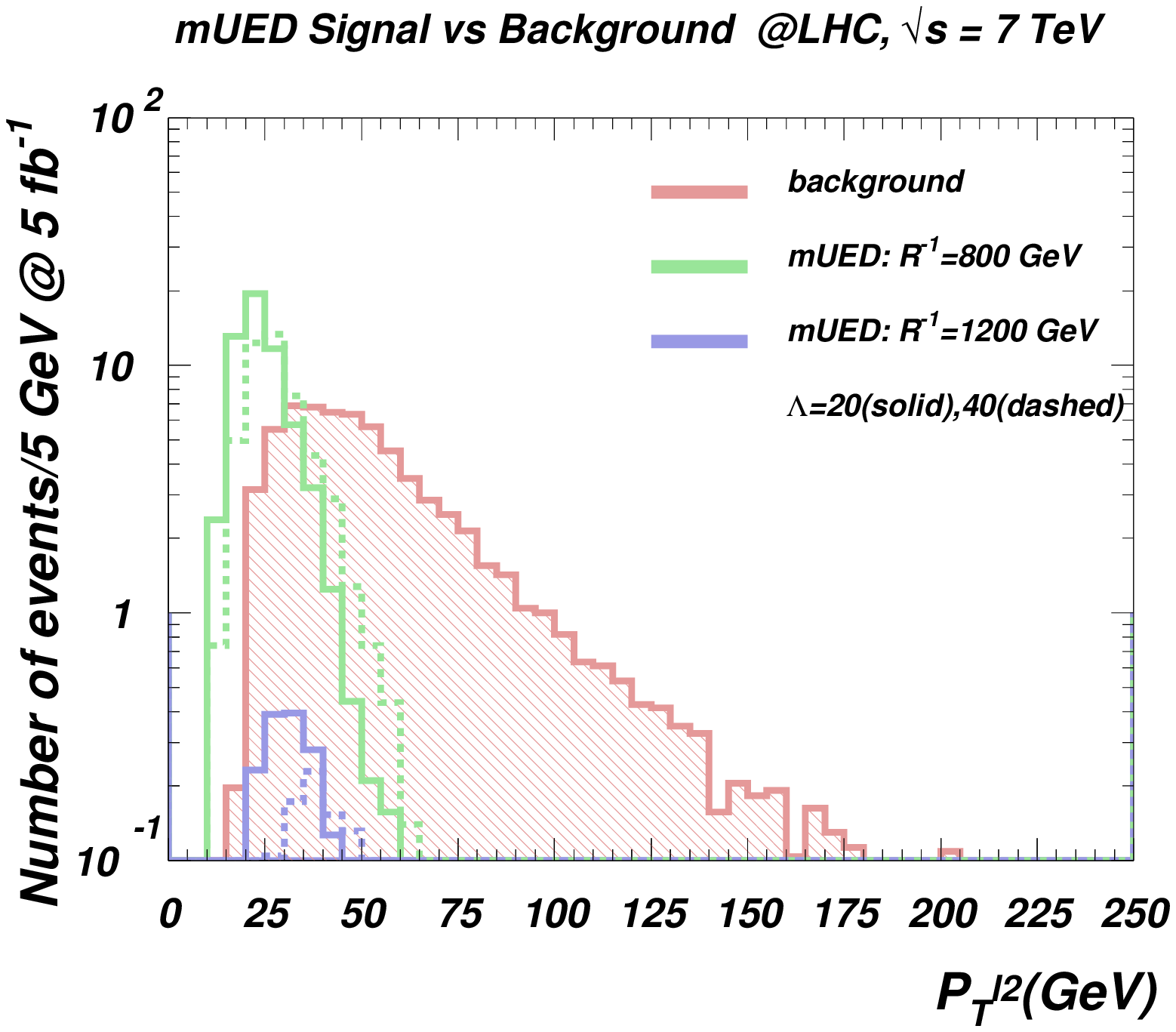}%
\includegraphics[width=0.52\textwidth]{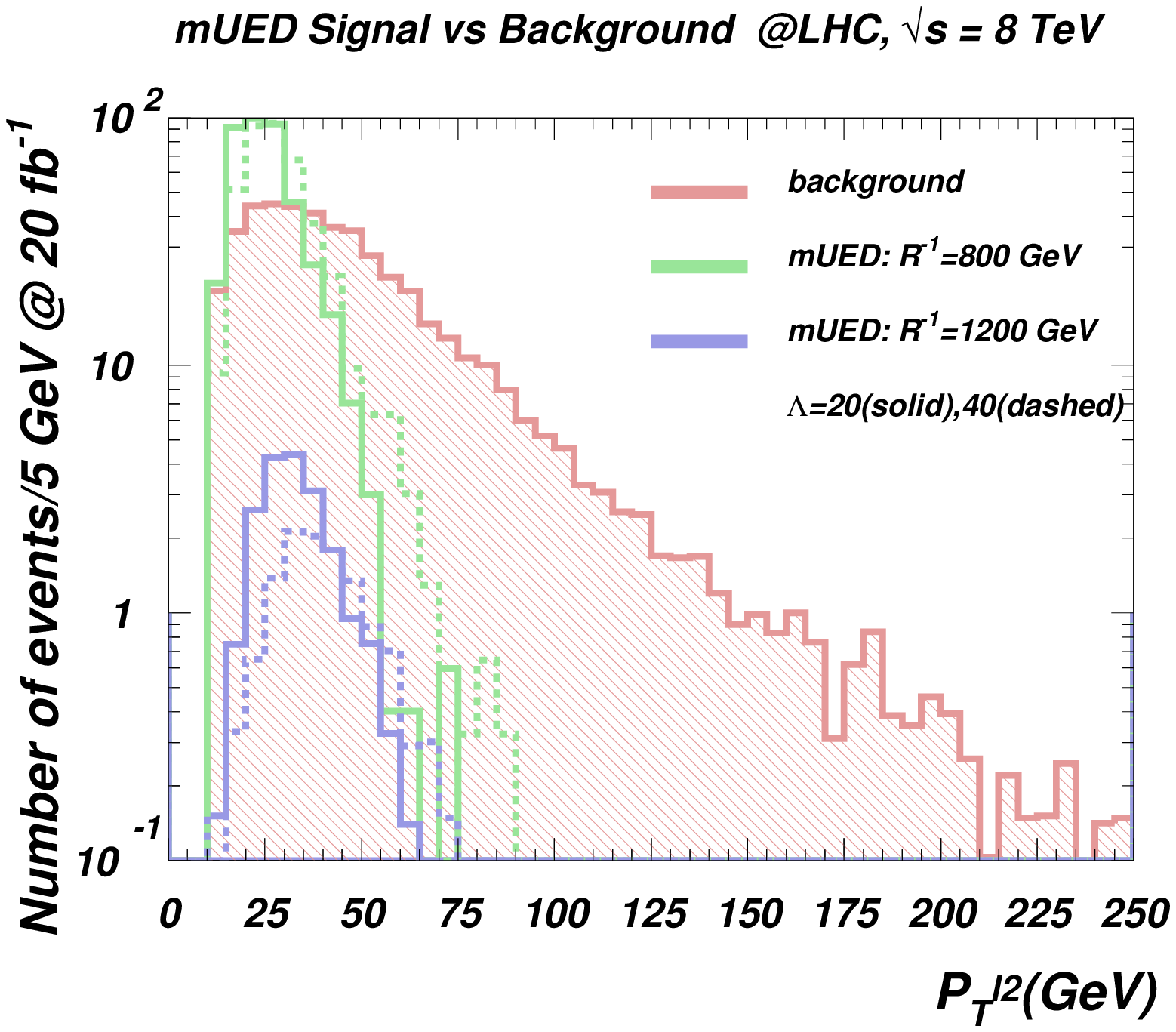}\\
\includegraphics[width=0.52\textwidth]{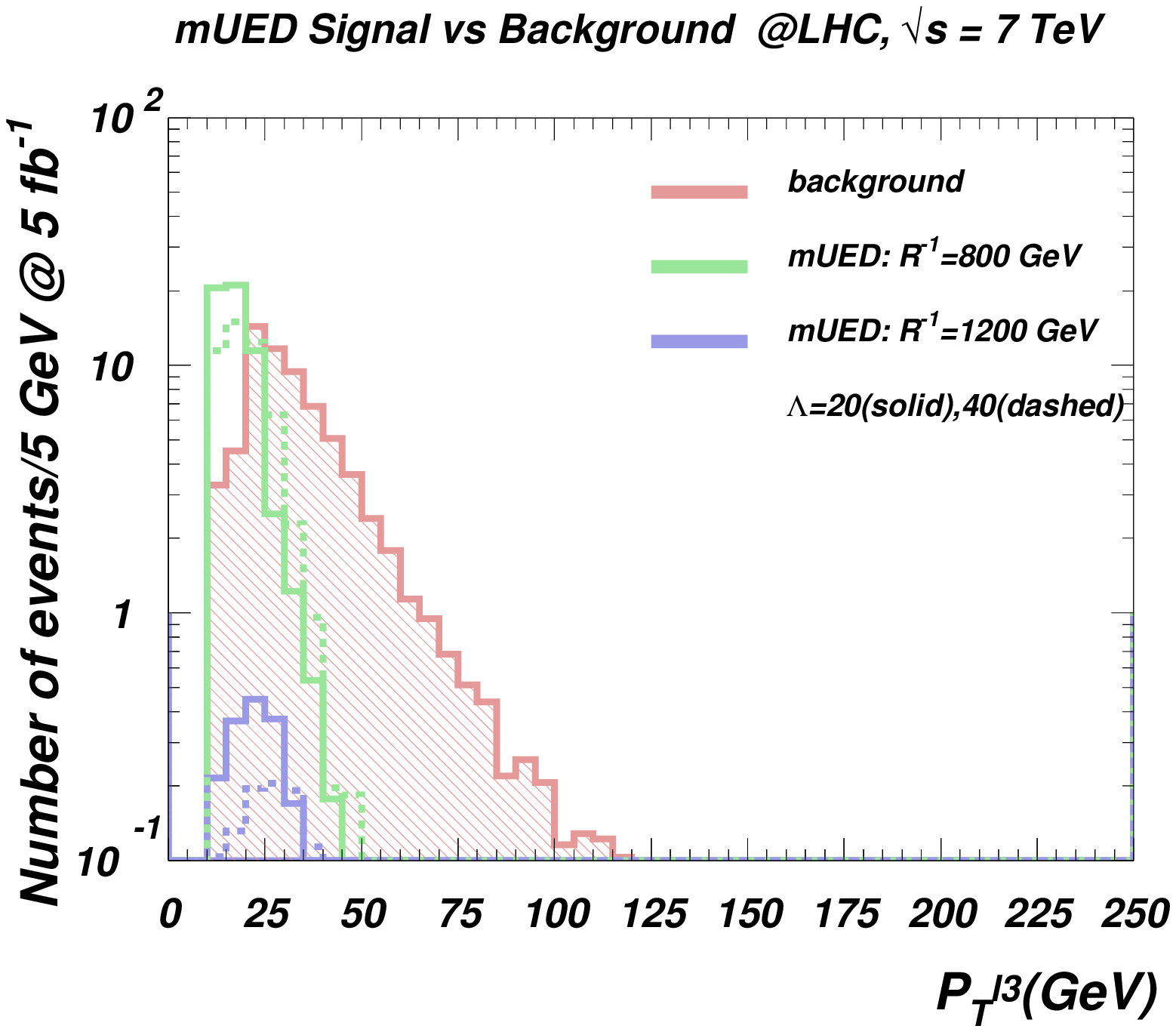}%
\includegraphics[width=0.52\textwidth]{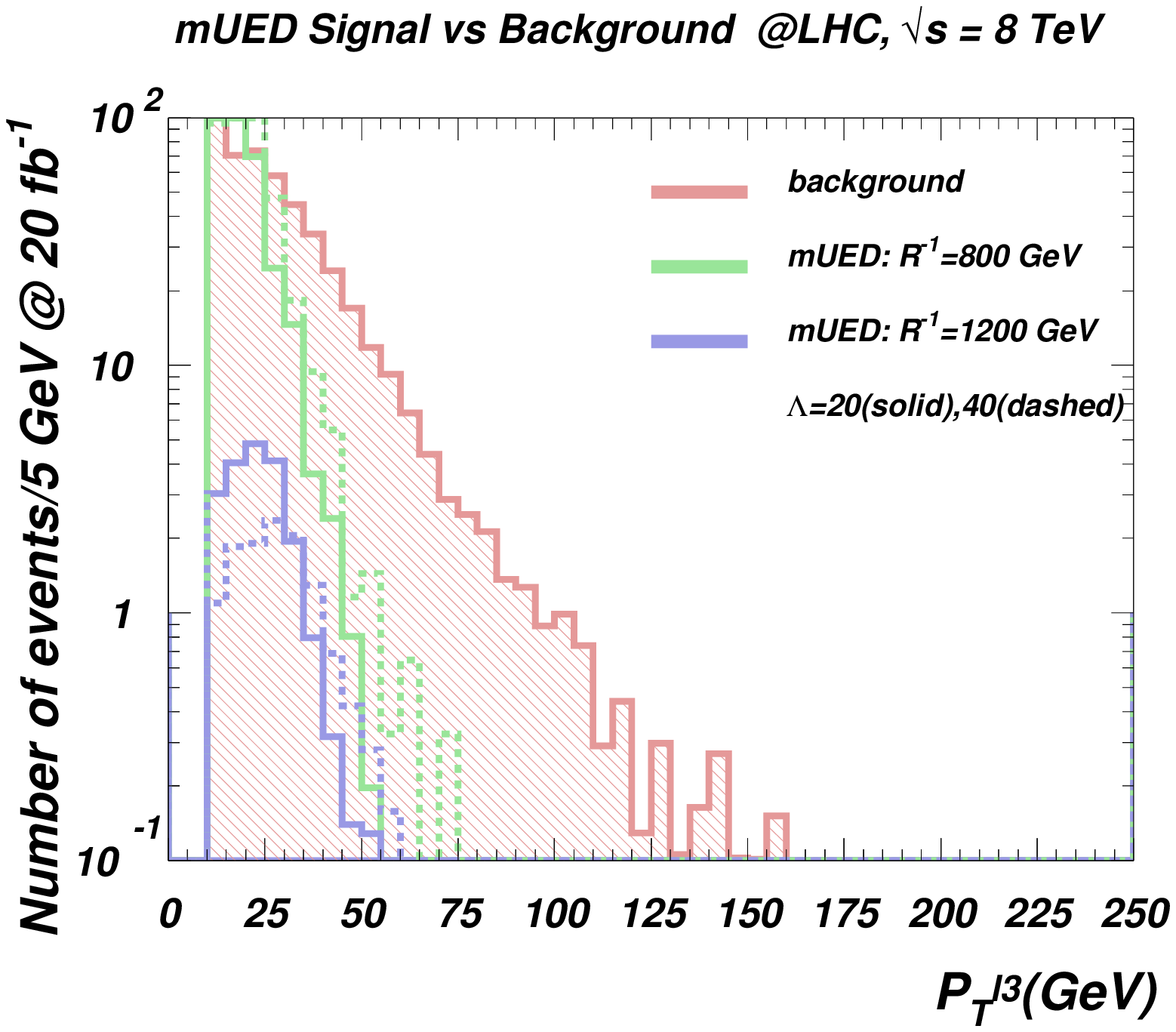}\\
\caption{Transverse momentum of the leptons with the first, second  and the third highest $P_T$, 
 after cuts in \eqref{eq:cut-ptlmax}--\eqref{eq:cut-wc},
 for the 
background and signal.}
\label{fig:ptl-sb}
\end{figure}
To further increase the signal significance we use the properties of the 
transverse momentum of leptons $\ell_1$, $\ell_2$, $\ell_3$ with the first-, second-  and third-highest values of $P_T$ respectively in each event. In figure~\ref{fig:ptl-sb} we plot the signal and background $P_T$ distributions for each of these leptons, $P_T^{\ell_1}$, $P_T^{\ell_2}$ and $P_T^{\ell_3}$.

\begin{figure}[htb]
\includegraphics[width=0.52\textwidth]{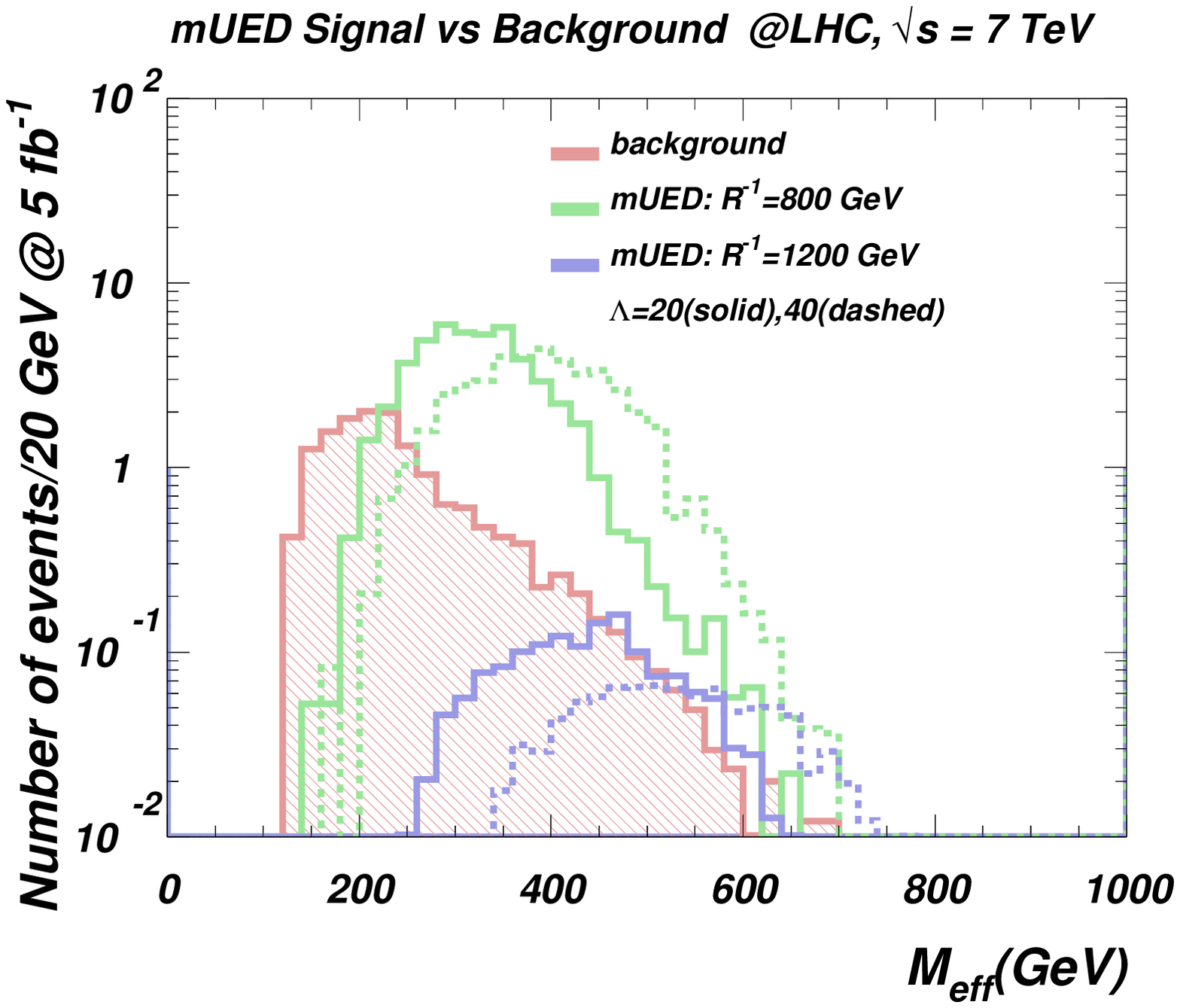}
\includegraphics[width=0.52\textwidth]{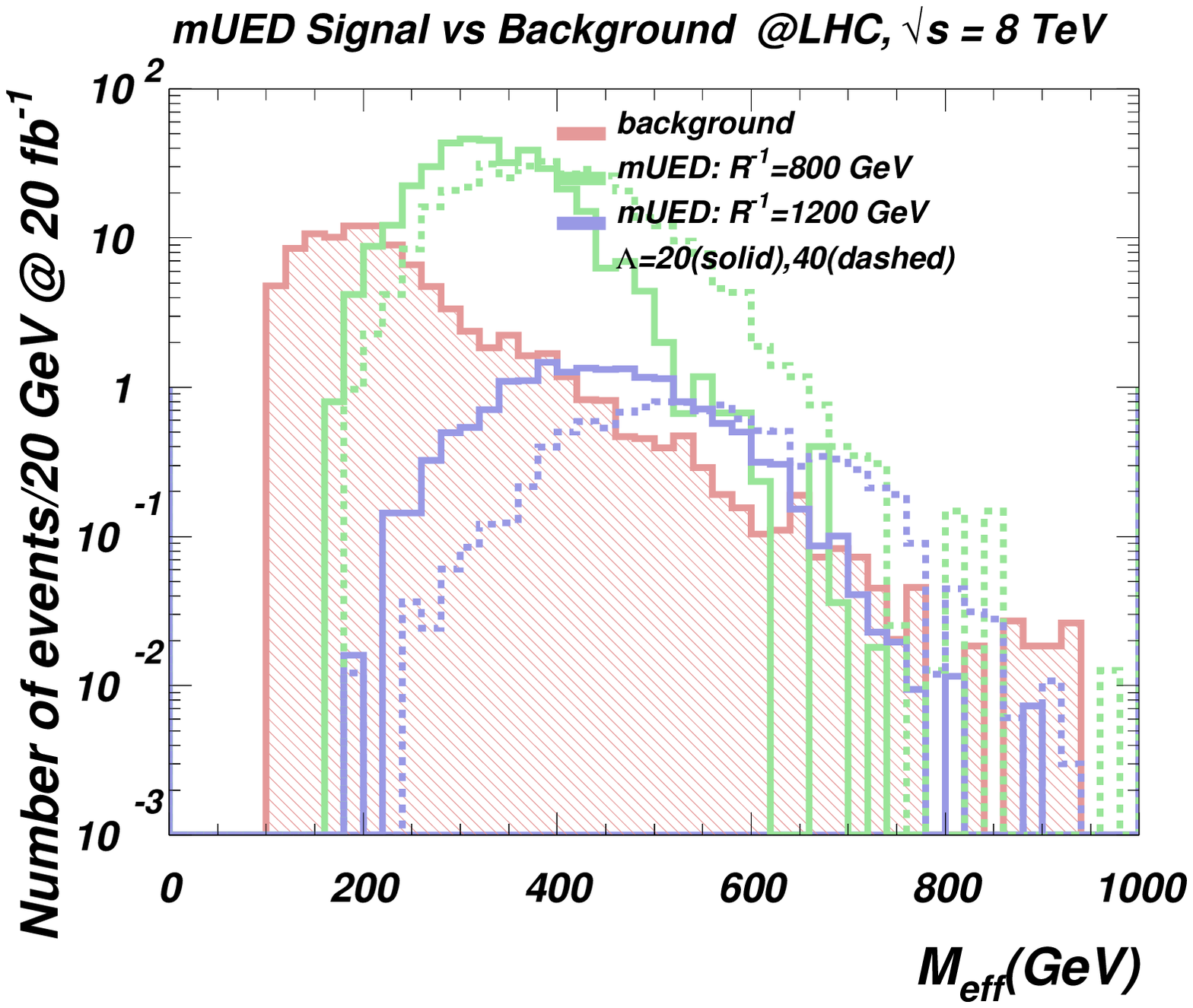}
\caption{Effective mass distribution 
for signal and background after  \eqref{eq:cut-ptlmax}--\eqref{eq:cut-ptl} cuts.}
\label{fig:meff-sb}
\end{figure}

Leptons are clearly  softer in signal events
so the following set of the {\it upper} cuts on the leptons' transverse momenta help us further:
\begin{equation}
 P_T^{\ell_1}<100\,\text{GeV};\quad
 P_T^{\ell_2}<70\,\text{GeV};\quad
 P_T^{\ell_3}<50\,\text{GeV}.
\label{eq:cut-ptl}
\end{equation}
We also  make use of the effective mass $M_{\text{eff}}=\ptmis + \sum_{\ell,j} P_T$ distribution (the sum is over all final-state leptons and jets)
presented  in figure~\ref{fig:meff-sb}
and apply the empirical cut
\begin{equation}
 M_{\text{eff}}> R^{-1}/5
\label{eq:cut-meff}
\end{equation}
as our final selection.
The effective mass $M_{\text{eff}}$ is correlated with the mass of the final state particles,
so it is harder for the MUED signal than for the background.
This distribution is probably the most sensitive to the value of $\Lambda R$
 and could be further used for determination of the  MUED 
parameter space in the case that we observe the signal.
This cut is especially important for optimisation of the search for large 
values of $R^{-1}$ -- it allows us to suppress the background by a further factor  of 2 or 3, leaving the signal almost intact.

\begin{figure}[htb]
\includegraphics[width=0.52\textwidth]{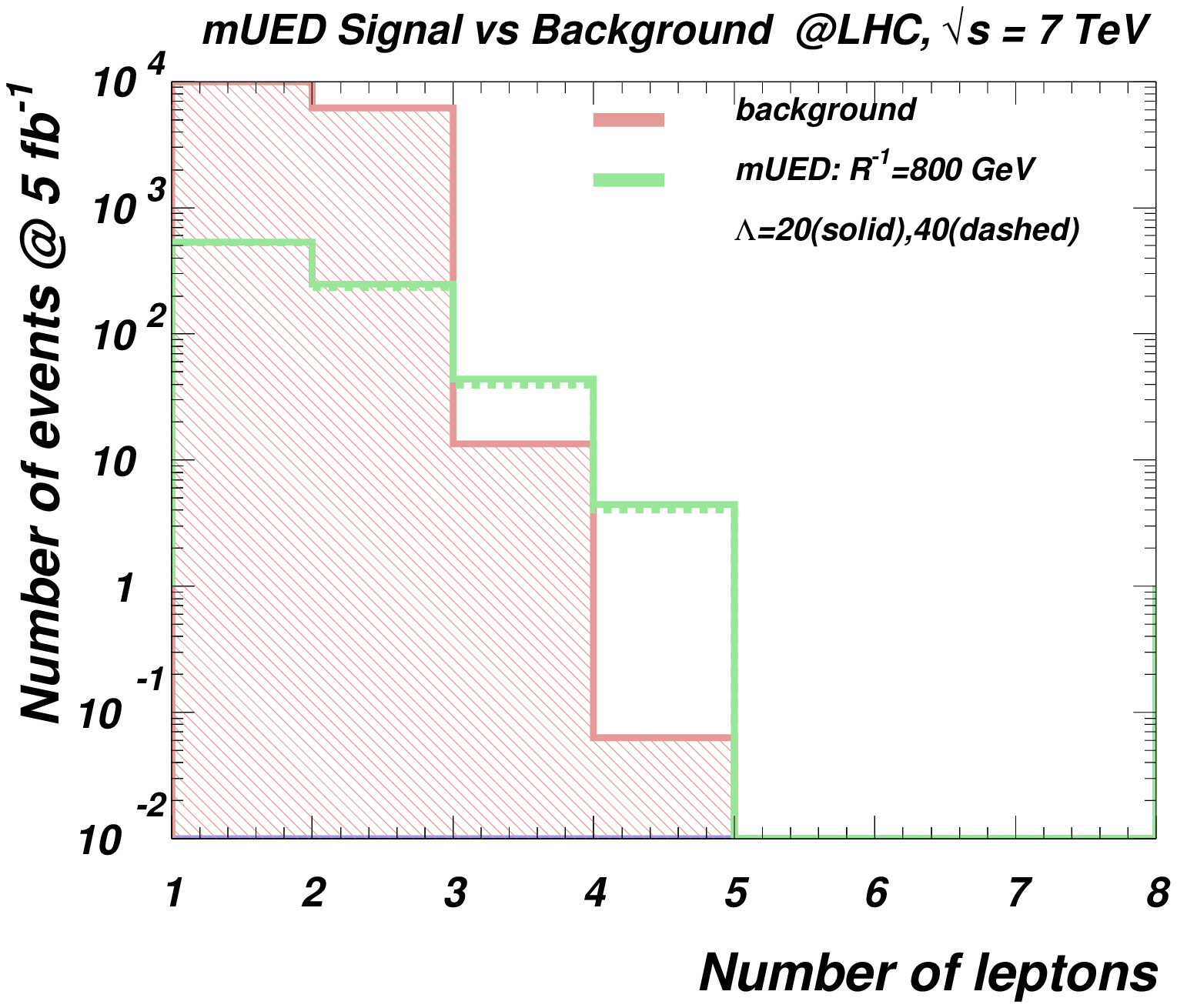}%
\includegraphics[width=0.52\textwidth]{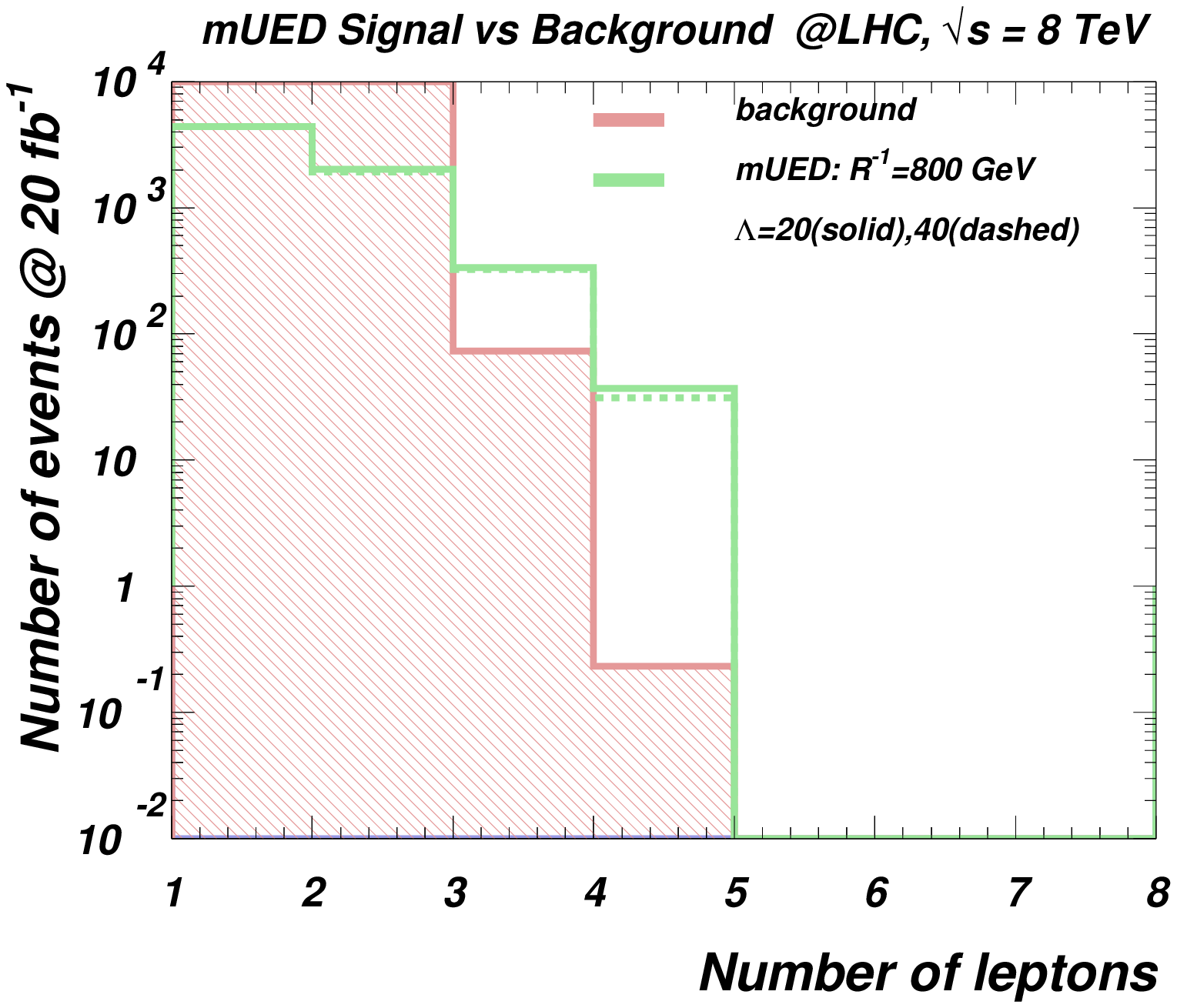}\\
\includegraphics[width=0.52\textwidth]{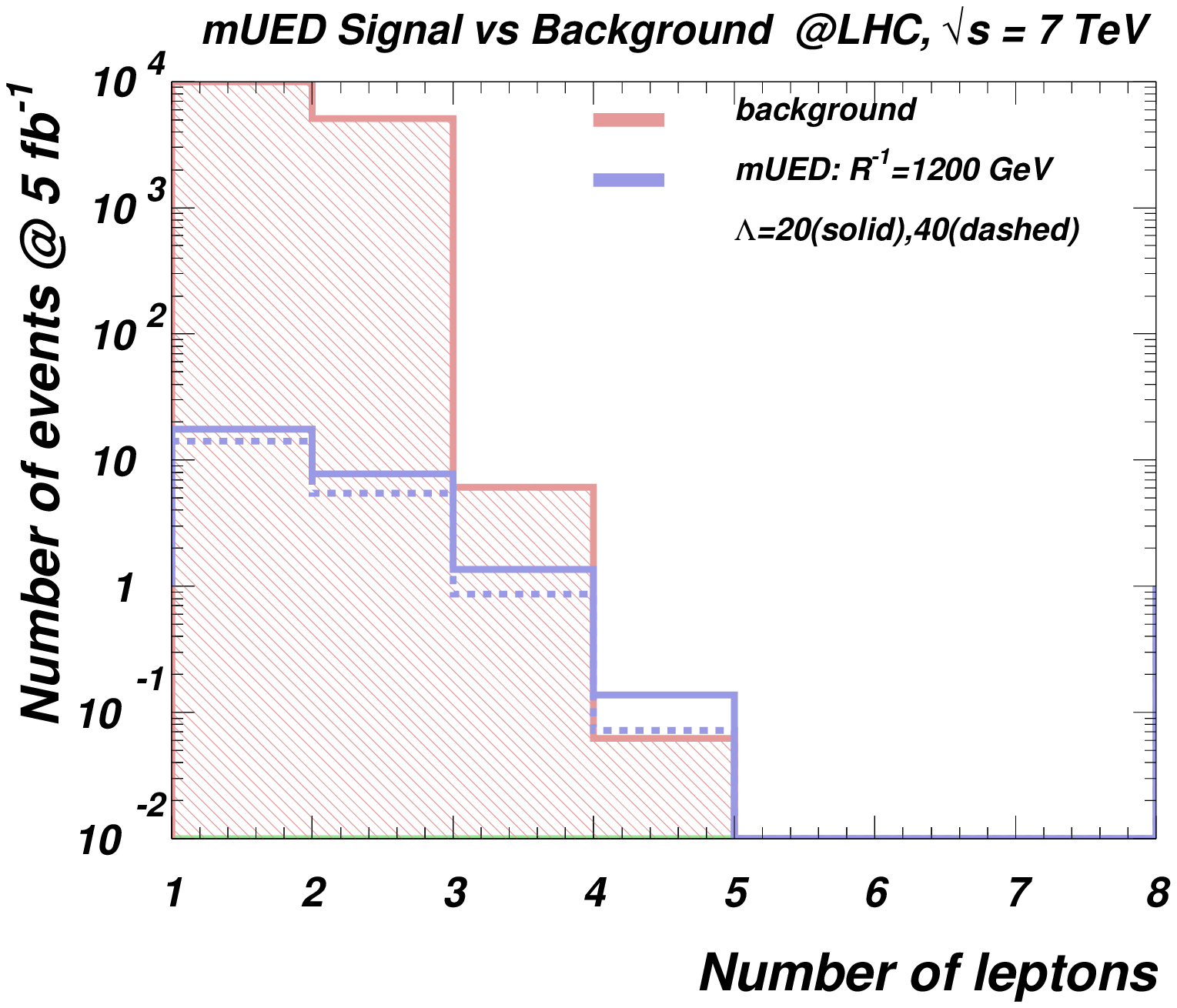}%
\includegraphics[width=0.52\textwidth]{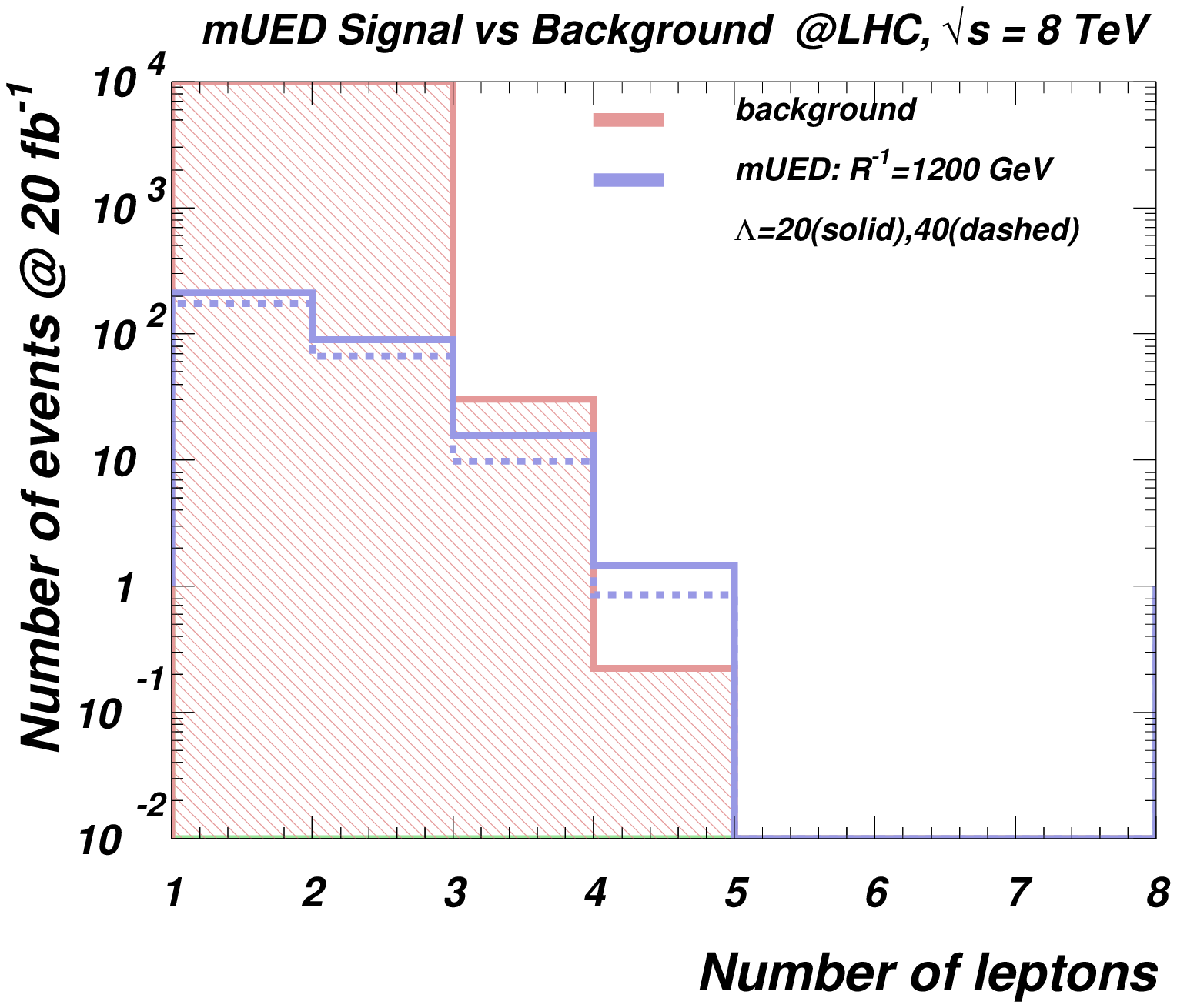}\\
\caption{Lepton multiplicity distribution for  background versus signal after all cuts -- \eqref{eq:cut-ptlmax},
(\ref{eq:cut-acc}), (\ref{eq:cut-wc}), (\ref{eq:cut-ptm}), (\ref{eq:cut-ptl}) and (\ref{eq:cut-meff}) --
for LHC @ 7~TeV (left) and LHC @ 8~TeV (right).}
\label{fig:nl2-sb}
\end{figure}

Finally, in figure~\ref{fig:nl2-sb}
we present lepton multiplicity distributions 
after application of all the cuts -- \eqref{eq:cut-ptlmax}, \eqref{eq:cut-acc}, \eqref{eq:cut-wc}, \eqref{eq:cut-ptm}, \eqref{eq:cut-ptl}
 and \eqref{eq:cut-meff} -- together.
One can see that, after all the cuts, the number of signal events for $R^{-1}=800$~GeV in case of the tri-lepton signature is 
about a factor of 4  above the background for the $\sqrt{s}=7$~TeV
case
($\approx 40$ signal events \emph{vs} ten background events for 5~fb$^{-1}$) and 
about a factor of 5 above the background for  $\sqrt{s}=8$~TeV
($\approx 150$ signal events versus 30 background events for 20~fb$^{-1}$).
For the $R^{-1}=1200$~GeV case, the signal is 
at about the single-event level versus six background events for  $\sqrt{s}=7$~TeV,  5~fb$^{-1}$
and at about the six-event level versus 15 background events for  $\sqrt{s}=8$~TeV,  20~fb$^{-1}$.

We should point out that for larger $R^{-1}$ (above around $1200$~GeV) the
sensitivity of number of events to variations in $\Lambda R$ increases. This is
because the cross-section for producing the strongly-interacting KK particles
falls off increasingly rapidly with mass due to PDF suppression, so small
changes in the mass caused by varying $\Lambda R$ lead to large changes in
the production cross-section.

Another remark is in order.
Looking at  figure~\ref{fig:nl2-sb} one could get the impression
that the di-lepton signature is a much more promising signal for MUED,
but this is not necessarily true.
Here we have explored and presented only the 
complete set of backgrounds relevant for signatures
with three or more leptons.
In the case of the di-lepton signature, one should go beyond parton-level simulations, and beyond backgrounds such as $W/Z+\text{jets}$, $WW/WZ/ZZ+\text{jets}$, 
and explore the so-called ``QCD backgrounds''. These come from, for example,
$W\to \ell\nu +\text{jets}$ production, where the second lepton is  faked
in the detector by jets (different flavours will have different fake probabilities) and/or photon conversion.
Such study requires simulation probably
even beyond the ``fast detector simulation'' level and so is outside the scope of the present paper.
The previous study on the LHC phenomenology of MUED~\cite{Bhattacherjee:2010vm}
has actually shown that the LHC reach for the tri-lepton signature in MUED
is  comparable to, or even better than, the di-lepton signature.
We should stress that, in comparison with ref.~\cite{Bhattacherjee:2010vm}
(the results of which we have reproduced quite closely for the tri-lepton signature when applying the same cuts
to the same set of backgrounds)
we suggest a more sophisticated and optimal set of selection cuts and
consider a more complete set of backgrounds.
We comment on previous studies in more detail at the end of the next subsection.

\subsection{LHC discovery reach}

Now we are in a position to produce results for the LHC reach.
We have performed a Monte Carlo signal simulation for the grid  in the $(R^{-1},\Lambda R)$
plane using the power of the High Energy Physics Model Database
(HEPMDB) [\url{http://hepmdb.soton.ac.uk/}]\cite{Brooijmans:2012yi}
and have applied the combined set of cuts \eqref{eq:cut-ptlmax}--\eqref{eq:cut-meff} described above.
\begin{figure}[htb]
\includegraphics[width=0.54\textwidth]{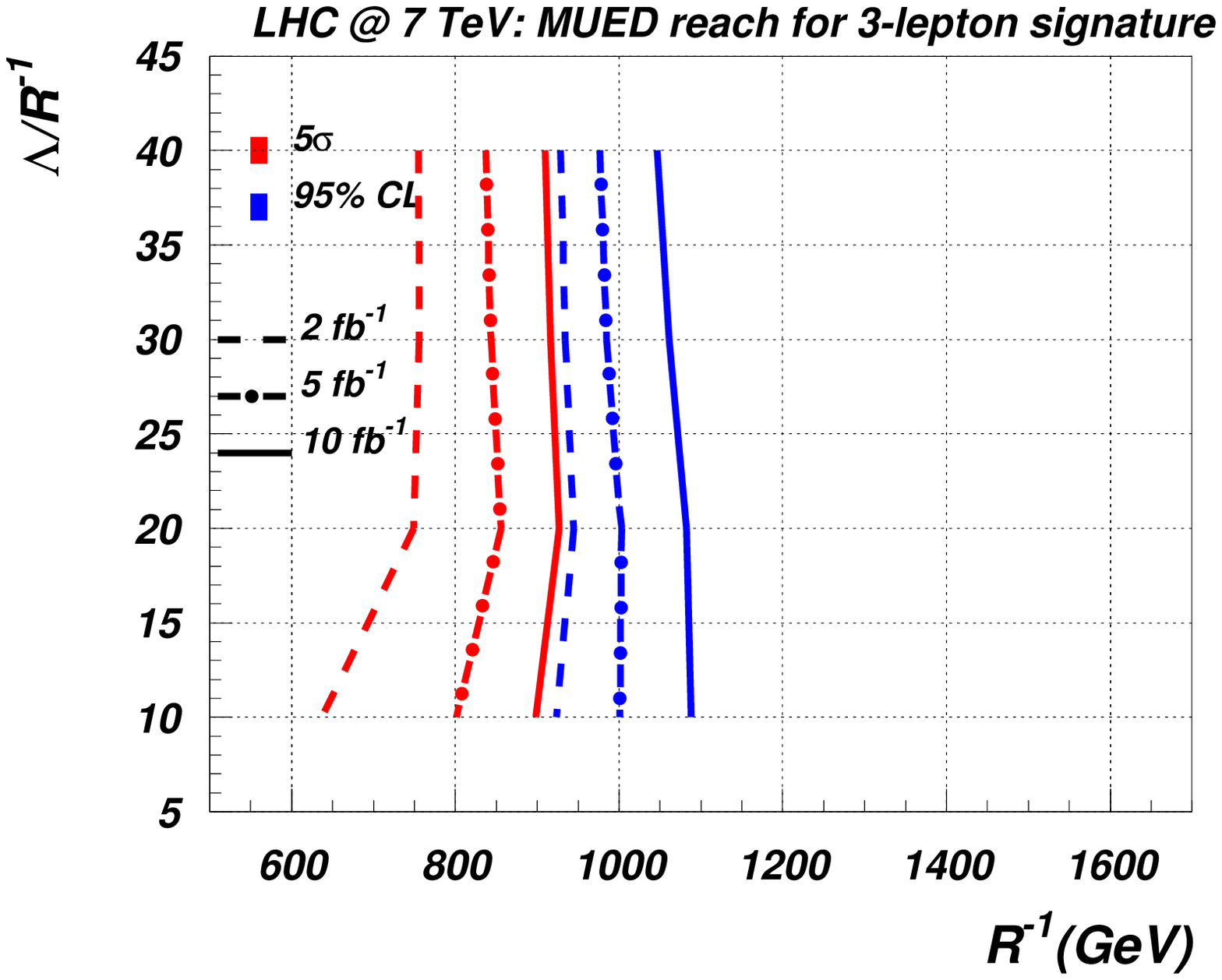}%
\hspace*{-0.8cm}%
\includegraphics[width=0.54\textwidth]{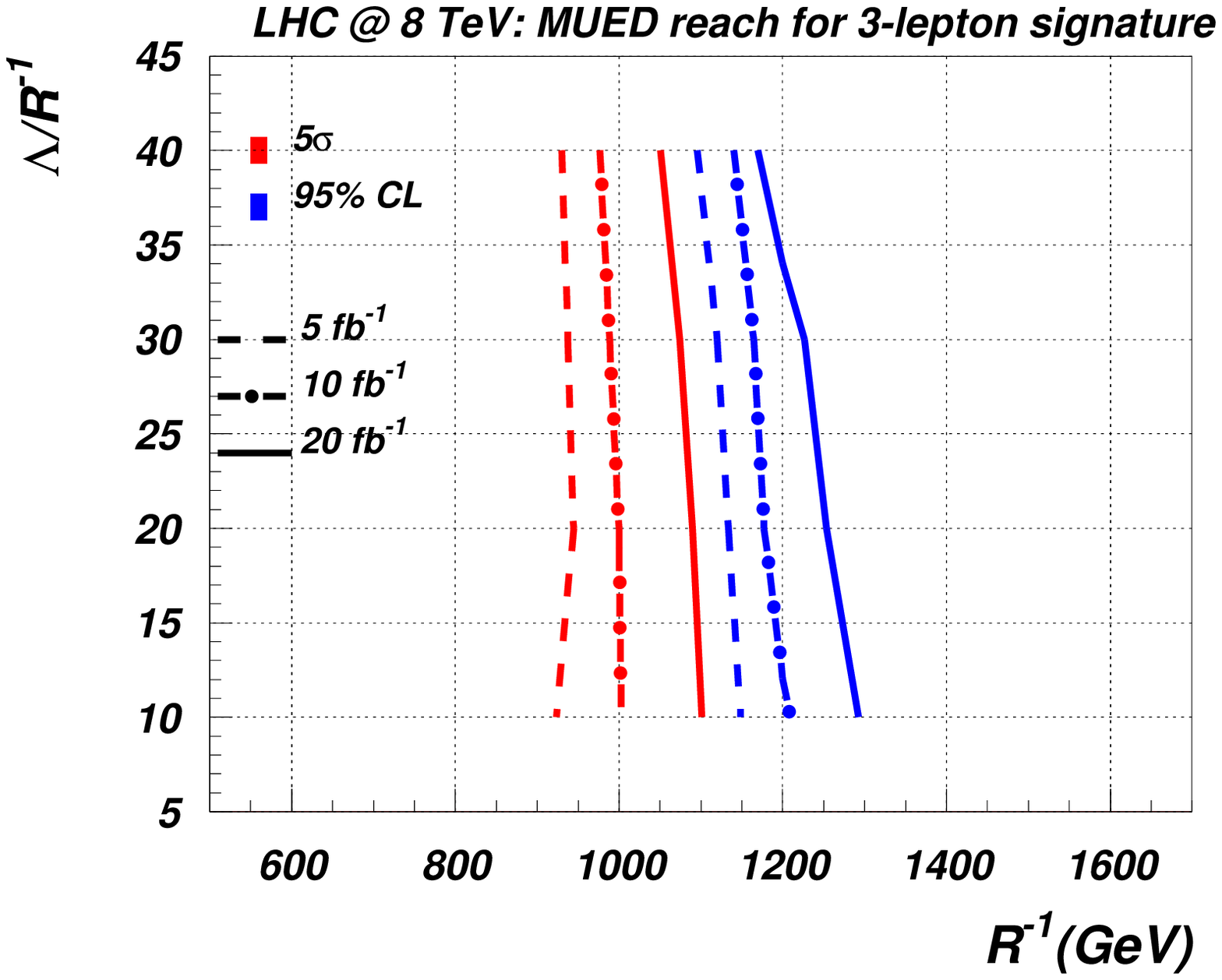}%
\caption{LHC @ 7~TeV (left) and LHC @ 8~TeV (right) exclusion and discovery potential for MUED for different luminosities.
}
\label{fig:lhc-disc}
\end{figure}
\begin{figure}[htb]
    \begin{center}
\includegraphics[width=0.45\textwidth]{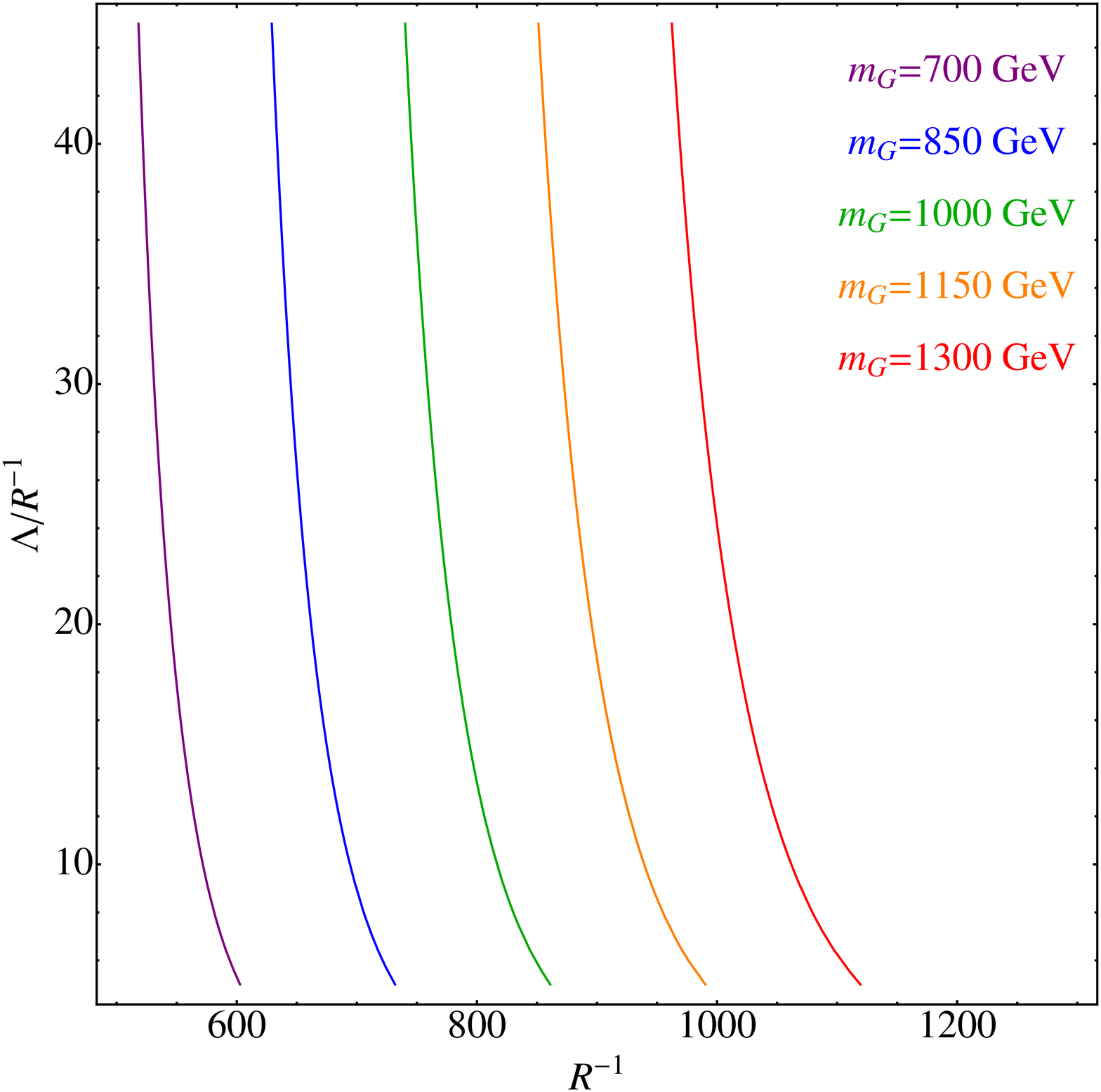}%
\includegraphics[width=0.45\textwidth]{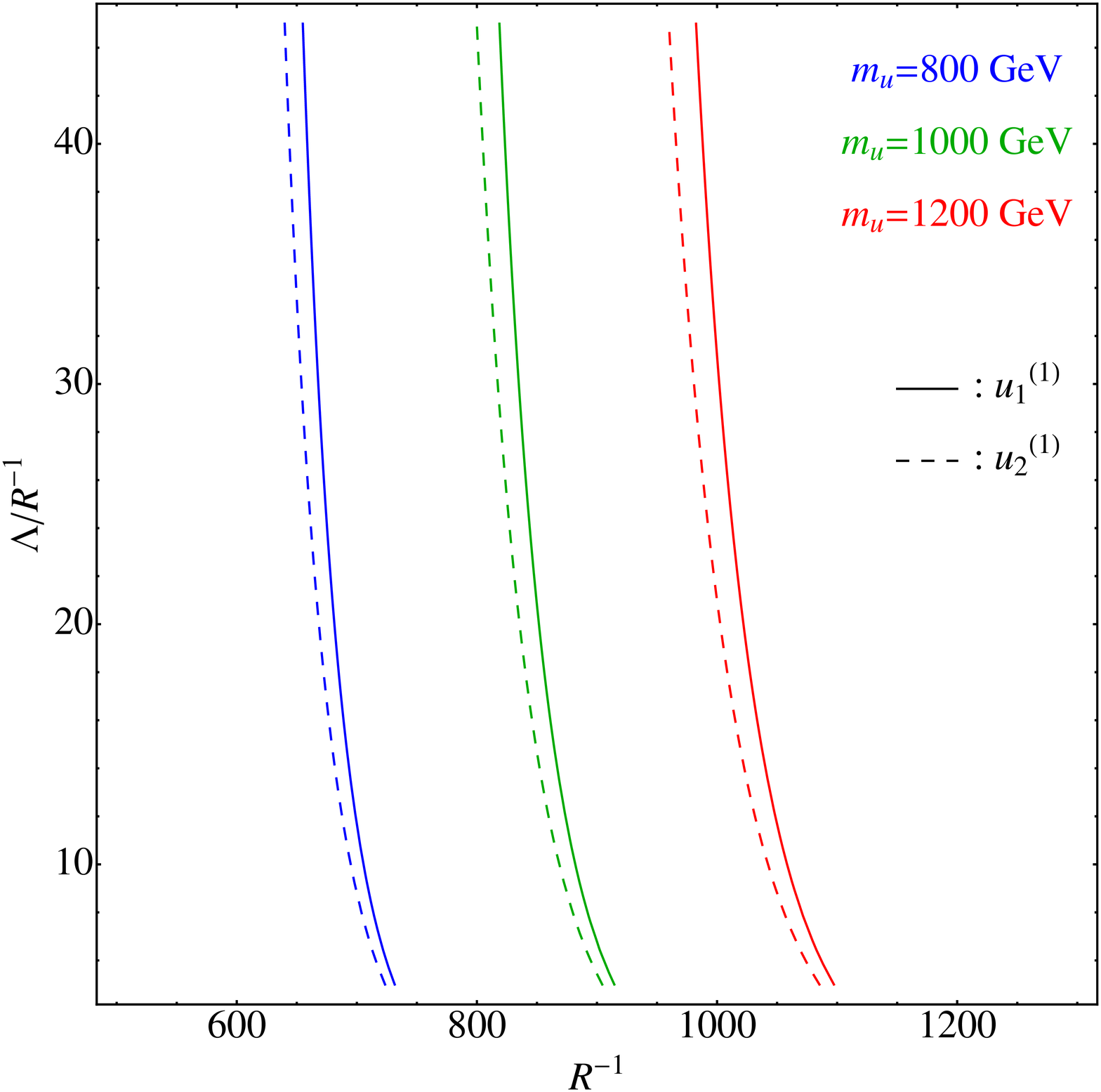}%
\caption{Constant mass contours for $n=1$ KK gluons (left) and KK quarks (right), the latter using $u^{(1)}_{1,2}$ for illustration. (KK indices are suppressed.)}
\label{fig:mass-cont}
    \end{center}
\end{figure}

The results are shown in figure~\ref{fig:lhc-disc}
in terms of exclusion (at 95\% CL) and disovery ($5\sigma$) 
contours for $\sqrt{s}=7$ and 8~TeV and different luminosities.
For both criteria, exclusion and discovery,
we define the statistical signal significance $\alpha$ as
\begin{equation}
 {\alpha}= \frac{N_S}{\sqrt{N_B+N_S}}
\label{eq:signif}
\end{equation}
and require $\alpha\ge 2$ for exclusion region and $\alpha\ge 5$
for the discovery region. The $N_{S(B)}= \sigma_{S(B)} \mathcal{L}$ 
denotes the number of signal (background) events for an integrated luminosity
$\mathcal{L}$.
The region on the left-hand side of the blue contour is excluded
at 95\% CL, while the  region on the left-hand side of the red contour
is discoverable at the LHC.

It is worth commenting on the shape of these contours.
For the low values of $\Lambda R$, the LHC sensitivity contours
bend to the left for low $R^{-1}$ values 
as one can see by observing red discovery contours for $\sqrt{s}=7$ 
for all quoted luminosities, and also the red discovery contour for $\sqrt{s}=8$
for the 5 fb~$^{-1}$ luminosity.
This behaviour is related to the fact that for low values of $\Lambda R$ 
the  mass split of the KK particles is quite small,
especially for low $R^{-1}$ values, as one can see in  figure~\ref{fig:lambda-dependence}.
The decrease of this mass split is correlated with the softness of the final
state leptons which, in turn, is correlated with lower cut efficiency
and respective LHC sensitivity.
On the other hand, one can see that the LHC sensitivity contours
start bending to the \emph{right} in the larger  $R^{-1}$ region,
even for low values of $\Lambda R$. This happens because
an increase in $R^{-1}$ leads to a large-enough mass split 
to make the leptons efficiently pass the selection cuts.

One can also see that, for large values of $\Lambda R$,
LHC sensitivity also drops (i.e. the contours bend to the left).
This is related to the fact that as $\Lambda R$ increases
the KK spectrum hardens which leads to a drop in the signal production rates,
especially for large values of $R^{-1}$.

For a total integrated luminosity of 20~fb~$^{-1}$ at $\sqrt{s}=8$~TeV,
which was reached at the end of 2012,
the best sensitivity of the LHC is actually  for
{\it lower} values of $\Lambda R$.
We would like to stress that this is a new finding
that was not noticed in previous studies, which generically state that there is a decrease of sensitivity
in the lower $\Lambda R$ region. 

In fact, the shape of LHC sensitivity contours --
governed mainly by the values of the signal cross section
for large values of $R^{-1}$
and $\mathcal{L}$ -- is  quite consistent with the shape of
contours of constant mass of KK gluons and KK quarks: these contours are shown
in figure~\ref{fig:mass-cont} in the same plane of parameters.

To conclude, we have found that the $\sqrt{s}=8$~TeV LHC 
with 20 fb~$^{-1}$ of data can discover MUED in the tri-lepton mode
with $R^{-1}\leq 1050$ GeV
and exclude the region with  $R^{-1}\lesssim 1200$~GeV
at 95\%CL with $\Lambda R$ in the whole range under study.

{Using this approach, 
we  estimate optimal 
LHC limits for present and near-future energies and luminosities within a consistent model of MUED.
}

We would like  to comment  here on the results of previous studies.
First of all, the closest phenomenological analysis to ours was performed
in ref.~\cite{Bhattacherjee:2010vm}. We have improved it by adding 
additional backgrounds (most importantly $t\bar{t}\ell\bar{\ell}$ and $t\bar{t}\ell\bar{\nu} (t\bar{t}\bar{\ell}\nu)$)
as well as suggesting a more optimal set of selection cuts which include the following:
relaxation of the upper cut on the leptons' transverse momentum;
application of an upper cut for the transverse momentum of the third lepton;
and the addition of a final $R^{-1}$-dependent $M_{\text{eff}}$ cut.
These steps allow us to extend the LHC limit further than in ref.~\cite{Bhattacherjee:2010vm}. For example, at the 7~TeV LHC
with an integrated luminosity of $\mathcal{L} = 2\,\text{fb}^{-1}$ and $\Lambda R=20$, the expected lower limit on $R^{-1}$ from our studies is about 750 GeV,
versus about 700 GeV from  ref.~\cite{Bhattacherjee:2010vm}. For higher energies and luminosities
the improvement due to the cuts we have suggested is even bigger.

We should also mention the results of ref.~\cite{Murayama:2011hj}, 
which explored the potential of using the $MT2$ variable
for the $\text{leptons}+\text{jets}\,+\etmis$ signature.
This method definitely improved on the results of the previous study 
of the same signature~\cite{Bhattacharyya:2009br}, which used conventional cuts and observables.
The authors of ref.~\cite{Murayama:2011hj} also demonstrated the
advantage of their method over the four-lepton signature with conventional cuts.
We would like to note that the tri-lepton signature that we use in our study 
also has a clear advantage over the four-lepton signature (which has a lower rate by about one order of magnitude).
Moreover, as one can see from our results, the tri-lepton signature is also more powerful than the  $\text{leptons}+\text{jets}\,+\etmis$ signature with $MT2$ variable in ref.~\cite{Murayama:2011hj}, which predicts the LHC lower limit on  $R^{-1}$ to be less than 500~GeV 
for $\mathcal{L}=2\,\text{fb}^{-1}$ and $\Lambda R=20$ at the 7~TeV LHC.

In the most recent study of MUED collider phenomenology~\cite{Datta:2011vg} the authors successfully used so-called $R_T$ and $\alpha$ shape
variables, which take into account the topology of the of $\text{jets}\,+\etmis$ signature (there are necessarily at least two quarks and two LKP particles in the final state at the parton level). This study has greatly optimised the LHC sensitivity to MUED for the $\text{jets}\,+\etmis$ signature.
For example, the authors showed that the 7 TeV LHC can exclude $R^{-1}$ up to 700~GeV
with $\mathcal{L}=2\,\text{fb}^{-1}$. On the other hand, the overall cut efficiency 
is as low as $10^{-3}$, which leaves the tri-lepton signature as the best for setting limits on -- or discovering -- MUED.
For example, using the tri-lepton signature and cuts that we suggest, for the same LHC energy and luminosity as above the LHC can exclude $R^{-1}$ up to 850~GeV; this can be seen from figure~\ref{fig:lhc-disc}.

Finally, we mention ref.~\cite{Choudhury:2009kz} which has studied the MUED signal using the top-quarks signature. This channel, although not being `number one' for {\it discovery},
could be very useful for understanding if the properties of the underlying theory are consistent with MUED.




\section{Conclusions\label{sec:conclu}}

 In this paper we have first worked out a consistent model of Minimal Universal Extra Dimensions and  have implemented it in CalcHEP. The model is available at the \emph{High Energy Model DataBase} (HEPMDB) in both Feynman-'t Hooft  and unitary gauges at \href{https://hepmdb.soton.ac.uk/}{https://hepmdb.soton.ac.uk/} under the name ``MUED-BBMP''. 
It  has been validated against other models for the sectors consistent with our approach~\cite{Datta:2010us} and has been demonstrated to  predict different results for the Higgs sector (which was not complete in the previous implementations). The first part of this article, along with the appendices, is a self-contained review of the model and can be used as comprehensive  documentation for the code implementation.

In the second part of this paper we have focussed on possible signatures of the model at the LHC.  The special MUED mass pattern (relatively compressed, as compared e.g. to typical SUSY models) provides an enhancement of the lepton signals that can be used as a model ``fingerprint''. Taking advantage of this, we have performed a  phenomenologically-realistic study of the  MUED tri-lepton signature at the LHC. Several comments are in order. In contrast with the previous analysis, we have taken into account the complete set of backgrounds  relevant to the signature under the study. We have also worked out a comprehensive set of cuts (\ref{eq:cut-acc})--(\ref{eq:cut-meff}) which lead to higher signal-to-background ratio and signal significance.

The main results of our the study are given in figure~\ref{fig:lhc-disc}. In the left-hand plot we show the first estimation of the LHC  sensitivity to MUED using the tri-lepton signature
for the present LHC energy and luminosity. In the right-hand plot the LHC sensitivity to MUED using the tri-lepton signature is presented, 
which  in particular tells us that the $\sqrt{s}=8$~TeV LHC with 20 fb~$^{-1}$ of data can discover MUED in the tri-lepton mode with $R^{-1}\leq 1050$ GeV and exclude the region with $R^{-1}\lesssim 1200$~GeV at 95\% CL.

Experimentalists have already started to analyse events with three leptons and missing transverse momentum, searching for production of neutralinos and charginos~\cite{:2012ewa,:2012ku}. We would like to encourage them to extend their analysis and interpret their results  in the MUED framework (using our proposed strategy to optimise the cuts) to establish  limits on direct production of KK quarks and KK gluons.
It is worth mentioning that our preliminary estimations \cite{14TeV-Mued} show that the tri-lepton signature would make it possible to close the whole MUED parameter space at the 14~TeV LHC, which is exciting news.

\section*{Acknowledgments}

The authors thank Beatriz de Carlos for initiating this collaboration at the early 
stages of this work as well as enlightening discussions. Thanks also to Steven West for useful comments.
The authors acknowledge support from  Royal Society grant JP090146. 
AB acknowledges partial support from the STFC Consolidated ST/J000396/1. 
This work has been partially supported by the MICINN, Spain, under contract FPA2010-17747; Consolider-Ingenio Multi-Dark CSD2009-00064, CPAN CSD2007-00042. We thank as well the Comunidad de Madrid through Proyecto HEPHACOS S2009/ESP-1473 and the European Commission under contract PITN-GA-2009-237920. 

\clearpage

\appendix


\section{Towards MUED at one loop: the case of U(1)\texorpdfstring{$\times$}{ x }U(1)\label{app:u1u1}}


In this appendix, we focus on the example of two decoupled U(1)s and a Higgs field $\Phi$ which acquires a vacuum expectation value $v$. We perform in detail the diagonalisation and gauge fixing of the Lagrangian. The starting action is gauge invariant but not 5D Lorentz covariant.
\be 
\Lag_5  =  - \frac{1}{4} F^{\mu \nu} F_{\mu \nu} + \frac{Z_B}{2} F^{\mu}_{\ 5} F_{\mu \, 5} - \frac{1}{4} F^{\prime \, \mu \nu} F^{\prime}_{\mu \nu} + \frac{Z_W}{2} F^{\prime \, \mu}_{\ \, 5} F^{\prime}_{\mu \, 5} + \al D_{\mu} \Phi \ar^2 - Z_H \al D_5 \Phi \ar^2 - V (\Phi) \ , \ \label{L51L}
\ee
where $F$ and $F'$ are the field-strengths of the two gauge bosons $B$ and $W$ respectively. Once again, the $Z$ functions are intended to model the mass corrections in a gauge-invariant way. The covariant derivative is
\be
D_M  \Phi \ = \ \dd_M - i \frac{g^{(5)}_1}{2} B_M + i \frac{g^{(5)}_2}{2} W_M \Phi  \quad \quad \text{for}\ M = \mu, 5 \quad , \label{covderiv}
\ee
where we have chosen the normalisation of the second U(1) so as to reproduce the SM mixing.

The Higgs field is expanded about its vev as $\Phi =  \rl  v + h + i \chi \rr$, with $h$ and $\chi$ even under $\ZZ$. At the n$^{\text{th}}$ KK level the quadratic 4D Lagrangian is
\ba
\Lag_4^{(n)} &=& - \frac{1}{4} F^{(n) \, 2}  - \frac{1}{4} F^{\prime \, (n) \, 2} + \frac{1}{2} \rl \dd h^{(n)} \rr^2  + \frac{1}{2} \rl \dd \chi^{(n)} \rr^2 + \frac{Z_B}{2}  \rl \dd B_5^{(n)} \rr^2+ \frac{Z_W}{2} \rl \dd W_5^{(n)} \rr^2 \nonumber \\
&&+  \frac{1}{2} \ccl Z_B \frac{n^2}{R^2} + \frac{1}{4} g_1^2 v^2 \ccr B_{\mu}^{(n) \, 2}  +  \frac{1}{2} \ccl Z_W \frac{n^2}{R^2} + \frac{1}{4} g_2^2 v^2 \ccr  W_{\mu}^{(n)\, 2}  - \frac{1}{2} g_1 g_2 v^2 B_{\mu}^{(n)} W^{\mu \, (n)}  \nonumber \\
&&- \frac{Z_H}{2}  \ccl -\frac{n}{R} \chi^{(n)} - \frac{g_1 v}{2} B_5^{(n)} + \frac{g_2 v}{2} W_5^{(n)} \ccr^2 - \frac{1}{2} Z_H \frac{n^2}{R^2} \rl h^{(n)} \rr^2 \label{L41L} \\
&&+ B_{\mu}^{(n)}\dd^{\mu} \rl \frac{g_1 v}{2} \chi^{(n)} + \frac{n}{R} Z_B B_5^{(n)} \rr + W_{\mu}^{(n)}\dd^{\mu} \rl -\frac{g_2 v}{2} \chi^{(n)} + \frac{n}{R} Z_W W_5^{(n)} \rr  \quad . \nonumber
\ea
The first line of the Lagrangian lists the kinetic terms. The second and third line are mass mixings. The last line is the usual derivative coupling between gauge and would-be Goldstone bosons.

In order for all the fields to be canonically normalised, we perform a rescaling $B_5^{(n)} \rightarrow Z_B^{-1/2} B_5^{(n)}$ and $W_5^{(n)} \rightarrow Z_W^{-1/2} W_5^{(n)}$ :
\ba
\Lag_4^{(n)} &=& - \frac{1}{4} F^{(n) \, 2}  - \frac{1}{4} F^{\prime \, (n) \, 2} + \frac{1}{2} \rl \dd h^{(n)} \rr^2 + \frac{1}{2} \rl \dd \chi^{(n)} \rr^2 + \frac{1}{2}  \rl \dd  B_5^{(n)} \rr^2+ \frac{1}{2} \rl \dd  W_5^{(n)} \rr^2 \nonumber \\
&&+  \frac{1}{2} \ccl Z_B \frac{n^2}{R^2} + \frac{1}{4} g_1^2 v^2 \ccr B_{\mu}^{(n) \, 2}  +  \frac{1}{2} \ccl Z_W \frac{n^2}{R^2} + \frac{1}{4} g_2^2 v^2 \ccr  W_{\mu}^{(n)\, 2}  - \frac{1}{2} g_1 g_2 v^2 B_{\mu}^{(n)} W^{\mu \, (n)}   \nonumber   \\
&&- \frac{Z_H}{2}  \ccl -\frac{n}{R} \chi^{(n)} - \frac{g_1 v}{2 \sqrt{Z_B}} B_5^{(n)} + \frac{g_2 v}{2\sqrt{Z_W}} W_5^{(n)} \ccr^2 - \frac{1}{2} Z_H \frac{n^2}{R^2} \rl h^{(n)} \rr^2 \label{L41Lcan} \\
&&+ B_{\mu}^{(n)}\dd^{\mu} \rl \frac{g_1 v}{2} \chi^{(n)} + \frac{n}{R} \sqrt{Z_B} B_5^{(n)} \rr + W_{\mu}^{(n)}\dd^{\mu} \rl -\frac{g_2 v}{2} \chi^{(n)} + \frac{n}{R} \sqrt{Z_W} W_5^{(n)} \rr \quad .  \nonumber
\ea
We now proceed with the diagonalisation.


\subsection{Gauge boson mixing}


The mass matrix of the vectors,
\be 
\mat_{v}^{\, 2} = \begin{pmatrix} Z_B \frac{n^2}{R^2} + \frac{1}{4} g_1^2 v^2 & - \frac{1}{4} g_1 g_2 v^2 \cr \cr - \frac{1}{4} g_1 g_2 v^2 & Z_W \frac{n^2}{R^2} + \frac{1}{4} g_2^2 v^2 \end{pmatrix} \quad , \label{M2V}
\ee
is diagonalised by
\be
\mathcal O_{v} = \begin{pmatrix} \cos \theta_{v} & - \sin \theta_{v} \cr \cr \sin  \theta_{v} & \cos  \theta_{v}  \end{pmatrix} \ \text{with} \ \tan 2  \theta_{v} = \frac{\frac{1}{2} g_1 g_2 v^2}{v^2/4 \rl g_2^2 - g_1^2\rr + n^2/R^2 \rl Z_W - Z_B\rr} \quad . \label{rotV}
\ee
We call $P_{\mu}^{(n)}$ and $V_{\mu}^{(n)}$ the mass eigenstates given by

\be 
\begin{pmatrix} P_{\mu}^{(n)} \cr \cr V_{\mu}^{(n)} \end{pmatrix} = \begin{pmatrix} \cos \theta_{v} & \sin \theta_{v} \cr \cr - \sin \theta_{v} & \cos  \theta_{v}  \end{pmatrix} \ \begin{pmatrix} B_{\mu}^{(n)} \cr \cr W_{\mu}^{(n)} \end{pmatrix} \label{PV} \quad ,
\ee
with masses
\ba
m^2_{\left\{P, V\right\}(n)} &=&  \frac{1}{2} \left[ \frac{v^2}{4} \rl g_1^2 + g_2^2 \rr + \frac{n^2}{R^2}  \rl Z_B + Z_W \rr \right.  \nonumber \\ &\quad& \quad \quad \left. \mp \sqrt{ \rl \frac{v^2}{4} \rl g_2^2 - g_1^2 \rr + \frac{n^2}{R^2}  \rl Z_W - Z_B \rr \rr^2 + \frac{1}{4} g_1^2 g_2^2 v^4} \, \right] \quad . \label{m2PV}
\ea


\subsection{Scalar mixing}


The scalar mass matrix is
\be 
\mat_{s}^{\, 2} = Z_H \, \begin{pmatrix} \frac{n^2}{R^2} &  - \frac{g_1 v}{2 \sqrt{Z_B}} \frac{n}{R} & \frac{g_2 v}{2 \sqrt{Z_W}} \frac{n}{R} \cr \cr  - \frac{g_1 v}{2 \sqrt{Z_B}} \frac{n}{R} &    \frac{g_1^2 v^2}{4 Z_B}  &  - \frac{g_1 g_2 v^2}{4 \sqrt{Z_B Z_W}} \cr \cr  \frac{g_2 v}{2 \sqrt{Z_W}} \frac{n}{R} & - \frac{g_1 g_2 v^2}{4 \sqrt{Z_B Z_W}} &  \frac{g_2^2 v^2}{4 Z_W} \end{pmatrix} \label{M2S}
\ee
in the $\rl \chi^{(n)}, B_5^{(n)}, W_5^{(n)}\rr$ basis. This matrix has a zero determinant which confirms that at least one of the eigenstates is a Goldstone boson.

It is diagonalised by two rotations. We choose them to be
\be
\mathcal O_{23} = \begin{pmatrix} 1 & 0 & 0 \cr \cr 0 & \cos \theta_{23} & - \sin \theta_{23}  \cr \cr 0 & \sin  \theta_{23} & \cos  \theta_{23}  \end{pmatrix} \ \text{with} \ \tan 2  \theta_{23} = \frac{2 g_1 g_2 / \sqrt{Z_B Z_W}}{g_2^2 / Z_W - g_1^2 / Z_B} \quad , \label{rot23S}
\ee
and
\be
\mathcal O_{13} = \begin{pmatrix} \cos \theta_{13} & 0 & \sin \theta_{13} \cr \cr 0 & 1 & 0  \cr \cr - \sin  \theta_{13} & 0 & \cos  \theta_{13}  \end{pmatrix} \ \text{with} \ \tan 2  \theta_{13} = 2 \frac{\frac{n}{R} \frac {v}{2} \sqrt{\frac{g_1^2}{Z_B} + \frac{g_2^2}{Z_W}}}{\frac{v^2}{4} \rl \frac{g_1^2}{Z_B} + \frac{g_2^2}{Z_W} \rr - \frac{n^2}{R^2}} \quad , \label{rot13S}
\ee
We call $G_1^{(n)}$, $G_2^{(n)}$ and $a^{(n)}$ these mass states. It turns out that both $G_{1, \, 2}^{(n)}$ are massless, while $a^{(n)}$ has a mass
\be 
 m^2_{a(n)} = Z_H \ccl \frac{n^2}{R^2} + \frac{v^2}{4} \rl \frac{g_1^2}{Z_B} + \frac{g_2^2}{Z_W} \rr \, \ccr \label{man} \quad .
\ee
and the diagonal mass matrix is ${\widetilde\mat}_s^{\, 2} = \mathcal O_{13}^{-1} \mathcal O_{23}^{-1} \, \mat_{s}^{\, 2} \, \mathcal O_{23} \, \mathcal O_{13} =  \begin{pmatrix} 0 & 0 & 0 \cr \cr 0 & 0 & 0  \cr \cr 0 & 0 & m^2_{a(n)} \end{pmatrix}$.


\subsection{Derivative coupling}


In a gauge theory with Goldstone bosons, the derivative coupling has the form $m A^{\mu} \dd_{\mu} G$ with $m$ the mass of the boson and $G$ the Goldstone boson. This coupling is uniquely determined by gauge invariance. The gauge fixing Lagrangian cancels out the derivative coupling and gives a mass $m_G = \sqrt{\xi}\, m$ to the Goldstone boson in the $R_{\xi}$-gauge.

In our case, the derivative couplings are
\be
B^{\mu \, (n)}\dd_{\mu} \rl \frac{g_1 v}{2} \chi^{(n)} + \frac{n}{R} \sqrt{Z_B} B_5^{(n)} \rr + W^{\mu \, (n)}\dd_{\mu} \rl -\frac{g_2 v}{2} \chi^{(n)} + \frac{n}{R} \sqrt{Z_W} W_5^{(n)} \rr \label{dercoupl}
\ee
which after diagonalisation of the scalars reads
\be
B^{\mu \, (n)}\dd_{\mu} \rl a \, G_1^{(n)} + b \, G_2^{(n)} \rr + W^{\mu \, (n)}\dd_{\mu} \rl c \, G_1^{(n)} + d \, G_2^{(n)} \rr \label{derG}
\ee
with some lengthy $a$, $b$, $c$, $d$. The rotation matrix $\mathcal O_G$
\be
\begin{pmatrix} G_1^{(n)} \cr \cr G_2^{(n)} \end{pmatrix} = \begin{pmatrix} \cos \theta_{G} & - \sin \theta_{G} \cr \cr  \sin \theta_{G} & \cos  \theta_{G}  \end{pmatrix} \ \begin{pmatrix} G_P^{(n)} \cr \cr G_Q^{(n)} \end{pmatrix} \nonumber 
\ee
that diagonalises the coupling \eqref{derG} is determined by gauge invariance. We have introduced $G_P^{(n)}$ and $G_Q^{(n)}$ which are the diagonal Goldstone bosons of $P_{\mu}^{(n)}$ and $Q_{\mu}^{(n)}$.

If we call $\mathcal D$ the initial coupling matrix, which has rank $2 \times 3$, the diagonalisation yields
\be 
\begin{pmatrix} B_{\mu}^{(n)} & W_{\mu}^{(n)} \end{pmatrix} \cdot \, \mathcal D \, \cdot \begin{pmatrix} \chi^{(n)} \cr \cr B_5^{(n)} \cr \cr W_5^{(n)} \end{pmatrix} = \begin{pmatrix} P_{\mu}^{(n)} & V_{\mu}^{(n)} \end{pmatrix} \,  \cdot \, \ccl \mathcal O_v^{\, T} \cdot \, \mathcal D \cdot \,  \mathcal O_{23} \mathcal O_{13} \mathcal O_G \ccr \, \cdot \, \begin{pmatrix} G_P^{(n)} \cr \cr G_V^{(n)} \cr \cr a^{(n)} \end{pmatrix} \label{derG2} \quad . \nonumber
\ee
The mass condition
\be
\mathcal D_{\rm diag}  \equiv \mathcal O_v^{\, T}  \cdot \, \mathcal D \cdot \,  \mathcal O_{23} \mathcal O_{13} \mathcal O_G = \begin{pmatrix} m_{P(n)} & 0 & 0 \cr \cr 0 & m_{V(n)} & 0 \end{pmatrix}  \nonumber
\ee
allows us to uniquely determine the rotation angle
\be
\theta_G = \arctan \rl \frac{b \cos \theta_v + d \sin \theta_v}{a \cos \theta_v + c \sin \theta_v} \rr \quad , \label{thetaG}
\ee
where $a$, $b$, $c$, $d$ were defined in \eqref{derG}.

In this way, the gauge fixing Lagrangian is valid in any basis, as it should be. Indeed, the equality
\be 
\mathcal D_{\rm diag} \cdot \mathcal D_{\rm diag}^T =  \mathcal {\widetilde M}_v^{\, 2} \equiv \mathcal O_{v}^{-1}  \mathcal M_{v}^{\, 2} \, \mathcal O_{v} \nonumber
\ee
reduces to $\mathcal D \cdot \mathcal D^{\, T} = \mathcal M_{v}^{\, 2}$. It is easy to check, from \eqref{dercoupl} and \eqref{M2V}, that this equality indeed holds.


\subsection{Diagonalised Lagrangian}


We are now ready to proceed with the gauge fixing and write our final Lagrangian. The mass eigenstates are the two massive gauge bosons $P^{(n)}$ and $V^{(n)}$ ; two massless unphysical scalars $G_P^{(n)}$, $G_V^{(n)}$ which are the Goldstone bosons ; two scalars $h^{(n)}$ and $a^{(n)}$. In this basis, the Lagrangian reduces to
\ba
\Lag_4^{(n)} &=& - \frac{1}{4} \rl F_{\mu \, \nu}^{\ (n)} \rr^2  - \frac{1}{4} \rl F_{\mu \, \nu}^{\prime \ (n)} \rr^2 + \frac{1}{2} \rl \dd_{\mu} a^{(n)} \rr^2 + \frac{1}{2} \rl \dd_{\mu} h^{(n)} \rr^2 \nonumber \\
&& +  \frac{1}{2} m_{P(n)}^2 \rl P_{\mu}^{(n)} \rr^2   +  \frac{1}{2} m_{V(n)}^2 \rl V_{\mu}^{(n)} \rr^2 -  \frac{1}{2} m_{a(n)}^2  \rl a^{(n)} \rr^2 -  \frac{1}{2} m_{h(n)}^2  \rl h^{(n)} \rr^2 \nonumber   \\
&&+ m_{P(n)} P_{\mu}^{(n)} \dd^{\mu} G_P^{(n)} + m_{V(n)} V_{\mu}^{(n)} \dd^{\mu} G_V^{(n)}  \quad ,  \label{L41Ldiag} 
\ea
where $m_{P(n)}$, $m_{V(n)}$ and $m_{a(n)}$ are given in equations \eqref{m2PV} and \eqref{man}, and $m_{h(n)} = \sqrt{Z_H}\frac{n}{R}$.

The gauge fixing Lagrangian in the $R_{\xi}$ gauge is meant to cancel out the last line of the above Lagrangian 
\be
\Lag_{GF}^{(n)} \ = \  - \frac{1}{2 \xi} \ccl \,  \rl \dd^{\mu} P_{\mu}^{(n)} - \xi m_{P(n)} G_P^{(n)} \rr^2 + \rl \dd^{\mu} V_{\mu}^{(n)} - \xi m_{V(n)} G_V^{(n)} \rr^2 \, \ccr \quad . \label{GF} 
\ee

The total Lagrangian is then
\ba
\Lag_4^{(n)} + \Lag_{GF}^{(n)} &=& - \frac{1}{4} \rl F_{\mu \, \nu}^{\ (n)} \rr^2  - \frac{1}{4} \rl F_{\mu \, \nu}^{\prime \ (n)} \rr^2 +  \frac{1}{2} m_{P(n)}^2 \rl P_{\mu}^{(n)} \rr^2   +  \frac{1}{2} m_{V(n)}^2 \rl V_{\mu}^{(n)} \rr^2 \nonumber \\
&&  + \frac{1}{2} \rl \dd_{\mu} a^{(n)} \rr^2 + \frac{1}{2} \rl \dd_{\mu} h^{(n)} \rr^2 -  \frac{1}{2} m_{a(n)}^2  \rl a^{(n)} \rr^2 -  \frac{1}{2} m_{h(n)}^2  \rl h^{(n)} \rr^2 \nonumber   \\
&& - \frac{ \xi }{2}  \, m_{P(n)}^2 G_P^{(n)} - \frac{ \xi }{2} \, m_{V(n)}^2 G_V^{(n)}  \quad .  \label{L41Ldiaggf} 
\ea
In the unitary gauge, $\xi \rightarrow \infty$, the Goldstone bosons decouple while in the Feynman-'t Hooft gauge, $\xi = 1$, they have the same mass as the gauge bosons.

\clearpage


\section{Splitting of four-gluon vertices in CalcHEP/CompHEP\label{app:fourgluon}}


In CalcHEP, it is important to take care of the vertices involving four gluons because of their colour structure. It was shown in \cite{Belyaev:2005ew} how to solve this issue by introducing Lagrange multipliers. The kinetic term $F_{\mu \nu} F^{\mu \nu}$ of $\SU3$ contains
\be
\Lag \ = \ - \frac{1}{4} g_3^2 f^{abc} f^{ade} G_{\mu}^b G_{\nu}^c G^{\mu \, d} G^{\mu \, e} \quad . \label{4g}
\ee
The trick is to replace it by
\be
\Lag \ = \ - \frac{1}{2} t^a_{\mu \nu} t^{a \, \mu \nu} + \frac{i}{\sqrt{2}} \, g_3 \, f^{abc} \, t^{a \, \mu \nu} \, G_{\mu}^b G_{\nu}^c \label{auxtens}
\ee
which reduces to \eqref{4g} once the equations of motion for the tensor $t^{a}_{\mu \nu}$ are solved and plugged back into \eqref{auxtens}.

In the case of MUED at one-loop with two KK modes, the 5D Lagrangian is
\be
\Lag_5 \ = \ - \frac{1}{4} g^{(5) \, 2}_3 f^{abc} f^{ade} G_{\mu}^b G_{\nu}^c G^{\mu \, d} G^{\nu \, e} + \frac{Z_G}{2}  g^{(5) \, 2}_3 f^{abc} f^{ade} G_{\mu}^b G_{5}^c G^{\mu \, d} G_{5}^{e} \quad . \label{4g5d} 
\ee


\subsection{Tensorial splitting}


We focus on the first term of \eqref{4g5d}. The expansion in KK modes and compactification yields
\ba
\Lag_4 &=& - \frac{g^2_3}{4}  f^{abc} f^{ade} \ccl G_{\mu}^{b \, (0)} G_{\nu}^{c \, (0)} G^{\mu \, d \, (0)} G^{\nu \, e \, (0)} + \sum_{n \, \geq \, 1}  \sigma_{4} \rl G_{\mu}^{b \, (n)} G_{\nu}^{c \, (n)} G^{\mu \, d \, (0)} G^{\nu \, e \, (0)} \rr \right. \nonumber \\
&& \quad \quad \quad \quad  \quad \quad  + \, \frac{1}{\sqrt{2}} \sum_{k, \, l, \, m \, \geq \, 1} \Delta^1_{klm} \, \sigma_{4} \rl G_{\mu}^{b \, (0)} G_{\nu}^{c \, (k)} G^{\mu \, d \, (l)} G^{\nu \, e \, (m)} \rr \nonumber \\
&& \quad \quad \quad \quad  \quad \quad \left. +  \, \frac{1}{2} \sum_{k, \, l, \, m, \, n \, \geq \, 1} \Delta^2_{klmn} \, G_{\mu}^{b \, (k)} G_{\nu}^{c \, (l)} G^{\mu \, d \, (m)} G^{\nu \, e \, (n}  \ccr \quad , \label{4g4d} 
\ea
where $\Delta^{1}_{klm}$ and $\Delta^{2}_{klmn}$ are sums of Kronecker deltas arising from the integration of various cosines, see Appendix B in \cite{Datta:2010us}.

The function $\sigma_4$ denotes permutations over the KK index. For example 
\ba
\sigma_{4} \rl G_{\mu}^{b \, (0)} G_{\nu}^{c \, (k)} G^{\mu \, d \, (l)} G^{\nu \, e \, (m)} \rr &=& G_{\mu}^{b \, (0)} G_{\nu}^{c \, (k)} G^{\mu \, d \, (l)} G^{\nu \, e \, (m)} + G_{\mu}^{b \, (k)} G_{\nu}^{c \, (0)} G^{\mu \, d \, (l)} G^{\nu \, e \, (m)} \nonumber \\
&&+ \, G_{\mu}^{b \, (k)} G_{\nu}^{c \, (l)} G^{\mu \, d \, (0)} G^{\nu \, e \, (m)} +G_{\mu}^{b \, (k)} G_{\nu}^{c \, (l)} G^{\mu \, d \, (m)} G^{\nu \, e \, (0)} \quad. \nonumber
\ea

We now truncate the tower and consider two KK levels. The Lagrangian \eqref{4g4d} becomes
\ba
\Lag_4 &=& - \frac{g^2_3}{4}  f^{abc} f^{ade} \ccl G_{\mu}^{b \, (0)} G_{\nu}^{c \, (0)} G^{\mu \, d \, (0)} G^{\nu \, e \, (0)} + \sum_{n \, =\, 1, \, 2}  \sigma_{4} \rl G_{\mu}^{b \, (n)} G_{\nu}^{c \, (n)} G^{\mu \, d \, (0)} G^{\nu \, e \, (0)} \rr \right. \nonumber \\
&& \quad \quad \quad \quad  \quad \quad  + \, \frac{3}{2} \rl G_{\mu}^{b \, (1)} G_{\nu}^{c \, (1)} G^{\mu \, d \, (1)} G^{\nu \, e \, (1)} + G_{\mu}^{b \, (2)} G_{\nu}^{c \, (2)} G^{\mu \, d \, (2)} G^{\nu \, e \, (2)} \rr \nonumber \\
&& \quad \quad \quad \quad  \quad \quad \left. +  \, \frac{1}{\sqrt{2}} \, \sigma_{4} \rl G_{\mu}^{b \, (0)} G_{\nu}^{c \, (1)} G^{\mu \, d \, (1)} G^{\nu \, e \, (2)} + G_{\mu}^{b \, (1)} G_{\nu}^{c \, (1)} G^{\mu \, d \, (2)} G^{\nu \, e \, (2)} \rr \ccr \ . \nonumber 
\ea
It turns out that the following five auxiliary tensors are needed
\ba
&&s^{a}_{\mu \nu} = \frac{i}{\sqrt{2}} \, g_3 \, f^{abc}  \, \rl G_{\mu}^{b \, (0)} G_{\nu}^{c \, (0)} + G_{\mu}^{b \, (1)} G_{\nu}^{c \, (1)} + G_{\mu}^{b \, (2)} G_{\nu}^{c \, (2)}  \rr \nonumber \\
&&t^{a}_{\mu \nu} = \frac{i}{\sqrt{2}} \, g_3 \, f^{abc}  \, \rl G_{\mu}^{b \, (0)} G_{\nu}^{c \, (1)}  + G_{\mu}^{b \, (1)} G_{\nu}^{c \, (0)} + \frac{1}{\sqrt{2}} \, G_{\mu}^{b \, (1)} G_{\nu}^{c \, (2)} + \frac{1}{\sqrt{2}} \, G_{\mu}^{b \, (2)} G_{\nu}^{c \, (1)} \rr \nonumber \\
&&u^{a}_{\mu \nu} = \frac{i}{\sqrt{2}} \, g_3 \, f^{abc}  \, \rl G_{\mu}^{b \, (0)} G_{\nu}^{c \, (2)} + G_{\mu}^{b \, (2)} G_{\nu}^{c \, (0)} + \frac{1}{\sqrt{2}} \, G_{\mu}^{b \, (1)} G_{\nu}^{c \, (1)} \rr \label{auxtens2kk} \\
&&v^{a}_{\mu \nu} = \frac{i}{\sqrt{2}} \, g_3 \, f^{abc}  \, \rl \frac{1}{\sqrt{2}} \, G_{\mu}^{b \, (1)} G_{\nu}^{c \, (2)} + \frac{1}{\sqrt{2}} \, G_{\mu}^{b \, (2)} G_{\nu}^{c \, (1)} \rr \nonumber \\
&&w^{a}_{\mu \nu} = \frac{i}{\sqrt{2}} \, g_3 \, f^{abc}  \, \rl + \frac{1}{\sqrt{2}} \, G_{\mu}^{b \, (2)} G_{\nu}^{c \, (2)}  \rr \nonumber
\ea


\subsection{Vectorial splitting}


We now turn to the second term in \eqref{4g5d}. The expansion in KK modes and compactification yield
\ba
\Lag_4 &=& \frac{Z_G}{2} \, g^2_3  f^{abc} f^{ade} \ccl  \sum_{n \, \geq \, 1}  \,G_{\mu}^{b \, (0)} G_{5}^{c \, (n)} G^{\mu \, d \, (0)} G_{5}^{e \, (n)}  \right. \nonumber \\
&& \quad \quad \quad \quad  \quad \quad  + \, \frac{1}{\sqrt{2}} \sum_{k, \, l, \, m \, \geq \, 1} \Delta^4_{klm} \,G_{\mu}^{b \, (0)} G_{5}^{c \, (k)} G^{\mu \, d \, (l)} G_{5}^{e \, (m)}   \nonumber \\
&& \quad \quad \quad \quad  \quad \quad  + \, \frac{1}{\sqrt{2}} \sum_{k, \, l, \, m \, \geq \, 1} \Delta^4_{klm}  \, G_{\mu}^{b \, (k)} G_{5}^{c \, (l)} G^{\mu \, d \, (0)} G_{5}^{e \, (m)}   \nonumber \\
&& \quad \quad \quad \quad  \quad \quad \left. +  \, \frac{1}{2} \sum_{k, \, l, \, m, \, n \, \geq \, 1} \Delta^5_{klmn} \,G_{\mu}^{b \, (k)} G_{5}^{c \, (l)} G^{\mu \, d \, (m)} G_5^{e \, (n)}   \ccr \quad , \label{4g4d5} 
\ea
where again $\Delta^4_{klm}$ and $\Delta^5_{klmn}$ are combinations of deltas. With two KK levels, the Lagrangian \eqref{4g4d5} becomes
\ba
\Lag_4 &=& \frac{Z_G}{2} \, g^2_3  f^{abc} f^{ade} \ccl  \sum_{n \, = \, 1, \, 2}  \rl G_{\mu}^{b \, (0)} G_{5}^{c \, (n)} G^{\mu \, d \, (0)} G_{5}^{e \, (n)} \rr \right. \label{4g4dvect2kk} \\
&& \quad \quad \quad \quad  \quad \quad  + \, \frac{1}{2}   \rl G_{\mu}^{b \, (1)} G_{5}^{c \, (1)} G^{\mu \, d \, (1)} G_{5}^{e \, (1)} +G_{\mu}^{b \, (2)} G_{5}^{c \, (2)} G^{\mu \, d \, (2)} G_{5}^{e \, (2)} \rr \nonumber \\
&& \quad \quad \quad \quad  \quad  \quad + \,  G_{\mu}^{b \, (1)} G_{5}^{c \, (2)} G^{\mu \, d \, (1)} G_{5}^{e \, (2)} + G_{\mu}^{b \, (2)} G_{5}^{c \, (1)} G^{\mu \, d \, (2)} G_{5}^{e \, (1)}  \nonumber \\
&& \quad \quad \quad \quad  \quad  +  \, \frac{1}{\sqrt{2}} \rl G_{\mu}^{b \, (0)} G_{5}^{c \, (1)} G^{\mu \, d \, (1)} G_5^{e \, (2)} + G_{\mu}^{b \, (1)} G_{5}^{c \, (2)} G^{\mu \, d \, (0)} G_5^{e \, (1)} + G_{\mu}^{b \, (1)} G_{5}^{c \, (1)} G^{\mu \, d \, (0)} G_5^{e \, (2)} \right. \nonumber \\
&& \quad \quad \quad \quad  \quad \quad  + \left. \left. \,  G_{\mu}^{b \, (0)} G_{5}^{c \, (2)} G^{\mu \, d \, (1)} G_5^{e \, (1)} - G_{\mu}^{b \, (0)} G_{5}^{c \, (1)} G^{\mu \, d \, (2)} G_5^{e \, (1)} - G_{\mu}^{b \, (2)} G_{5}^{c \, (1)} G^{\mu \, d \, (0)} G_5^{e \, (1)} \rr \ccr \quad .  \nonumber 
\ea

In this case we will need vectorial auxiliary fields with the following kind of Lagrangian
\be
\Lag \ = \ -\frac{1}{2} V^a_{\mu} V^{a \, \mu} + Z_G \, g_3 \, f^{abc} \, V^{a \, \mu} \, G_{\mu}^{b \, (n)} G_{5}^{c \, (m)} \quad . \label{auxvect}
\ee
We need four auxiliary vectors
\ba
&&S^{a}_{\mu} = Z_G \, g_3 \, f^{abc}  \, \rl G_{\mu}^{b \, (0)} G_{5}^{c \, (1)} +  \frac{1}{\sqrt{2}} \,  G_{\mu}^{b \, (1)} G_{5}^{c \, (2)} -  \frac{1}{\sqrt{2}} \,  G_{\mu}^{b \, (2)} G_{5}^{c \, (1)}  \rr \nonumber \\
&&T^{a}_{\mu} = Z_G \, g_3 \, f^{abc}  \, \rl G_{\mu}^{b \, (0)} G_{5}^{c \, (2)}  + \frac{1}{\sqrt{2}} \, G_{\mu}^{b \, (1)} G_{5}^{c \, (1)}  \rr \nonumber \\
&&U^{a}_{\mu} = Z_G \, g_3 \, f^{abc}  \, \rl \frac{1}{\sqrt{2}} \, G_{\mu}^{b \, (2)} G_{5}^{c \, (2)} \rr \label{auxvect2kk} \\
&&V^{a}_{\mu} = Z_G \, g_3 \, f^{abc}  \, \rl \frac{1}{\sqrt{2}} G_{\mu}^{b \, (1)} G_{5}^{c \, (2)} + \frac{1}{\sqrt{2}} G_{\mu}^{b \, (2)} G_{5}^{c \, (1)}  \rr \nonumber
\ea
which, once replaced into \eqref{auxvect} and eliminated, reproduce the Lagrangian \eqref{4g4dvect2kk}.

\bibliographystyle{JHEP}  
\bibliography{mued}   
\end{document}